%% file: main.tex
\documentclass[11pt, a4paper, oneside]{Thesis} 

\graphicspath{{Pictures/}} 

\usepackage[square,sort,compress,numbers]{natbib}
\usepackage{bigints}
\usepackage{amsmath,latexsym} 
\usepackage{subfigure}
\usepackage{float}
\usepackage[T1]{fontenc}
\usepackage{fancyhdr}
\title{\ttitle} 
\DeclareUnicodeCharacter{2212}{-}
\DeclareUnicodeCharacter{2212}{+}
\begin{document}

\setstretch{1.3} 

\fancyhead{} 
\rhead{\thepage} 
\lhead{} 

\input{variables}

\maketitle

\clearpage
\setstretch{1.3} 

\pagestyle{empty} 
\pagenumbering{gobble}

\addtocontents{toc}{\vspace{2em}} 
\frontmatter 
\Certificate
\Declaration
\begin{acknowledgements}
With heartfelt joy and sincere gratitude, I acknowledge the continuous guidance, support, and inspirational influences of many individuals without whom this journey to pursue my doctorate would not have been possible.

I express my profound gratitude to my supervisor, \textbf{Prof. Pradyumn Kumar Sahoo}, Professor, Department of Mathematics, BITS-Pilani, Hyderabad Campus, Hyderabad, Telangana. His unwavering dedication to research, expertise in executing research projects, keen attention to emerging advancements, and, above all, his significant contributions to the development of cosmology and inspiring young mathematicians to explore this field have been the primary motivation for my work and many others. His timely advice, meticulous scrutiny, scholarly guidance, and scientific approach have inspired me and contributed significantly to my research.

I sincerely thank my members of the Doctoral Advisory Committee (DAC) members, \textbf{Prof. Bivudutta Mishra}, and \textbf{Prof. Kota Venkata Ratnam}, for their valuable suggestions and constant encouragement to improve my research work.

I am privileged to thank HoD, the DRC convener, faculty members, my colleagues, and the Department of Mathematics staff for supporting this fantastic journey of my Ph.D. career.

I sincerely thank my coauthors, \textbf{Dr. Raja Solanki}, \textbf{Dr. Sanjay Mandal}, and \textbf{Dr. Moreshwar Tayde}, for their valuable suggestions, discussions, encouragement and collaborations.

I acknowledge \textbf{BITS-Pilani, Hyderabad Campus} for providing me with the necessary facilities, and the \textbf{University Grant Commission (UGC), India}, for providing a Research Fellowship (UGC Reference No. 191620024300) to carry out my research works.

I would like to thank my \textbf{colleagues} and \textbf{friends}, especially \textbf{Sunita}, and  \textbf{Kartik}, for their support, encouragement, and companionship throughout this journey.

Most importantly, I would like to thank my \textbf{mother, sister} and \textbf{family members} for their love, care, and support throughout my journey.

To my wife, \textbf{Shubhangi} - your patience, strength, and unconditional love have been my foundation. Through every late night, every setback, and every small victory, you stood by me with quiet courage and steadfast belief. I could not have done this without you.

And to my dearest son, \textbf{Virat} - your laughter, wonder, and innocent presence have been my light on the darkest days. You remind me daily of what truly matters. This achievement is as much yours as it is mine, and I hope it one day makes you proud.

\vspace{2.5 cm}
Jaybhaye Lakhan Valmik,\\
ID: 2020PHXF0477H.

\end{acknowledgements}

\begin{abstract}
Over the past century, GR has been the most successful theory of gravitation, yet its limitations become evident in the face of recent cosmological observations. In the last two decades, precision cosmology has revealed the accelerated expansion of the Universe. This phenomenon challenges the completeness of GR and the standard $\Lambda$CDM model, which explains this acceleration through the cosmological constant. However, the $\Lambda$CDM model faces unresolved issues, prompting the search for alternative explanations. Researchers have proposed various approaches, from dynamical DE models to large-scale modifications of gravity, aiming to provide a more comprehensive understanding of the late-time cosmic acceleration and address the persistent challenges of modern cosmology.

Chapter-\ref{Chapter1} explores our evolving understanding of the Universe, from ancient perspectives to modern advancements. It provides a foundation in GR, the standard cosmological model, and the standard $\Lambda$CDM model, addressing key features and challenges. It examines modified gravity theories to tackle modern cosmological issues, focusing on $f(R)$ gravity and its extensions, mainly the $f(R,L_m)$ gravity framework, which introduces curvature-matter coupling. This discussion focuses on empirical validations, including tests conducted within the solar system, the geodesic deviation equation, and the Raychaudhuri equation. Together, these elements enhance our understanding of spacetime structure and the behavior of particles in the context of modified gravity. Finally, the chapter discusses key Cosmological Observations that provide insights into the nature of the Universe and validate theoretical models.

In Chapter-\ref{Chapter2}, we explore the scenario of cosmic expansion within the framework of $f(R,L_m)$ gravity theory, focusing on a non-linear $f(R,L_m)$ model. We derive the equations of motion for the flat FLRW metric and find the precise solutions to the related field equations. By employing updated $H(z)$ datasets along with Pantheon datasets, we determine the optimal ranges for the model parameters. Furthermore, we examine the physical characteristics of the density and deceleration parameters. The analysis of the deceleration parameter reveals a shift from deceleration to acceleration in the expansion of the Universe. Additionally, we assess the stability of our cosmological solution against observational constraints by considering linear perturbations. Finally, our examination of the Om diagnostic parameter indicates that our model exhibits quintessence-like behavior.

In Chapter-\ref{Chapter3}, we introduce a parametrization of the Hubble parameter that is not tied to any particular cosmological model and utilize it to explore the Friedmann equations within the framework of the FLRW background. By employing an MCMC method, we estimate the model parameters through the analysis of a comprehensive dataset that includes CC, BAO, and Pantheon+SH0ES datasets. Our investigation of the deceleration parameter's trajectory reveals a transition from a decelerating to an accelerating phase in the evolution of the Universe. Furthermore, we examine the behavior of key cosmological variables such as the density parameter, pressure, EoS parameter, and energy conditions for two non-linear $f(R,L_m)$ models.

In Chapter-\ref{Chapter4}, we focus on addressing the late-time cosmic acceleration observed in the Universe through the framework of $f(R,L_m)$ gravity, incorporating an effective EoS that accounts for bulk viscosity. We derive the exact solution for our model, which is dominated by bulk viscous matter. Utilizing the combined CC + Pantheon+SH0ES datasets, we estimate the optimal values for the free parameters in our model. We also analyze the density parameter, pressure, and EoS parameter for a fluid exhibiting viscosity. Furthermore, we examine the statefinder parameters associated with the proposed $f(R,L_m)$ model. Our findings indicate that the evolutionary trajectory of this model falls within the quintessence region. Additionally, we apply the Om diagnostic test, which supports the quintessence behavior of our model. Finally, we verify the energy condition criteria, revealing that while there were violations of the SEC in the past, the NEC and DEC upheld the positivity requirements. Overall, our $f(R,L_m)$ cosmological model, which incorporates bulk viscosity effects, aligns well with recent observational data and offers a comprehensive understanding of the cosmic expansion phenomenon.

In Chapter-\ref{Chapter5}, we delve into the concept of matter bounce non-singular cosmology within the framework of $f(R,L_m)$ gravity. We derive the relevant Friedmann equations for the proposed models, set against the FLRW background. The discussion encompasses the influence of model parameters, including the bouncing scale factor, on key physical quantities such as the EoS parameter, pressure, and density parameters. Furthermore, we analyze the dynamical aspects of cosmographic parameters, including jerk, lerk, and snap. We also address the violation of the NEC and the SEC, which indicate the non-singular and accelerating nature of the models corresponding to the selected non-linear $f(R,L_m)$ functions. Finally, we investigate the adiabatic speed of sound to assess the feasibility of the proposed cosmological bouncing scenario.

Chapter-\ref{Chapter6} focuses on exploring the gravitational baryogenesis epoch within the framework of $f(R,L_m)$ gravity. In our analysis, we operate under the premise that the Universe is filled with DE and a perfect fluid, along with a non-zero baryon-to-entropy ratio during the radiation-dominated era. We limit our investigation to the gravitational baryogenesis scenario, highlighting the relevant model parameters that align with observed baryon-to-entropy ratio data. Our findings indicate that $f(R,L_m)$ gravity can play a significant and consistent role in facilitating gravitational baryogenesis.

In Chapter-\ref{Chapter7}, we offer a thoughtful summary of the outcomes presented in this thesis and explore potential future directions for research.
\end{abstract} 
\Dedicatory{\bf \begin{LARGE}
Dedicated to
\end{LARGE} 
\\
\vspace{3cm}
\bf \begin{LARGE}
 The loving memory of my father\\
 \end{LARGE}}



\lhead{\emph{Contents}} 
\tableofcontents 
\addtocontents{toc}{\vspace{1em}}
\lhead{\emph{List of Tables}}
 \listoftables 
\addtocontents{toc}{\vspace{1em}}
\lhead{\emph{List of Figures}}
\listoffigures 
\addtocontents{toc}{\vspace{1em}}



\lhead{\emph{List of symbols and Abbreviations}}
\listofsymbols{ll}{
\begin{tabular}{lcl}
$g_{\mu \nu}$&:& Metric tensor\\
$g$&:& Determinant of $g_{\mu \nu}$\\
${\Gamma}^{\alpha}_{\mu\nu}$&:& Christoffel symbol\\
$\nabla$&:&Covariant derivative\\
$R^\mu _{\nu\sigma \lambda}$ &:& Reimann tensor\\
$R_{\mu \nu}$ &:& Ricci tensor\\
$R$&:& Ricci scalar \\
$G_{\mu \nu}$ &:&Einstein tensor\\
$ T_{\mu \nu}$&:& Energy momentum tensor\\
$T $&:& Trace of energy momentum tensor\\ 
$z $&:& Redshift\\
$H $&:& Hubble parameter\\
$a$&:& Scale factor\\
$H_0$&:& Present value of the Hubble parameter\\
$\rho$&:& Density parameter\\
$p $&:&  Pressure\\
$\omega$&:& EoS parameter\\
$L_m$&:& Matter Lagrangian density\\
$\square$&:& d'Alembert operator\\
GR &:&General relativity\\
$\Lambda$CDM &:&$\Lambda$ Cold Dark Matter\\
FLRW &:&Friedmann-Lemaitre-Robertson-Walker\\
CC &:&Cosmic Chronometer\\
BAO &:&Baryon acoustic oscillations\\
SNeIa &:&Type Ia Supernovae\\ 
MCMC &:&Markov chain monte carlo\\
CMB &:&Cosmic microwave background\\
DM &:&Dark matter\\
DE &:&Dark energy\\
EoS &:&Equation of state\\
SEC &:&Strong Energy Condition\\
NEC &:&Null Energy Condition\\
DEC &:&Dominant Energy Condition\\
BA &:&Baryon asymmetry\\
CP &:&Charge parity\\
\end{tabular}
}

\addtocontents{toc}{\vspace{2em}}

%
%


\clearpage 





\mainmatter 

\pagestyle{fancy} 

\input{Chapters/Introduction}
\input{Chapters/Chapter2}
\input{Chapters/Chapter3}

\input{Chapters/Chapter4}
\input{Chapters/Chapter5}
\input{Chapters/Chapter6}
\input{Chapters/Conclusion}






\addtocontents{toc}{\vspace{2em}} 

\backmatter


\label{References}
\lhead{\emph{References}}
\input{References.tex}
\cleardoublepage
\pagestyle{fancy}

\label{Publications}
\lhead{\emph{List of Publications}}

\chapter{List of Publications}
\section*{Thesis Publications}
\begin{enumerate}

\item \textbf{Lakhan V. Jaybhaye}, R. Solanki, S. Mandal, P. K. Sahoo, \textit{Cosmology in $f(R,L_m)$ gravity}, \textcolor{blue}{Physics Letters B}, \textbf{831}, 137148 (2022).

\item \textbf{Lakhan V. Jaybhaye}, R. Solanki, P. K. Sahoo, \textit{Late time cosmic acceleration through parametrization of Hubble parameter in $f(R,L_m)$ gravity}, \textcolor{blue}{Physics of the Dark Universe}, \textbf{46}, 101639 (2024).

\item \textbf{Lakhan V. Jaybhaye}, R. Solanki, S. Mandal, P. K. Sahoo, \textit{Constraining viscous dark energy equation of state in $f(R,L_m)$) gravity}, \textcolor{blue}{Universe}, \textbf{9}, 163 (2023).

\item \textbf{Lakhan V. Jaybhaye}, R. Solanki, P. K. Sahoo, \textit{Bouncing cosmological models in $f(R,L_m)$ gravity}, \textcolor{blue}{Physica Scripta}, \textbf{99}, 065031 (2024).

\item \textbf{Lakhan V. Jaybhaye}, S. Bhattacharjee, P. K. Sahoo, \textit{Baryogenesis in $f(R,L_m)$ gravity}, \textcolor{blue}{Physics of the Dark Universe}, \textbf{40}, 101223 (2023).

\end{enumerate}
\section*{Other Publications}
\begin{enumerate}

\item \textbf{Lakhan V. Jaybhaye}, S. Mandal, P. K. Sahoo, \textit{Constraints on Energy Conditions in $f(R,L_m)$ Gravity}, \textcolor{blue}{International Journal of Geometric Methods in Modern Physics}, \textbf{19}, 2250050 (2021)

\item  R. Solanki, B. Patel, \textbf{Lakhan V. Jaybhaye}, P. K. Sahoo, \textit{Cosmic acceleration with bulk viscosity in an anisotropic $f(R,L_m)$ background}. \textcolor{blue}{Communications in Theoretical Physics}, \textbf{75(7)}, 075401 (2023). 

\item \textbf{Lakhan V. Jaybhaye}, M. Tayde, P. K. Sahoo, \textit{Wormhole solutions under the effect of dark matter in $f(R,L_m)$ gravity}, \textcolor{blue}{Communications in Theoretical Physics}, \textbf{76(5)}, 055402 (2024). 
 
\end{enumerate}

\cleardoublepage
\pagestyle{fancy}
\lhead{\emph{Paper presented at conferences}}
\chapter{Paper presented at conferences}
\label{Paper presented at conferences}
\begin{enumerate}
\item Presented research paper entitled “\textit{Cosmology in $f(R,L_m)$ gravity}” at the conference “\textbf{International Conference on Mathematical Sciences and Its Applications}” organized by the School of Mathematical Sciences, Swami Ramanand Teerth Marathwada University, Nanded, Maharashtra, during the period $28^{th}-30^{th}$ July, $2022$.
\item Presented research paper poster with flash talk entitled “\textit{Constraining viscous dark energy equation of state in $f(R, L_m)$ gravity}” at the conference “\textbf{32nd meeting of Indian Association for General Relativity and Gravitation (IAGRG32)}” organized by the Indian Institute of Science Education and Research, Kolkata, during the period $19^{th}-21^{th}$ December, $2022$.
\item Presented research paper entitled “\textit{Baryogenesis in $f(R, L_m)$ gravity}” at the conference “\textbf{Envisioning Mathematics Education in Line with NEP 2020}” organized by the Mahatma Gandhi Mahavidyalaya, Ahmedpur, Maharashtra, during the period $21^{th}$ December, $2023$.
\item Presented research paper entitled “\textit{Late time cosmic acceleration through parametrization of Hubble parameter in $f(R,L_m)$ gravity}” at the conference “\textbf{31st International Conference of International Academy of Physical Sciences}” organized by the Pt. Ravishankar Shukla University, Raipur, during the period $20^{th}-21^{th}$ December, $2024$.
\end{enumerate}

\cleardoublepage
\pagestyle{fancy}
\lhead{\emph{Biography}}

\chapter{Biography}

\section*{Brief Biography of the Candidate:}
\textbf{Mr. Jaybhaye Lakhan Valmik} received his Bachelor's degree in 2016 and his Master's degree in 2018 in Mathematics from Dayanand Science College, Latur, under Swami Ramanand Teerth Marathwada University, Nanded. He qualified for the Council for Scientific and Industrial Research (CSIR), the National Eligibility Test (NET) for the Lectureship (LS), as well as the Junior Research Fellowship (JRF) in 2019 with a 99.75 percentile (All India Rank-67). He has published eight research articles in renowned international journals during his Ph.D. research career. He has presented his research at many National and International conferences. 

\section*{Brief Biography of the Supervisor:}
\textbf{Prof. P. K. Sahoo} completed his Ph. D. from Sambalpur University, Odisha, India, in Jan 2004. He is currently a Professor in the Department of Mathematics at the Birla Institute of Technology and Science-Pilani, Hyderabad Campus, where he served as the Head of the Department from October 2020 to September 2024. He has organized several academic events, including the 89th Annual Conference of the Indian Mathematical Society in 2023. He has contributed significantly to BITS through five sponsored research projects funded by the University Grants Commission (UGC, 2012-2014), DAAD Research Internships in Science and Engineering (RISE) Worldwide (2018, 2019, 2023 and 2024), Council of Scientific and Industrial Research (CSIR) (2019–2022), National Board for Higher Mathematics (NBHM) (2022-2025) and Science and Engineering Research Board (SERB), Department of Science and Technology (DST) (2023–2026). In addition, he is an expert reviewer for Physical Science Projects under SERB, DST, and UGC research schemes, Government of India.

\textbf{Prof. Sahoo} has published more than 250 research articles in prestigious national and international journals. He has been an invited speaker at numerous international and national conferences and maintains research collaborations nationally and internationally. For five consecutive years, he has been ranked among the top 2\% of scientists worldwide in the field of Nuclear and Particle Physics, according to a Stanford University, USA survey. He is an expert reviewer and editorial board member for several renowned scientific journals and serves as a Ph.D. examiner at multiple universities. He is guiding 18 Ph.D. students as supervisor and co-supervisor (7 completed and 11 ongoing) and many M.Sc. theses. He has been honored with a visiting professor fellowship at Transilvania University of Brasov, Romania. He has received the Science Academics Summer Research Fellowship, UGC Visiting Fellowship, and several prestigious fellowships, including a Fellow of the Institute of Mathematics and its Applications (FIMA), London, Fellow of the Royal Astronomical Society (FRAS), London, and a foreign member of the Russian Gravitational Society. Additionally, He is a member of COST (CA21136), a project focused on addressing observational tensions in cosmology through systematics and fundamental physics. He also had the opportunity to visit the European Organization for Nuclear Research (CERN), Geneva, Switzerland, as a visiting scientist.

\printglossary
\end{document}

%% file: variables.tex
%

\thesistitle{Study of Curvature-Matter Coupling in Modified Gravity}
\documenttype{\textbf{THESIS}}
\supervisor{\textbf{Prof. Pradyumn Kumar Sahoo}}
\supervisorposition{\textbf{Professor}}
\supervisorinstitute{\textbf{BITS-Pilani, Hyderabad Campus}}
\examiner{}
\degree{Ph.D. Research Scholar}
\coursecode{\textbf{DOCTOR OF PHILOSOPHY}}
\coursename{THESIS}
\authors{\textbf{JAYBHAYE LAKHAN VALMIK}}
\IDNumber{2020PHXF0477H}
\addresses{}
\subject{}
\keywords{}
\university{\texorpdfstring{\href{http://www.bits-pilani.ac.in/} 
                {Birla Institute of Technology and Science, Pilani}} 
                {Birla Institute of Technology and Science, Pilani}}
\UNIVERSITY{\texorpdfstring{\href{http://www.bits-pilani.ac.in/} 
                {BIRLA INSTITUTE OF TECHNOLOGY AND SCIENCE, PILANI}} 
                {BIRLA INSTITUTE OF TECHNOLOGY AND SCIENCE, PILANI}}



\department{\texorpdfstring{\href{http://www.bits-pilani.ac.in/pilani/Mathematics/Mathematics} 
                {Mathematics}} 
                {Mathematics}}
\DEPARTMENT{\texorpdfstring{\href{http://www.bits-pilani.ac.in/pilani/Mathematics/Mathematics} 
                {Mathematics}} 
                {Mathematics}}
\group{\texorpdfstring{\href{Research Group Web Site URL Here (include http://)}
                {Research Group Name}} 
                {Research Group Name}}
\GROUP{\texorpdfstring{\href{Research Group Web Site URL Here (include http://)}
                {RESEARCH GROUP NAME (IN BLOCK CAPITALS)}}
                {RESEARCH GROUP NAME (IN BLOCK CAPITALS)}}
\faculty{\texorpdfstring{\href{Faculty Web Site URL Here (include http://)}
                {Faculty Name}}
                {Faculty Name}}
\FACULTY{\texorpdfstring{\href{Faculty Web Site URL Here (include http://)}
                {FACULTY NAME (IN BLOCK CAPITALS)}}
                {FACULTY NAME (IN BLOCK CAPITALS)}}

%% file: Chapters/Introduction.tex

\chapter{A brief overview of the preliminaries} 
\label{Chapter1}

\lhead{Chapter 1. \emph{A brief overview of the preliminaries}} 

\clearpage
\pagebreak
The study of the Universe has covered a broad spectrum of subjects, frequently crossing paths with religious beliefs. Although some elements of the Universe may be unreachable to scientific inspection, they can still be explored using alternative philosophical techniques, such as logical arguments. According to the platonic Universe, the earth is stationary at the center, surrounded by perfect circles of moon, sun, planets, and fixed stars in a particular order determined by the Demiurge's design. These complex movements repeat themselves periodically. Nicolaus Copernicus ($1473-1543$) introduced a revolutionary model of the Universe, placing the Sun at its center instead of the Earth, challenging the ancient Greek two-sphere system. He argued that the Universe became more conceptually straightforward and elegant when the earth was considered to orbit the Sun. The prevailing model of the Universe, known as the hot Big Bang theory \citep{lem}, asserts that the Universe began in an extremely hot and dense state and has since expanded and cooled over time. This expansion continues even today. The field of cosmology delves into the complete expanse of space, time, and every event that takes place within it. The study of the cosmos, or the Universe as a whole, is known as cosmology. The structures in the Universe are incredibly complex and exist on various scales: stars form galaxies, planets orbit them, clusters form within galaxies, and clusters are part of larger superclusters.

The field of modern cosmology is deeply rooted in the framework of General Relativity (GR), a theory that has revolutionized our understanding of gravitational interactions on cosmic scales. Proposed by Albert Einstein in $1915$, GR has withstanded the test of time, offering precise explanations for a wide range of phenomena, from the motion of planets to the enigmatic behavior of black holes and the intricate large-scale structure of the Universe \citep{Einstein:1915ca}. At its core, GR is governed by the Einstein field equations, which elegantly describe how matter and energy shape the curvature of spacetime, giving rise to the gravitational effects we observe \citep{Misner:1973prb}. One of the most profound applications of GR lies in cosmology, where it forms the backbone of the standard model of evolution of the Universe. The Friedmann-Lemaitre-Robertson-Walker (FLRW) metric, derived from Einstein’s equations under the assumptions of homogeneity and isotropy, provides a mathematical description of the large-scale structure of the Universe \citep{Friedmann1922,Lemaitre1927}. This metric is the cornerstone of the $\Lambda$CDM model, the prevailing cosmological paradigm that incorporates dark energy ($\Lambda$) and cold dark matter (CDM) to account for the observed accelerated expansion of the universe and the formation of cosmic structures \citep{Peebles1993,Riess,Perlmutter}.

Despite the remarkable success of the $\Lambda$CDM model in explaining key observational data such as cosmic microwave background (CMB) radiation, baryon acoustic oscillations (BAO), and Type Ia supernovae (SNeIa), it is not without its challenges. The fundamental nature of dark matter (DM) and dark energy (DE) remains elusive, and persistent discrepancies in measurements of the Hubble parameter, known as the Hubble tension, suggest potential gaps in our understanding of the expansion history of the Universe \citep{planck_collaboration/2020,rie}. These unresolved issues have stimulated significant interest in exploring alternative theories of gravity, including modified gravity models such as $f(R)$ gravity, curvature-matter coupling theories, and other extensions of GR \citep{Clifton2012,Nojiri2011}. Modified gravity theories aim to address the limitations of GR by introducing additional degrees of freedom or modifying the Einstein-Hilbert action. These theories offer alternative explanations for cosmic acceleration without necessarily invoking DE \citep{Capozziello2011}. Over the years, various modified gravity models have been rigorously tested against observational data and theoretical constraints, including Solar System tests, geodesic deviation equations, and the Raychaudhuri equations, which describe the evolution of spacetime congruences \citep{Will2014,Wald1984}. A crucial aspect of evaluating cosmological models lies in their ability to align with observational data. This requires sophisticated statistical techniques, such as $\chi^2$ minimization and Markov Chain Monte Carlo (MCMC) methods, to constrain model parameters and compare competing theories \citep{Hobson2006}. Observations from cosmic chronometers (CC), BAO, and SNeIa provide stringent tests for these models, helping to refine our understanding of the dynamics of the Universe \citep{Weinberg2013,Scolnic/2018}. 

This thesis seeks to provide a thorough and original exploration of GR and its extensions in the context of cosmology. We begin with a detailed overview of GR, introducing essential mathematical tools such as the metric tensor, Christoffel symbols, geodesic equations, and curvature tensors \citep{Carroll2004}. Building on this foundation, we delve into the standard cosmological model, critically examining its successes and limitations \citep{Weinberg2008}. Finally, we investigate modified theories of gravity, with a particular focus on curvature-matter coupling in the framework of $f(R, Lm)$ gravity \citep{hark/10}. By integrating theoretical insights with observational constraints, this work aims to contribute to the ongoing quest to deepen our understanding of gravitational interactions and the evolution of the Universe, offering fresh perspectives on some of the most pressing questions in modern cosmology.

\section{General Relativity}
\justifying
Albert Einstein's General Theory of Relativity is widely recognized as one of the most significant discoveries. It transforms our understanding of gravity by using the spacetime metric $g_{\mu \nu}$. Gravity is described by the curvature of spacetime, where physical processes modify this geometry, moving away from the concept of fixed space and time. GR introduces a non-Euclidean, four-dimensional spacetime, making traditional Euclidean geometry obsolete. In $1917$, Einstein proposed a cosmological model for a static and uniform universe \citep{ein}. He eventually included a cosmic constant to solve the gravitational instability, but small-scale perturbations continued. Willem de Sitter expanded on this idea by solving Einstein's equations for a Universe without matter for the first time \citep{des,des1}. GR defines physical rules in any reference frame employing a thorough mathematical framework that includes scalars, vectors, tensors, and differential operators. Unlike Newton's theory, GR explains strong gravitational fields around objects such as neutron stars and black holes. Although gravitational redshift has verified GR, understanding DM, DE, and singularities such as black holes and the Big Bang remains difficult. Despite these limitations, GR is still the most successful theory for defining gravity as a geometric characteristic of curved spacetime, particularly for explaining the large-scale evolution of the Universe. 

\subsection{Metric tensor}
A differential manifold is defined as a type of Riemannian space, where each point on the manifold is associated with a tensor \citep{dub}
\begin{equation}\label{1}
g_{\mu\nu}(x)=g_{\mu\nu}(x^1,x^2,x^3,...,x^n).
\end{equation}
This tensor is both symmetric and nondegenerate, with two covariant indices, and is referred to as the metric tensor. Thus, we have
\begin{equation}\label{2}
g_{\mu\nu}=g_{\nu\mu},\,\,g=\text{det}|g_{\mu\nu}|.
\end{equation}
The primary conditions for the function $g_{\mu\nu}$ are that it must be continuous and its derivatives with respect to all coordinates $(x^1,x^2,x^3,\dots,x^n)$ must also be continuous. This metric tensor allows the formulation of an invariant second-order differential equation in the Riemannian space, written as
\begin{equation}\label{3}
ds^2=g_{\mu\nu}dx^{\mu}dx^{\nu}.
\end{equation}

\subsection{Christoffel symbol}
The curvature of a manifold is influenced by the metric that defines its geometry, although determining the curvature associated with a specific metric is not straightforward. In mathematics, it is often crucial to be careful when converting our intuitive insights into a precise mathematical structure. The various manifestations of curvature are linked to a connection that relates vectors within the tangent spaces of nearby points. From the metric, one can derive a unique connection, represented by an entity known as the Christoffel symbol, which can be written as
\begin{equation}\label{4}
{\Gamma}^{\alpha}_{\mu\nu}=\frac{1}{2}g^{\alpha \lambda}\left(\partial_{\mu}g_{\nu\lambda}+\partial_{\nu}g_{\lambda \mu}-\partial_{\lambda}g_{\mu \nu}\right).
\end{equation}
The notation presents ${\Gamma}^{\alpha}_{\mu\nu}$ as if it were a tensor, yet it is not; hence, we refer to it as an object or symbol.

\subsection{Covariant derivative}
In flat space using inertial coordinates, the operator $\partial_{\mu}$ transforms a $(k,l)$ tensor field into a $(k,l+1)$ tensor field. It operates linearly on its inputs and adheres to the Leibniz rule when applied to tensor products. These characteristics remain valid even in broader contexts, which we will examine next. However, the behavior of the partial derivative changes with the coordinate system employed. To eliminate this dependence, we propose the introduction of a new operator known as the covariant derivative $\nabla$, which will perform similarly to the partial derivative, but will not be tied to particular coordinates. We begin by stipulating that $\nabla$ should also convert a $(k,l)$ tensor field into a $(k,l+1)$ tensor field and that it fulfills the following two essential properties:
\begin{itemize}
    \item Linearity: $\nabla(T+S)=\nabla T+\nabla S$.
    \item Leibniz rule: $\nabla(T*S)=(\nabla T)*S+(\nabla S)*T$.
\end{itemize}
If the operator $\nabla$ is expected to follow the Leibniz rule, it can be consistently represented as combining a partial derivative and a linear transformation. In other words, to compute the covariant derivative, we start with the partial derivative and then apply a correction to ensure that the result remains covariant.
Examine the consequences of the covariant derivative of a vector $V^\nu$. For every direction indicated by $\mu$, the covariant derivative $\nabla _\mu$ includes the partial derivative $\partial _\mu$ along with a correction defined by a set of $n$ matrices $(\Gamma_\mu)^\alpha _\nu$. In practice, parentheses are usually omitted, and these matrices, called connection coefficients, are typically notated with varying index positions as ${\Gamma}^{\alpha}_{\mu\nu}$. Thus, we arrive at the following expression
\begin{equation}\label{5}
\nabla_{\mu} W^\nu = \partial_{\mu}W^\nu +{\Gamma}^{\nu}_{\mu\lambda} W^\nu.
\end{equation}

\subsection{Parallel transport}
Parallel transport is the transport of a vector along a trajectory without changing it. A connection is necessary to properly define parallel transport. Our intuitive handling of vectors in flat space implicitly utilizes the Christoffel connection. The key distinction between flat and curved spaces is that in curved spaces, the outcome of parallel transporting a vector from one point to another is influenced by the particular path taken. Let $x^\mu (\lambda)$ denote a trajectory with a tangent vector $\frac{dx^\mu}{d\lambda}$, and let $W^\mu$ signify the vector being moved. We state that $W^\mu$ is being parallel transported along this trajectory if
\begin{equation}\label{6}
\frac{dx^\mu}{d\lambda}.\nabla_{\mu} W^\nu = 0.
\end{equation}

\subsection{Geodesics}
A geodesic is a mathematical idea that generalizes the concept of straight-line movement in a flat space to travel through curved space. When a curve joins two points, the length of the arc can determine the distance between them along that curve. The curves that minimize this distance are particularly significant. A parameterized curve $x^\mu(\lambda)$ is classified as a geodesic if it meets the following condition
\begin{equation}\label{7}
\frac{d^2 x^\mu}{d\lambda^2}+ {\Gamma}^{\mu}_{\delta\sigma} \frac{dx^\delta}{d\lambda}\frac{dx^\sigma}{d\lambda} = 0.
\end{equation}

\subsection{Riemann tensor}
Ultimately, the technical representation of curvature is contained in the Riemann tensor, which is a $(1,3)$ tensor derived from the connection by
\begin{equation}\label{8}
R^\mu _{\nu\sigma \lambda}=\partial_\sigma {\Gamma}^{\mu}_{\lambda\nu} - \partial_\lambda {\Gamma}^{\mu}_{\sigma\nu} + {\Gamma}^{\mu}_{\sigma\delta} {\Gamma}^{\delta}_{\lambda\nu} - {\Gamma}^{\mu}_{\lambda\delta} {\Gamma}^{\delta}_{\sigma\nu}.
\end{equation}

\subsection{Ricci tensor}
The Ricci tensor is a $(0, 2)$ tensor that results from contracting the Riemann tensor. That is,
\begin{equation}\label{9}
R_{\mu \nu}=R^\sigma _{\mu\sigma \nu}.
\end{equation}
The Ricci tensor related to the Christoffel connection exhibits symmetry, which means \( R_{\mu \nu} = R_{\nu \mu} \). The Ricci scalar is defined as the trace of the Ricci tensor,
\begin{equation}\label{10}
R=R^\mu _\mu =g_{\mu \nu} R_{\mu \nu}.
\end{equation}

\subsection{Einstein tensor}
The Ricci tensor is determined by the metric tensor, allowing the Einstein tensor to be defined purely in terms of the metric tensor as
\begin{equation}\label{11}
G_{\mu \nu} =  R_{\mu \nu} −\frac{1}{2} g_{\mu \nu}R,
\end{equation}
which is a symmetric tensor central to GR, characterizing the gravitational field. It adheres to the equation $\nabla ^\mu G_{\mu \nu}=0$. This tensor describes a vital connection between the geometric nature of spacetime and gravitational characteristics. Up to this point, we have examined essential mathematical concepts and methods that underlie the formal framework of the GR.

\subsection{Energy-momentum tensor}
The energy-momentum tensor, commonly represented as $ T_{\mu \nu} $, characterizes the attributes of a matter source by detailing its energy, momentum, and fluxes across spacetime. Being a second-rank tensor, it illustrates quantities such as energy density, momentum density, and energy flux within a given volume. This $4\times 4$ matrix enables us to represent the physical laws governing the source of matter in a mathematical context, each component providing specific information on the distribution and movement of energy and momentum.
\begin{itemize}
\item $T_{00}$= energy density
\item $T_{0 \mu}=$ density of $\mu$th component of momentum ($\mu=1,\,2,\,3$)
\item $T_{\nu 0}=$ energy flux in the $\nu$th direction ($\nu=1,\,2,\,3$)
\item $T_{\mu\nu}=$ flux in the $\mu$th direction of $\nu$th component of momentum ($\mu,\,\nu=1,\,2,\,3$).
\end{itemize}
The energy-momentum tensor is locally conserved, expressed by $\nabla^{\mu}T_{\mu\nu}=0$, a condition with several interpretations. For any type of matter, the Einstein equation satisfies $\nabla^{\mu}G_{\mu\nu}=0$, implying the conservation of energy and momentum. Furthermore, $\nabla^{\mu}T_{\mu\nu}=0$ can be derived directly from the Einstein equation.

\subsection{Energy conditions}
Energy conditions are widely recognized as fundamental elements in the study of cosmology. Before delve into specific cosmological models, it is essential to explore the broader concept of energy conditions. Consider an observer moving with a velocity $u^{\mu}$, who measures the energy-momentum density as $T_{\mu\nu}u^{\mu}$. Furthermore, the observed energy density is given by $T_{\mu\nu}u^{\mu}u^{\nu}$. 

Among these conditions, the Null Energy Condition (NEC) asserts that
\begin{equation}\label{27EC}
T_{\mu\nu}u^{\mu}u^{\nu}\geq 0.
\end{equation}
For any null vector field $u^{\mu}$, the NEC holds significant importance. In the context of an ideal fluid, the NEC simplifies to the inequality
\begin{equation}\label{28EC}
\rho + p \geq 0.
\end{equation}
This condition is crucial in cosmology, as it provides insights into whether the Universe experiences inflation, super-inflation, or whether it will form a singularity or exhibit a bounce solution. Additionally, the NEC acts as a boundary in cosmological studies, its violation could lead to the breakdown of numerous physical laws. Although the NEC is considered relatively weak, it can be combined with the Weak Energy Condition (WEC) to strengthen its implications in theoretical frameworks.

The WEC states that for any time-like vector $u^{\mu}$ (representing an observer), the inequality $T_{\mu\nu}u^{\mu}u^{\nu} \geq 0$ must hold. In the context of an ideal fluid, the WEC simplifies to $\rho + p \geq 0$, and it also requires that the energy density must be non-negative, i.e., $\rho \geq 0$. This latter condition strengthens the WEC, as it rules out certain intriguing geometries in string theory, such as anti-de Sitter (AdS) spacetime, which is regarded as one of the most well-understood examples in the study of quantum gravity.

The Dominant Energy Condition (DEC) generalizes the WEC by incorporating its principles and adding an additional constraint: $\rho \geq |p|$. This further refines the conditions under which energy and momentum are distributed in physical systems, ensuring a more stringent framework for cosmological and theoretical analyses.

The Strong Energy Condition (SEC) states that
\begin{equation}\label{29EC}
\left(T_{\mu\nu}-\frac{1}{2}g_{\mu\nu}T\right)u^{\mu}u^{\nu}\geq 0.
\end{equation}
For any time-like vector $u^{\mu}$, the SEC applicable to an ideal fluid distribution is expressed as 
\begin{equation}\label{30EC}
\rho+3p\geq 0.
\end{equation}
This condition can be derived from Einstein’s equations and is also represented by $R_{\mu\nu}u^{\mu}u^{\nu} \geq 0$. The SEC suggests that gravity acts as an attractive force. However, recent evidence shows that the Universe is currently experiencing accelerated expansion, a phenomenon confirmed by observations of the CMB \citep{Caldwell}. Although the implications of DE appear to challenge the SEC, it is essential to recognize that the SEC remains a robust principle regarding gravitational interactions.

\subsection{Einstein field equation}
The fascinating aspect of GR is that spacetime is curved and dynamic. In simpler terms, the movement of matter is influenced by the curvature of space, and simultaneously, matter generates curvature within that space. The source of spacetime can be expressed through Einstein equation,
\begin{equation}\label{12}
G_{\mu\nu}\equiv R_{\mu\nu}-\frac{1}{2}g_{\mu\nu}R=8\pi G T_{\mu\nu}.
\end{equation}
It is important to note that the Einstein equation can be derived through the variation principle by varying the Einstein-Hilbert action,
\begin{equation}\label{13}
S=\frac{M_P^2}{2}\int d^4x\sqrt{-g} R+\int d^4x\sqrt{-g} L_m.
\end{equation} 
The reduced Planck mass is $M_P\equiv 1/\sqrt{8\pi G}$. In the action presented, the initial term represents the gravitational component, while the subsequent term signifies the matter component.

\section{Standard cosmological model}
\justifying
This section outlines the standard cosmological model, covering key concepts like the FLRW metric, the redshift, the Hubble parameter, and the Friedmann equations. We also discuss the strengths and limitations of the $\Lambda$CDM model. For a deeper exploration of cosmology, refer to Weinberg \citep{wei}.

\subsection{FLRW metric}
To describe the real world, we need to go beyond the idealized Copernican principle, which assumes uniformity throughout all space and time and instead embraces a more flexible framework. It appears that a straightforward proposal that fits with observations is that the Universe is spatially homogeneous and isotropic while also evolving. The scalar curvature must possess a finite value at every point in a Universe exhibiting such homogeneity and isotropy. Therefore, in the early $1920$s, Friedmann and Lemaitre anticipated, while Robertson \citep{robert} and Walker \citep{walker} rigorously confirmed, that the metric capable of describing this type of Universe in spherical coordinates $(r, \theta, \phi)$ is as follows
\begin{equation}\label{14}
ds^2=-c^2dt^2+a^2(t)\left[\frac{dr^2}{1-k r^2}+r^2(d\theta^2+sin^2\theta d\phi^2)\right].
\end{equation}
This particular metric is known as the FLRW metric. In this context, $a(t)$ denotes the scale factor, while curvature $k$ is a constant with dimensions of $1/\text{length}^2$. It can be either negative, positive, or zero. The FLRW metric defines the spacetime interval between any two distinct events in the Universe, making it ideal for a comoving coordinate system. When $k$ is set to zero, Eq. \eqref{14} reduces to
\begin{equation}\label{15}
ds^2=-c^2dt^2+a^2(t)\left[dr^2+r^2(d\theta^2+sin^2\theta d\phi^2)\right].
\end{equation}

\subsection{Redshift}
Almost everything in the Universe seems to be moving away from us, and this movement seems to accelerate as the distance increases, which is a crucial element in the study of cosmology. As a photon moves through the expanding Universe, it becomes associated with the expansion of space itself. This stretch in space elongates the wavelength of photons, resulting in a loss of energy over time. This diminishing energy gradually transitions from a high-energy blue color to a low-energy red hue for photons that originated in the visible spectrum. This transition, known as redshift $z$, explains why galaxies appear to be receding and why their motion away from us is described by the term redshift, which is defined by
\begin{equation}\label{16}
z = \frac{\lambda_{ob} - \lambda_{em}}{\lambda_{em}}.
\end{equation}
To express redshift in terms of observed and emitted wavelengths, we consider the relationship between these wavelengths and the scale factor of the Universe. Let $\lambda_{ob}$ and $\lambda_{em}$ represent the observed and emitted wavelengths, respectively. According to the Doppler effect, the relative change in wavelength due to redshift can be expressed as
\begin{equation}\label{17}
\frac{\lambda_{ob} - \lambda_{em}}{\lambda_{em}} = \frac{dv}{c}.
\end{equation}

Here, $v$ is the velocity of the recession and $c$ is the speed of light. Since $\lambda_{ob} > \lambda_{em}$ light appears redshifted. In a cosmological context, the time it takes for light to travel from the point of emission to the observer can be described by $dt=\frac{dr}{c}$. This leads to
\begin{equation}\label{18}
\frac{\lambda_{ob} - \lambda_{em}}{\lambda_{em}} = \frac{\dot{a}}{a} \frac{dr}{c} = \frac{\dot{a}}{a} dt = \frac{da}{a},
\end{equation}
where $a$ is the scale factor, and $\dot{a}$ is its rate of change. Consequently, the wavelength $\lambda$ is proportional to the scale factor $a$, meaning that as the Universe expands, so does the photon's wavelength. This relationship gives the redshift $z$ in terms of the scale factor,
\begin{equation}\label{19}
1+z = \frac{\lambda_{ob}}{\lambda_{em}} = \frac{a(t_{ob})}{a(t_{em})},
\end{equation}
where $a(t_{ob})$ and $a(t_{em})$ are the values of the scale factor at the times of observation and emission, respectively.

\subsection{Hubble parameter}
The rate at which distance changes, an essential aspect of cosmic expansion, can be described by the equation $\vec{r} = a(t)\,\vec{x}$. By differentiating with respect to time, we obtain
\begin{equation}\label{20}
\dot{\vec{r}} = \dot{a(t)} \vec{x} = \dot{a(t)} \frac{\vec{r}}{a(t)},
\end{equation}
resulting in the formula
\begin{equation}\label{21}
\dot{\vec{r}}= \frac{\dot{a(t)}}{a(t)} \vec{r} = H(t)\,\vec{r}.
\end{equation}
In this context, the dot signifies the differentiation of time, while $H(t)$, known as the Hubble parameter, describes the rate at which the Universe expands at a given time. The value of the Hubble parameter can be expressed as
\begin{equation}\label{22}
H(t) = \frac{\dot{a(t)}}{a(t)}.
\end{equation}
This equation demonstrates how the rate of expansion of the Universe, represented by $ H(t) $, evolves with changes in the scale factor $ a(t) $. At the current moment, this expression simplifies to the Hubble law
\begin{equation}\label{23}
\vec{v} = H_{0}\,\vec{r},
\end{equation}
where  $H_0$ denotes the current Hubble constant, and $\vec{v}$ indicates the speed at which distant objects move away from the observer. Edwin Hubble's 1929 discovery revealed that the speed of a galaxy recession is directly related to its distance from the observer, which provided the initial observational support for the expanding Universe. This concept remains fundamental to comprehending the formation and development of the Universe.

\subsection{Friedmann equations}
To describe the Universe, we treat it as homogeneous and isotropic, indicating that it looks the same in every direction and location. Additionally, we presume that it consists of a perfect fluid distribution of matter, characterized by an idealized form of matter with uniform properties. The energy-momentum tensor, which contains details about the density parameter $\rho$, pressure $p$, is represented as follows
\begin{equation}\label{24}
T_{\mu \nu} = \left(\rho + p \right) u_{\mu} u_{\nu} + p g_{\mu \nu},
\end{equation}
where $u_{\mu} = (1,0,0,0)$ indicates the 4-velocity vector of the fluid. This tensor includes stresses and provides information about the density parameter and flow of matter. Using the FLRW metric \eqref{14}, Einstein field Eq. \eqref{12}, and the energy-momentum tensor for a perfect fluid \eqref{24}, we derive the Friedmann equations, which govern the dynamics of an expanding Universe as
\begin{equation}\label{25}
3H^2+\frac{3\kappa c^2}{a^2}=8\pi G \rho(t),
\end{equation}
\begin{equation}\label{26}
2\dot{H}+3H^2+\frac{\kappa c^2}{a^2}=-\frac{8 \pi G}{c^2}p(t).
\end{equation}
Moreover, we can remember that the energy-momentum tensor satisfies the conservation equation $\nabla^{\mu} T_{\mu \nu} = 0$, which results in 
\begin{equation}\label{27}
\dot{\rho} + 3\left(\rho + p\right) \frac{\dot{a}}{a} = 0.
\end{equation}
Since the conservation law does not serve as an independent equation among the unknown variables, it is essential to introduce a closure relation to solve the system, namely an equation of state (EoS) that links the pressure of the barotropic fluid with its density for each component. The typical assumption is to consider a linear relationship 
\begin{equation}\label{28}
p = \omega \rho,
\end{equation}
where $\omega$ is a constant parameter that does not vary with time. 

The EoS parameter determines the relationship between pressure and the density parameter. It helps to classify the expansion characteristics of the Universe and distinguish between accelerated and deceleration phases. This parameter enables us to recognize various cosmic epochs on the basis of distinct values.

\begin{itemize}
\item The value of the EoS parameter $\omega=1$ signifies a stiff fluid.
\item When $\omega=\frac{1}{3}$, the model indicates a radiation-dominated phase.
\item The value of the EoS parameter $\omega=0$ characterizes a matter-dominated phase.
\item During the current rapid expansion of the Universe, the quintessence phase is described by the range $-1 < \omega < -\frac{1}{3}$.
\item The cosmological constant, represented by the value $\omega=-1$, forms the basis of the $\Lambda$CDM model.
\item The Universe enters the phantom energy era if $\omega < -1$.
\end{itemize}

Using this EoS \eqref{28} in the conservation Eq. \eqref{27} yields
\begin{equation}\label{29}
\rho \propto a^{-3(1+\omega)}.
\end{equation}
In the case of non-relativistic, pressureless ($p=0$) matter is commonly known as dust or cold matter. The EoS is $\omega =0$, indicating a Universe dominated by matter where matter primarily contributes to the density parameter as 
\begin{equation}\label{30}
\rho_{m}  = \rho_{m_{0}} a^{-3}(t).
\end{equation}
In a radiation-dominated Universe, where most of the density parameter consists of radiation, including photons and neutrinos, the EOS can be expressed as $\omega = \ \frac {1}{3}$. Under this condition, the density parameter as
\begin{equation}\label{31}
\rho_{r}  = \rho_{r_{0}} a^{-4}(t).
\end{equation}
In a vacuum-dominated Universe, where the density parameter remains constant over time, the EoS is characterized by $\omega = -1$, representing vacuum energy or the cosmological constant and defined by
\begin{equation}\label{32}
\rho_{\Lambda} = \rho_{\Lambda_{0}}.
\end{equation}
Here, $\rho_{m_{0}}$, $\rho_{r_{0}}$, and $\rho_{\Lambda_{0}}$ denote the corresponding densities at time $t=t_{0}$. The evolution of the Universe is typically divided into four distinct phases. The first is the inflationary era, which occurred immediately after the Big Bang and caused rapid expansion. This is followed by the radiation era, which represents the early stage when radiation predominates, the matter-dominated era during which cosmic structures begin to form, and the latest phase, where the cosmological constant is the main component.

\subsection{\texorpdfstring{$\Lambda$}{Lambda}CDM model}
Einstein's initial model of GR suggested that the Universe could not remain static, requiring it to either expand or contract. However, believing in a static Universe, he added $\Lambda$ as an antigravity term in his field equations to prevent expansion. Willem de Sitter solved Einstein's equations for the Universe without matter in the same year, demonstrating that $\Lambda$ could also drive cosmic inflation. Later, Alexander Friedmann recognized that a static solution was unstable and proposed an expanding Universe model. This idea gained observational support in 1929 when Edwin Hubble discovered the expansion of the Universe by studying galaxies. After abandoning $\Lambda$, Einstein reportedly referred to it as his greatest blunder. Nevertheless, the concept of the cosmological constant persisted in scientific discourse and modern cosmology. It has been revived to represent DE, a mysterious form of energy believed to be driving the current accelerated expansion of the Universe. In the $\Lambda$CDM model, which provides a framework to describe cosmic evolution, $\Lambda$ is incorporated as a form of DE with a constant negative EoS, $\omega_\Lambda=-1$. This model uses the Einstein-Hilbert action from which the field equations can be derived. The action for GR with $\Lambda$ is given by
\begin{equation}\label{33}
S = \int d^{4}x \, \sqrt{-g} \left[\frac{1}{2\kappa}(R-2\Lambda) + L_{m}\right],
\end{equation}
where $\kappa = \frac{8\pi G}{c^4}$. The Einstein field equations can be obtained by vanishing the variation of the action with respect to metric tensor $g_{\mu \nu}$, which is 
\begin{equation}\label{34}
R_{\mu\nu}-\frac{1}{2}g_{\mu\nu}R + \Lambda g_{\mu \nu} = \kappa\,T_{\mu \nu}.
\end{equation}
In a Universe described by the FLRW metric, these equations yield the Friedmann equations as
\begin{equation}\label{35}
H^{2} = \frac{8 \pi G}{3} \rho - \frac{\kappa}{a^{2}} + \frac{\Lambda}{3},
\end{equation}
\begin{equation}\label{36}
2 \dot{H} + 3 H^{2} =-8 \pi G p - \frac{\kappa}{a^{2}} + \Lambda.
\end{equation}
Merging Eqs. \eqref{35} and \eqref{36} results in
\begin{equation}\label{37}
\frac{\ddot{a}}{a} = -\frac{4 \pi G}{3} \left(\rho+3p\right) + \frac{\Lambda}{3},
\end{equation}
describes the conditions for cosmic acceleration $\ddot{a}>0$ achievable when $\Lambda $ dominates the equation, while the Universe is decelerating for $\ddot{a}<0$. Thus, $\Lambda$ serves as an independent energy component in $\Lambda$CDM cosmology, supplementing traditional matter-energy contributions and providing the simplest explanation for the observed accelerated expansion of the Universe.

\subsection{Problems with \texorpdfstring{$\Lambda$}{Lambda}CDM}
The $\Lambda$CDM model, which serves as the foundation of modern cosmology, offers a strong framework for following the evolution and composition of the Universe. However, various unresolved issues call into question its completeness and precision. These issues encompass the cosmological constant problem, the horizon problem, and the cosmological coincidence, each posing significant inquiries regarding the nature of DE and the organized structure of the Universe. Furthermore, the mysterious characteristics of DM and the ongoing observational discrepancies such as the $H_{0}$ tension and $\sigma_{8}$ tension underscore the need for a more profound theoretical basis or possible extensions to the model.
The Horizon problem highlights the unexpected uniformity of the Universe across vast distances \citep{Tsujikawa/2013}. Areas that appear consistent are causally separated according to the standard $\Lambda$CDM model, leading to inquiries about how their starting conditions might have been aligned. Cosmic inflation, which refers to a phase of rapid expansion in the early Universe, is effectively simplified by eliminating irregularities and guaranteeing the observed isotropy of the CMB \citep{Caldwell}.
Another central issue is the cosmological constant problem \citep{Joyce/2015}, which highlights an exceptional discrepancy between theoretical predictions and experimental observations. Calculating the vacuum energy from quantum field theory gives a $\Lambda$ inconsistent with observations. The discrepancy is of the order of $10^{120}$. This difference of nearly 120 orders of magnitude highlights a significant gap in understanding the relationship between quantum field theory and gravity.
The DM \citep{Astesiano/2021,Katsuragawa/2017,Zaregonbadi/2016} is a crucial part of the $\Lambda$CDM model, yet it is still an elusive aspect of the Universe. It is not composed of baryonic matter and interacts with other forms of matter only via gravitational influences, as suggested by observations such as galactic rotation curves and the formation of large-scale structures. Although theoretically significant, attempts to detect DM particles have been unsuccessful, leaving its nature uncertain and its existence inferred exclusively from indirect evidence.
The cosmological coincidence issue introduces additional complexity \citep{Velten/2014}. Observations show that the present densities of matter and DE are remarkably similar, even though their evolutionary paths are vastly different. This unusual equilibrium seems coincidental and prompts more profound inquiries into the fundamental processes that regulate the energy makeup of the Universe during this phase of accelerated expansion.
The discrepancies observed between different datasets add complexity to the cosmological picture. The $H_0$ tension \citep{Valentino/2021a} highlights a significant difference between local measurements of the Hubble constant ($H_0$) and the values derived from CMB observations within the $\Lambda$CDM model, with the current difference approximately at $4.4\sigma$. Similarly, the $\sigma_{8}$ tension \citep{Valentino/2021b} concerns inconsistencies in the rate at which cosmic structures develop. Although estimates from the Planck collaboration suggest $S_{8}=\sigma_{8}\sqrt{\Omega_{m}/0.3} = 0.834 \pm 0.016$, the KIDS-450 survey \citep{Joudaki/2017} finds $S_{8}= 0.745 \pm 0.039$, leading to a discrepancy of $2\sigma$. 
These limitations have prompted the exploration of other interpretations. The following sections will review several theories beyond $\Lambda$CDM.

\section{Modified theories of gravity}
\justifying
GR is an exceptionally successful theory with numerous advantages, yet it faces various theoretical problems that have led to the exploration of modified gravity theories. As mentioned earlier, GR struggles to address significant concerns and does not adequately address local conservation of energy and momentum. All other fundamental forces uphold this principle. A substantial drawback of GR is its inability to be quantized, which hinders the development of a quantum field theory for gravity, a crucial component in the quest to unify gravity with other fundamental forces.
Despite extensive attempts, this unification has yet to be achieved.

In response to these shortcomings, various modified gravity theories have been introduced as extensions of GR. These alternatives aim to establish a more comprehensive framework for addressing observational and theoretical discrepancies, including explaining the nature of DM and DE. They also strive to unify the inflation of the early Universe with the accelerated expansion observed in later times, explain transitions between different cosmic phases without the need for exotic matter, and account for the transition of the Universe from deceleration to acceleration. Moreover, modified gravity opens avenues for linking gravity with quantum mechanics and other fundamental interactions in the Universe. The following sections will examine these theories more closely, investigating their frameworks and relevance in modern cosmology.

\subsection{\texorpdfstring{$f(R)$}{f(R)} gravity: Linear curvature-matter coupling}
In 1970, Buchdahl \citep{buch} proposed a new method for understanding gravity by substituting the traditional Einstein-Hilbert action with a generalized function of the Ricci scalar $R$. This approach, referred to as $f(R)$ gravity, became significant in theoretical physics and cosmology following the important work of Starobinsky \citep{star}. This section summarizes the key mathematical framework of $f(R)$ gravity. Its action is expressed as follows \citep{barr}
\begin{equation}\label{38}
S=\int{\Big[\frac{1}{2\kappa}f(R)+L_m\Big]\sqrt{-g}\text{d}^4x}.
\end{equation}
Here, $g$ indicates the determinant of the metric tensor $g_{\mu \nu}$. The notation $f(R)$ represents a general analytical function derived from the Ricci scalar $R$ and $L_m$ stands for the matter Lagrangian density. The field equations can be derived by performing a variation of the action given in Eq. \eqref{38} with respect to $g_{\mu \nu}$ as
\begin{equation}\label{39}
f_R(R)R_{\mu\nu}-\frac{1}{2}g_{\mu\nu} f(R)-(\nabla_{\mu}\nabla_{\nu}-g_{\mu\nu}\square)f_R(R)=\kappa T_{\mu\nu},
\end{equation}
where $f_R=\frac{\partial f(R)}{\partial R}$. The symbol $\nabla$ denotes the covariant derivative, the operator $\square$ indicates the d'Alembert operator, and $T_{\mu\nu}$ is the energy-momentum tensor of the matter fields is precisely defined by the variational derivative of $L_m$ in terms of $g^{\mu \nu}$ as
\begin{equation}\label{40}
T_{\mu\nu}=-\frac{2}{\sqrt{-g}}\frac{\delta\left(\sqrt{-g}\,L_m\right)}{\delta g^{\mu\nu}}.
\end{equation}
The trace of Eq. \eqref{40} is clearly defined as
\begin{equation}\label{41}
R f_R(R)-2f(R)+3\square f_R(R)=T,
\end{equation}
where $T=g^{\mu \nu}T_{\mu \nu}$ is the stress of the energy-momentum tensor, and $\square f_R=\frac{1}{\sqrt{-g}}\partial _\mu(\sqrt{-g}g^{\mu \nu}\partial _\nu f_R)$.
In addition, it is essential to recall that the energy-momentum tensor satisfies the conservation equation $\nabla^{\mu} T_{\mu \nu} = 0$.

One of the most straightforward extensions of GR is $f(R)$ gravity, where the Lagrangian density $f$  is formulated as a general function of the Ricci scalar $R$ \citep{buch,berg/68,ruzm/70,brei/71}. A prominent example is the model defined by $f(R) = R+\alpha R^2$ ($\alpha > 0$), which incorporates the $\alpha R^2$ term. This additional term may facilitate the accelerated expansion of the Universe. In particular, this model signifies the first inflationary scenario, initially suggested by Starobinsky in 1980 \citep{star}. In $1998$, observations of distant SNeIa \citep{Riess} revealed that the expansion of the Universe is accelerating. This groundbreaking discovery led scientists to propose two possible explanations: either a mysterious form of energy, now called DE, permeates the cosmos, or our understanding of gravity, as described by GR, requires revision on cosmological scales. One of the most explored methods for understanding this late-time cosmic acceleration is $f(R)$ gravity, considered one of the more straightforward theoretical frameworks. A DE model with a Lagrangian density given by $f(R) = R-\frac{\alpha}{R^n}$ ($\alpha > 0, n>0$) has been suggested within the framework of the metric formalism \citep{capo1/02,capo2/03,carr/04,noji/03}. Moreover, due to a strong interaction between DM and DE, it does not exhibit a typical matter-dominated era, as seen in standard cosmological models \citep{amen1/07,amen2/07}. This underscores the challenges of developing a consistent $f(R)$ framework. Amendola et al. \citep{amen3/07} proposed essential conditions that $f(R)$ DE models must meet to remain cosmologically viable.

In regions with a much higher local density than the average cosmological density, $f(R)$ models must closely resemble GR to comply with local gravity constraints. Numerous $f(R)$ models have been proposed that satisfy both cosmological and local gravitational conditions \citep{amen3/07,amen4/08,appl/07,cogn/08,deru/08,hu/07,li/07,lind/09,tsuj/08}. Since these models modify gravity on large scales, they produce distinctive observational effects, such as variations in galaxy clustering patterns \citep{bean/07,carr1/06,faul/07,pogo/08,sawi/07,song/07,tsuj1/08}, changes in the CMB \citep{song/07,li1/07,song1/07,zhan/06}, and modification in weak gravitational lensing \citep{schm/08,tsuj2/08}.

In the metric formalism, $f(R)$ gravity can be considered a generalized version of the Brans-Dicke (BD) theory \citep{bran/61}, characterized by a BD parameter $\omega_{\text{BD}} = 0$ \citep{chib/03,ohan/72,teys/83}. In contrast to the original BD theory, this version includes an additional degree of freedom of the scalar called the scalaron \citep{star}, which arises from gravitational dynamics. If the mass of the scalaron stays comparable to the current Hubble constant $H_0$, it will not meet local gravitational constraints. It would generate a long-range fifth force with a coupling strength close to unity. However, strategically designing the field potential within $f(R)$ gravity makes it feasible to ensure that the scalaron attains a significant mass in regions of high density, thus mitigating the effects of the fifth force. The possible $f(R)$ models discussed earlier are designed to satisfy this condition. Thus, the effect of the fifth force weakens in regions of high density, which allows these models to align with local gravity tests. In the Einstein frame, the interaction with matter creates a local maximum in the effective potential, stabilizing the field around this point. The field remains primarily constant throughout most of the object for a spherically symmetric object with a thin shell near its surface. This process, called the chameleon effect, dramatically reduces the interaction between the field and non-relativistic matter outside the object \citep{khou/04,khou1/04}. Various experimental studies aimed at detecting violations of the equivalence principle, along with multiple solar system tests, impose stringent restrictions on DE models based on $f(R)$ \citep{hu/07,tsuj/08,faul/07,capo3/08,brax/08}.

In intense gravitational fields, like those surrounding neutron stars and white dwarfs, it is crucial to consider the effects of gravitational potentials on the field equations. The study of relativistic star structures within the framework of $f(R)$ gravity has been thoroughly investigated by several researchers \citep{babi/09,cart/00,koba1/08,koba2/09,tsuj3/09,upad/09}. The difficulties involved in modeling relativistic stars were initially pointed out in \citep{koba1/08}, especially about the singularity problems seen in $f(R)$ DE models under conditions of high curvature \citep{frol/08}. An analytical solution for a thin shell field configuration in chameleon models has been established in the Einstein frame \citep{tsuj3/09} for stars with uniform density. Additionally, numerical investigations have confirmed the presence of relativistic stars in $f(R)$ gravity, applicable to constant-density setups \citep{babi/09,upad/09}. 

\subsection{Linear non-minimal curvature-matter coupling}
The $f(R)$ modified gravity theory can be extended by incorporating a linear, non-minimal interaction between matter and geometry into the action. The following action represents this modified theory \citep{bert/07}
\begin{equation}\label{42}
S=\int{\Big[\frac{1}{2}f_1(R)+[1+\lambda f_2(R)]L_m\Big]\sqrt{-g}\text{d}^4x}.
\end{equation}
The matter Lagrangian density is indicated by $L_m$, and the components $f_i(R)$ (where i=1,2) are arbitrary functions of the Ricci scalar $R$. The strength of the interaction between $f_2(R)$ and the matter Lagrangian is described by the coupling constant $\lambda$. The resulting field equations are produced by changing the action about the metric $g_{\mu \nu}$ as
\begin{multline}\label{43}
F_1(R)R_{\mu\nu}-\frac{1}{2}g_{\mu\nu} f_1(R)-\nabla_{\mu}\nabla_{\nu}F_1(R)+g_{\mu\nu}F_1(R)=-2\lambda F_2(R)L_m R_{\mu \nu}\\
+2\lambda(\nabla_{\mu}\nabla_{\nu}-g_{\mu\nu}\square)L_m F_2(R)+[1+\lambda f_2(R)] T_{\mu\nu},
\end{multline}
where $F_i(R)=\frac{\partial f_i(R)}{\partial R}$.

\subsection{\texorpdfstring{$f(R,L_m)$}{f(R,Lm)} gravity}
To clarify, $f(R,L_m)$ gravity extends the framework of $f(R)$ gravity by suggesting that the gravitational Lagrangian can be formulated as a function of both the Ricci scalar $R$ and the matter Lagrangian density $L_m$ \citep{hark/10}. This idea represents one of the most comprehensive forms of the Einstein-Hilbert action. The term maximal extension is used in this context to refer to an element in a specific mathematical order that has no successors. The theory of $f(R,L_m)$ is the most general among gravitational theories, including standard GR, $f(R)$ gravity, linear curvature-matter coupling, and others. It encompasses all gravitational frameworks formulated in a Riemannian space, with the action explicitly relying on both the Ricci scalar and the matter Lagrangian. The theory acts as a full generalization of the Einstein-Hilbert action and can be expressed in the following way,
\begin{equation}\label{44}
S=\int {f(R,L_m) \sqrt{-g}\text{d}^4x}.
\end{equation}
The energy-momentum tensor associated with matter is described in Eq. \eqref{40}. By varying the action concerning the metric, we can obtain the following field equation
\begin{multline}\label{45}
f_R(R,L_m) R_{\mu\nu} + (g_{\mu\nu} \square - \nabla_\mu \nabla_\nu)f_R(R,L_m)- \frac{1}{2} (f(R,L_m)-f_{L_m}(R,L_m)L_m)g_{\mu\nu}\\
= \frac{1}{2} f_{L_m}(R,L_m) T_{\mu\nu}.
\end{multline}
When $f (R, L_m) = \frac{R}{2} + L_m$, the standard Einstein field equations of GR are obtained, as illustrated in Eq. \eqref{34}. In situations where the function $f(R,L_m)$ is defined as $ f_1(R) + f_2(R)G(L_m) $, $f_1$ and $f_2$ represent any functions dependent on the Ricci scalar. At the same time, $G$ is a function related to the matter Lagrangian density. These field equations illustrate a modified gravity framework that allows a flexible interaction between curvature and matter, a concept investigated in \citep{hark/08}.\\
By contracting Eq. \eqref{45}, we can derive a relationship that links the Ricci scalar, the matter Lagrangian density, and the trace of the energy-momentum tensor, expressed as follows
\begin{equation}\label{46}
R f_R(R,L_m) + 3\square f_R(R,L_m) - 2(f(R,L_m)-f_{L_m}(R,L_m)L_m) = \frac{1}{2} f_{L_m}(R,L_m) T.
\end{equation}
Taking the covariant divergence of Eq. \eqref{45} allows us to apply the mathematical identity given in \citep{koiv/06},
\begin{equation}\label{47}
\nabla^\mu \Bigg[f_R R_{\mu\nu}- \frac{1}{2} g_{\mu\nu} f+ (g_{\mu\nu} \square - \nabla_\mu \nabla_\nu)f_R \Bigg]=0.  
\end{equation}
We analyze the divergence of the energy-momentum tensor $T_{\mu \nu}$ and arrive at the following equation
\begin{equation}\label{48}
\nabla^\mu T_{\mu\nu} = 2\nabla^\mu ln(f_{L_m}) \frac{\partial L_m}{\partial g^{\mu\nu}}.
\end{equation}

\subsubsection{ Solar system tests of \texorpdfstring{$f(R,L_m)$}{f(R,Lm)} gravity}
\justifying
Modified gravity theories that include a coupling between curvature and matter predict the existence of an additional force. This force causes test particles to follow non-geodesic paths. Its effects can be investigated within the Solar System by analyzing how it influences the orbital parameters of planets around the sun. The impact of this extra force on planetary motion can be assessed using a straightforward approach that leverages the properties of the Runge–Lenz vector, defined as follows
\begin{equation}\label{49}
\vec{A}=\vec{v}\times \vec{L}-\alpha \vec{e_r}, 
\end{equation}
where $\vec{v}$ is the velocity of the planet with mass $m$ relative to the Sun, which has mass $M_\odot$, and $\alpha=GmM_\odot$ \citep{hark/08}. For an elliptical orbit defined by eccentricity $e$, a semi-major axis $a$, and an orbital period $T$, the equation that describes the orbit is given by $(L/\mu\alpha)r^{-1}=1+e cos\theta$. The angular momentum is represented by the variable $\vec{L}$, expressed as
\begin{equation}\label{50}
\vec{L}=\vec{r}\times \vec{p}=\mu r^2 \dot{\theta}\vec{k}. 
\end{equation}
The two-body position vector is given by $\vec{r}=r\vec{e_r}$, while the relative momentum is $\vec{p}=\mu\vec{v}$. The reduced mass is $\mu=\frac{mM_\odot}{m+M_\odot}$. Furthermore, the Runge–Lenz vector can be expressed as follows
\begin{equation}\label{51}
\vec{A}=\big(\frac{\vec{L}}{\mu r}\big)\vec{e_r}-\dot{r}L\vec{e_\theta}.
\end{equation}
The derivative with respect to the polar angle $\theta$ can be represented as 
$$\frac{\text{d}\vec{A}}{\text{d}\theta}=r^2\left[\frac{\text{d}V(r)}{\text{d}r}-\frac{\alpha}{r^2}\right]\vec{e_\theta},$$ where $V(r)$ denotes the potential related to the central force. In the case of a planet, the gravitational potential considers contributions from post-Newtonian corrections,
\begin{equation}\label{52}
V_{\text{P}\text{N}}(r)=-\frac{\alpha}{r}-3\frac{\alpha ^2}{m r^2}.
\end{equation}
We include the gravitational term that arises from matter-geometry coupling. So, we have
\begin{equation}\label{53}
\frac{\text{d}\vec{A}}{\text{d}\theta}= r^2 \Bigg[\frac{6\alpha^2}{mr^3}+m\vec{a_E}(r)\Bigg]\vec{e_\theta}.
\end{equation}
Assuming that $\mu$ is about the same as $m$, we can show how the change in angle $\theta$ by $2\pi$ affects the change in $\Delta \phi$ of the perihelion. This can be expressed as follows
\begin{equation}\label{54}
 \Delta \phi=\bigg(\frac{1}{\alpha e}\bigg) \bigintss _0 ^{2\pi}  \Bigg|\dot{\vec{L}} \times \frac{\text{d}\vec{A}}{\text{d}\theta}\Bigg| \text{d}\theta
\end{equation}
This can be calculated explicitly in the following manner
\begin{equation}\label{55}
 \Delta \phi=24\pi^3 \bigg(\frac{a}{T}\bigg)^2 \frac{1}{1-e^2}+\frac{L}{8\pi^3 m e} \frac{(1-e^2)^{3/2}}{(a/T)^3}
 \bigintss _0 ^{2\pi}  \frac{a_E\big[L^2(1+e cos \theta)^{-1}/m \alpha\big]}{(1+e cos \theta)^{2}} cos \theta\text{d}\theta.
\end{equation}
We used the relationship $\alpha/L=2\pi(a/T)/\sqrt{1-e^2}$. The first term in Eq. \eqref{55} represents the well-known perihelion precession predicted by GR. In contrast, the second term contributes to the perihelion precession due to the extra force resulting from the curvature-matter coupling.

We explore a compelling scenario in which the additional force can be effectively treated as a constant $a_E \approx\text{constant} $. This approximation is particularly relevant in specific regions of spacetime. In particular, a MOND-type acceleration, represented as $a_E \approx\sqrt{a_0 a_N}=\sqrt{GM_\odot a_0}/r$, where $a_0$ is a fixed acceleration, is pertinent in this setting. This model has been proposed as a robust dynamical explanation for DM \citep{milg/83,beke/04}. Moreover, a comparable framework can express the additional acceleration of the curvature-matter coupling in the Newtonian limit \citep{bert/07}. Importantly, Eq. \eqref{55} is a powerful tool to determine the perihelion precession of planets within our Solar System,
\begin{equation}\label{56}
 \Delta \phi=\frac{6\pi G M_\odot}{a(1-e^2)} +\frac{2\pi a^2 \sqrt{1-e^2}}{GM_\odot}a_E.
\end{equation}
We utilized Kepler's third law, which states that $T^2 = 4\pi^2 a^3/GM_\odot$. For Mercury, its orbital eccentricity is $e = 0.205615$, its semi-major axis is $a = 57.91 \times 10^{11} cm$, and the mass of the Sun is $M_\odot = 1.989 \times 10^{33} g$. Substituting these values produces a general relativistic prediction for the precession angle, $ (\Delta_\phi)_{GR} = 42.962$ $arcseconds / century$, derived from the first term in Eq. \eqref{56}. In contrast, Mercury's observed perihelion precession is $(\Delta_\phi)_{obs} = 43.11 \pm 0.21$ $arcseconds / century$ \citep{will/06}.
The difference between these values can be explained by additional physical mechanisms, resulting in $ (\Delta_\phi)_E = (\Delta_\phi)_{obs} - (\Delta_\phi)_{GR} = 0.17 $ $arcseconds / century$. To align with the observed perihelion precession of Mercury, the constant acceleration $ a_E $ due to curvature-matter coupling within the Solar System must be constrained to $ a_E \leq 1.28 \times 10^{-9} $ $cm/s^2$. In comparison, the acceleration $ a_0 \approx 10^{-8} $ $cm/s^2$, proposed to account for DM phenomena and the Pioneer anomaly, slightly exceeds $ a_E $, which is limited by precise measurements within the Solar System \citep{bert/07}. However, since the assumption of a constant additional acceleration may not be valid on larger cosmic scales, this does not exclude the possibility that curvature-matter coupling effects could influence solar system dynamics and galactic phenomena.

\subsubsection{Geodesic deviation equation and the Raychaudhury equation in \texorpdfstring{$f(R,L_m)$}{f(R,Lm)} gravity}
Assuming that the matter Lagrangian density depends solely on the rest density parameter $\rho$, we can derive the equations of motion for test particles in the $f(R,L_m)$ gravity model using Eq. \eqref{48}. This leads us to the following results
\begin{equation}\label{57}
 \frac{\text{d}^2 x^\mu}{\text{d}s^2}=U^\nu \nabla_\nu U^\mu =\frac{\text{d}^2 x^\mu}{\text{d}s^2}+\Gamma^\mu _{\nu \lambda}U^\nu U^\lambda=f^\mu .
\end{equation}
Here, $s$ represents the world-line parameter, $U^\mu=\text{d}x^\mu /\text{d}s$ denotes the particle's four-velocity, $\Gamma^\nu _{\sigma\beta}$ are the Christoffel symbols associated with the metric, and $f^\mu$ corresponds to the additional force as 
\begin{equation}\label{58}
f^\mu=-\nabla_\nu ln \Bigg[f_{L_m}(R,L_m)\frac{\text{d}L_m (\rho)}{\text{d}\rho}\Bigg] (U^\mu U^\nu-g^{\mu\nu}).
\end{equation}
It is essential to emphasize that,  similar to the scenario of linear curvature-matter coupling, the supplementary force $f^\mu$ resulting from the interaction between curvature and matter is orthogonal to the four-velocity, which means $f^\mu U_\mu = 0$. As a result, test particles in modified gravity theories with arbitrary curvature-matter coupling follow non-geodesic trajectories when $f^\mu$ is present.
We now introduce the Raychaudhuri equation and the geodesic deviation equation in the context of $f(R,L_m)$ gravity, considering the influences of additional forces and the interaction between curvature and matter \citep{hark/12}. Let's consider a family of curves parameterized by a single parameter, denoted as $x^\mu(s;\lambda)$, where for a constant value $\lambda = \lambda_0$, the path $x^\mu(s;\lambda_0)$ satisfies the equation represented in Eq. \eqref{57}. To facilitate this description, we define the tangent vector fields along the trajectories of the particles as $U^\mu=\partial x^\mu(s;\lambda)/\partial s$ and $n^\mu=\partial x^\mu(s;\lambda)/\partial \lambda$, under the assumption of a smooth parametric representation. Furthermore, we define the four-vector to complete the formulation as
\begin{equation}\label{59}
\eta^\mu= \Bigg[\frac{\partial x^\mu(s;\lambda)}{\partial \lambda}\Bigg] \delta\lambda \equiv n^\mu \delta\lambda.
\end{equation}
The parameters $\lambda$ and $\lambda + \delta\lambda$, which align with the same value of $s$, denote points situated on adjacent and infinitesimally close geodesics. By examining Eq. \eqref{57}, we can derive the equation for geodesic deviation. The second derivative of the deviation vector $\eta^\mu$, taken with respect to the parameter $s$, is expressed as follows \citep{land/98,hawk/73}
\begin{equation}\label{60}
 \frac{\text{d}^2 \eta^\mu}{\text{d}s^2}=R^\mu _{\nu \alpha \beta}\eta^\alpha U^\beta U^\nu +\eta^\alpha\nabla_\alpha f^\mu.
\end{equation}
In standard GR, the geodesic deviation equation, also known as the Jacobi equation, describes how test particles follow geodesics under specific conditions, where the force term $f^\mu$ is set to zero. The deviation vector $\eta^\mu$ plays a crucial role in this context; if a geodesic $x^\mu_0(s) = x^\mu(s; \lambda_0)$ satisfies the geodesic equation, then the perturbed geodesic $x^\mu_1(s) = x^\mu_0(s) + \eta^\mu$ remains a valid solution. This is due to the approximation $x^\mu(s; \lambda_1) \approx x^\mu(s; \lambda_0) + \eta^\mu(s; \lambda_0) \delta\lambda$, which implies $x^\mu(s; \lambda_0 + \delta\lambda)$. In the framework of $f(R,L_m)$ gravity, where curvature interacts with matter, the geodesic deviation equation is modified to include an additional force, as described by Eq. \eqref{58}. This results in a new formulation for the equation,
\begin{equation}\label{61}
\frac{\text{d}^2 \eta^\mu}{\text{d}s^2}=R^\mu _{\nu \alpha \beta}\eta^\alpha U^\beta U^\nu +\eta^\alpha\nabla_\alpha \left\{ \nabla_\nu ln \Bigg[f_{L_m}(R,L_m)\frac{\text{d}L_m (\rho)}{\text{d}\rho}\Bigg] (L_m g^{\mu\nu}-T^{\mu\nu}) \right\}.
\end{equation}
The equation for geodesic deviation is expressed clearly as
\begin{multline}\label{62}
\frac{\text{d}^2 \eta^\mu}{\text{d}s^2}=R^\mu _{\nu \alpha \beta}\eta^\alpha U^\beta U^\nu +\eta^\alpha \left\{\nabla_\alpha \nabla_\nu ln \Bigg[f_{L_m}(R,L_m)\frac{\text{d}L_m (\rho)}{\text{d}\rho}\Bigg]\right\} (L_m g^{\mu\nu}-T^{\mu\nu})\\
 + \eta^\alpha \nabla_\nu ln \Bigg[f_{L_m}(R,L_m)\frac{\text{d}L_m (\rho)}{\text{d}\rho}\Bigg] ( g^{\mu\nu}\nabla_\alpha L_m-\nabla_\alpha T^{\mu\nu}).
\end{multline}
The Raychaudhuri equation, which includes the influence of an additional force, is derived as shown in \citep{hawk/73},
\begin{equation}\label{63}
 \dot{\theta}+\frac{1}{3}\theta^2 + (\sigma^2 -\omega^2)=\nabla_\mu f^\mu +R_{\mu \nu}U^\mu U^\nu.
 \end{equation}
The expansion of a particle congruence is given by $\theta = \nabla_\nu U^\nu$, with $\omega^2 = \omega_{\mu\nu}\omega^{\mu\nu}$ and $\sigma^2 = \sigma_{\mu\nu}\sigma^{\mu\nu}$. In $f(R,L_m)$ modified gravity, the Raychaudhuri equation is generalized. This form is derived by utilizing the field Eq. \eqref{45} and incorporating the additional force described by Eq. \eqref{58} as
\begin{multline}\label{64}
\dot{\theta}=-\frac{1}{3}\theta^2 - (\sigma^2 -\omega^2)+\Lambda (R,L_m)+
 \nabla_\mu \left\{\nabla_\nu ln \Bigg[f_{L_m}(R,L_m)\frac{\text{d}L_m (\rho)}{\text{d}\rho}\Bigg] (L_m g^{\mu\nu}-T^{\mu\nu})\right\}\\
 + \frac{1}{f_{L_m}(R,L_m)}U^\mu U^\nu \nabla_\mu \nabla_\nu f_{R}(R,L_m)+\Phi(R,L_m)( T_{\mu\nu} U^\mu U^\nu -\frac{1}{3}T ).
\end{multline}
We have denoted the following
\begin{equation}\label{65}
\Lambda (R,L_m)=\frac{2f_{R}(R,L_m)R-f(R,L_m)+f_{L_m}(R,L_m)L_m}{6f_{R}(R,L_m)} ,   
\end{equation}
\begin{equation}\label{66}
\Phi(R,L_m)=\frac{f_{L_m}(R,L_m)}{f_{R}(R,L_m)}.    
\end{equation}

\section{Relevant physical background}
\justifying
Bulk viscosity is a phenomenon in fluid dynamics that arises from deviations from local thermodynamic equilibrium. In the field of cosmology, it is particularly relevant for understanding the behavior of cosmic fluids, especially in high-energy conditions such as those present in the early Universe. Unlike shear viscosity, which opposes variations in velocity, bulk viscosity resists changes in volume, thus affecting how the Universe expands. In scenarios involving the early Universe, bulk viscosity can change the effective pressure of cosmic fluids, leading to important consequences for the evolution of the Universe. For example, it can facilitate the production of entropy and influence the thermal history of the cosmos. Grasping the concept of bulk viscosity is vital for accurately modeling dissipative processes in cosmological contexts \citep{IB-1,IB-2}, including during inflation and the reheating phase that follows.

Bouncing cosmology provides an alternative perspective to conventional Big Bang theory by suggesting that the universe first underwent a contraction phase before reaching a minimum scale and subsequently expanding again \citep{MB-1,Haro}. This model circumvents the issue of an initial singularity and offers solutions to challenges such as horizon and flatness problems. The behavior of the bounce in these models is influenced by the characteristics of the cosmic fluid, including dissipative effects such as bulk viscosity. These viscous forces can help mitigate anisotropies during the contraction period and facilitate a smooth transition to the expansion phase. Furthermore, bouncing cosmologies often necessitate modifications to the traditional cosmological framework, making them a compelling area of research for examining the relationships between gravity, matter, and dissipative processes in the early stages of the Universe.

Baryogenesis refers to the mechanism that explains the imbalance between matter and antimatter in the Universe. This phenomenon, as described by the Sakharov conditions \citep{Sakh}, relies on three essential criteria: the violation of baryon number conservation, the break of the CP symmetry and a departure from thermal equilibrium. In the early stages of the Universe, these criteria can be satisfied in high-energy scenarios, such as during phase transitions or through effects like bulk viscosity. These viscous forces can create conditions that push the Universe away from equilibrium, which are necessary for baryogenesis to occur. Moreover, bouncing cosmologies present a distinctive context for baryogenesis. The nonsingular bounce and its related dynamics can also facilitate the nonequilibrium conditions needed for this process. Exploring how baryogenesis interacts with dissipative processes and cosmological models is vital for understanding the emergence of the matter-dominated Universe we see today.

\section{Cosmological Observations}
\justifying
Precise cosmological observations depend on effective modeling strategies and statistical techniques to analyze the evolution of the Universe. Theoretical frameworks provide insight into cosmic behavior, while statistical methods such as $\chi^2$ minimization and the MCMC approach facilitate efficient parameter estimation by aligning models with observational data. Critical observational tools, including the CC, BAO, and SNeIa, offer vital constraints for understanding cosmic expansion and the formation of structures. Collectively, these approaches deepen our comprehension of the essential characteristics of the Universe and evaluate the validity of modified gravity theories.

\subsection{Modelling and statistics}
Now, we explore how geometric structures describe the Universe and facilitate accurate predictions. By developing essential equations and utilizing statistical methods, we evaluated the evolution of the cosmos and verified alignment with observational data. This methodology deepens our understanding of cosmic behavior and supports modified gravity theories.
\subsubsection{Modelling the Universe}
To construct a cosmological model that describes the physical Universe, we follow key steps. First, we define the underlying geometry by specifying a spacetime manifold with an appropriate metric and connection. Then, we introduce realistic assumptions, such as selecting a functional form of modified gravity and a suitable equation of state to represent specific cosmological epochs. These assumptions help to derive analytical solutions using differential equation techniques. The model's free parameters are then determined using observational datasets and statistical methods. Finally, we analyze cosmological parameters like deceleration and the equation of state to predict the model's evolution and compare it with established cosmological models.
\subsubsection{\texorpdfstring{$\chi^2$}{Chi} minimization}
Consider a function $f_{\text{model}}(x, \theta)$ that is characterized by the independent variable $x$ and a set of parameters denoted as $\theta$. Suppose that we have a set of $n$ independent observations represented as $\{ f_{k, \text{obs}}(x_{k, \text{obs}}) \}_{k=1}^{n}$, where each observation comes with an associated standard deviation $\sigma_{k, \text{obs}}$. The corresponding value of $\chi^2$ can be defined as follows \cite{R52}
\begin{equation}\label{e108}
\chi^2(\theta) = \sum_{k=1}^n  \frac{\left( f_{model}(x_{k,obs},\theta) - f_{k,obs} \right)^2}{\sigma_{k,obs}^2} \text{.} 
\end{equation}
If the $n$ observations are dependent, a covariance matrix $C$ is used instead of the standard deviation. Under these conditions, the $\chi^2$ function is modified accordingly,
\begin{equation}\label{e109}
\chi^2(\theta) = \sum_{i,j=1}^n X_i C^{-1}_{ij} X_j . 
\end{equation}
In this context, we define $X_i$ to be the difference between the observed value $f_{i,obs}$ and the model prediction $f_{model}(x_{i,obs}, \theta)$. Moreover, an important statistical concept known as the likelihood function can be expressed in the following way
\begin{equation}\label{e110}
\mathcal{L}(\theta) = P(\theta| D = Data). 
\end{equation}
The following relationship can be established between the two statistical parameters mentioned earlier
\begin{equation}\label{e111}
\mathcal{L}(\theta) \propto exp\left( -\frac{1}{2} \chi^2(\theta) \right) \text{.}  
\end{equation}
Our main goal is to find the best values for the parameter set $\theta$ so that the function $f_{model}(x,\theta)$ closely matches the observed data. We do this by minimizing the $\chi^2(\theta)$ function, which is the same as maximizing the likelihood function. The parameter value that gives us the lowest $\chi^2$, referred to as $\theta_{min}$, gives us the best-fit solution. Several optimization techniques are available to solve this problem, including gradient descent, Newton's method, and the Random Walk algorithm \cite{R53,R54}. However, these methods often face challenges in parameter spaces with many local minima, leading to solutions that are only locally optimal rather than globally optimal. To address this issue, we can create a more effective method by combining the strengths of these algorithms. One popular technique used in computational science is the MCMC algorithm.

\subsubsection{The MCMC approach}
In the past decade, probabilistic data analysis, especially Bayesian statistical inference, has significantly impacted scientific research. This approach relies on the posterior probability density function (PDF) or the likelihood function. Although some algorithms efficiently find the optimal values for these functions, a deeper understanding of the posterior PDF is often necessary.  MCMC methods are specifically designed to enable efficient sampling from the posterior PDF, which makes them particularly useful for exploring high-dimensional parameter spaces \cite{R55}.
A Markov Chain represents a series of parameter values derived from a stochastic process, where each step is influenced solely by the current state and its immediate predecessor. Conversely, the Monte Carlo method allows for an extensive examination of the complete parameter space. The main principle behind the MCMC involves constructing a Markov Chain that samples a model's parameter space in accordance with a specified probability distribution. This Markov Chain is made up of a sequence of parameter values, where each value is generated based on the prior one through particular transition rules defined by a proposal distribution. This proposal distribution presents a new parameter value, and its acceptance is evaluated based on its posterior probability, which reflects both the observed data and the prior probability function. After the chain has sufficiently converged, the posterior distribution of the parameters can be inferred by observing the frequency of the sampled values. This posterior distribution is crucial for estimating optimal parameter values and their uncertainties, thereby facilitating predictions for various outcomes. Among the diverse MCMC techniques, the Metropolis-Hastings algorithm stands out as one of the most commonly utilized methods, due to its integration of a random-walk strategy.
Mackey \citep{mac} has recently introduced a new and more sophisticated algorithm called \textit{emcee}: the MCMC hammer. This algorithm brings notable improvements compared to conventional MCMC sampling techniques. One of its main strengths is the capacity to effectively handle a large number of walkers, often extending hundreds of feet. In principle, increasing the walker count does not result in any significant downsides, aside from possible performance issues. However, an extensive comparison between different MCMC algorithms and the \textit{emcee} sampling method is not the focus of this thesis.

\subsection{Observational data}
Observations in cosmology have played a vital role in exploring the history of the expansion of the Universe. In this overview, we summarize various discoveries related to cosmic evolution.

\subsubsection{Cosmic Chronometer}
The CC method is a vital technique for determining the Hubble constant by examining ancient galaxies that display gradual changes in their redshifts. This method utilizes the principle of differential aging, which enables researchers to estimate the ages of galaxies across various levels of redshift. In accordance with the FLRW background, the Hubble constant can be formulated as $H =\frac{-1}{1+z} \frac{\text{d}z}{\text{d}t}$. A key benefit of the CC method is its capacity to evaluate the Hubble parameter without needing to rely on definite cosmological assumptions. This feature offers an impartial estimate of the rate at which the Universe is expanding, serving as a useful resource for examining different cosmological models. For our study, we have assembled a dataset that includes $31$ observations sourced from various references \citep{H1,H2,H3,H4,H5,H6,H7}. The dataset spans a considerable range of redshifts, from $0.1$ to $2$, facilitating a comprehensive exploration of the expansion history of the Universe.

\subsubsection{Type Ia Supernovae}
The SNeIa are extraordinary stellar explosions resulting from a white dwarf star accumulating mass from a companion star until it reaches the Chandrasekhar mass limit. This accumulation triggers a thermonuclear explosion \citep{1}. SNeIa is vital in astrophysics as standard candles to determine luminosity distances \citep{2,3,4}, which are crucial for understanding the expansion of the universe. In $1998$, a groundbreaking study by Riess et al. revealed the accelerated expansion of the universe through the analysis of $16$ distant SNeIa and $34$ nearby ones observed by the Hubble Telescope \citep{Riess}. This finding was further validated in $1999$ by Perlmutter and his team, who examined $18$ nearby supernovae from the Calan-Tololo sample alongside $42$ high-redshift Supernovae \citep{Perlmutter}. Various research groups and surveys have significantly advanced our understanding of SNeIa over the years. Notable contributions have come from the Sloan Digital Sky Survey (SDSS) \citep{19.1,19}, the Lick Observatory Supernova Search (LOSS) \citep{18,18.1}, the Carnegie Supernova Project (CSP) \citep{17,17.1}, the Nearby Supernova Factory (NSF) \citep{16,16.1}, the Supernova Legacy Survey (SNLS) \citep{1314,1314.1}, and the Higher-Z Team \citep{sdss,1112}. Recently, expansive datasets, such as the Union $2.1$ compilation with $580$ SNeIa and the Pantheon compilation featuring $1048$ SNeIa across a wide redshift range $0.01 < z < 2.26$, have offered more accurate measurements related to cosmic expansion.

\subsubsection{Baryonic Acoustic Oscillations}
The BAO method is essential for investigating large-scale structures of the Universe. These oscillations originate from the interactions of baryonic matter with radiation within the photon-baryon fluid, a process initiated by acoustic waves in the early stages of the Universe. This traditional approach to measuring cosmic distances results in a distinct peak in the correlation function of galaxies because of compression effects. The sound horizon during the recombination phase, which is influenced by the temperature of the CMB and the density of baryons, determines the comoving position of the BAO peak. For any specific redshift $z$, the angular separation can be expressed as $\Delta \theta =r_d /((1+z)\text{D}_\text{A} (z))$, reflecting the coordinates of the BAO peak. In the radial direction, the redshift separation is given by $\nabla z = r_d / \text{D}_\text{H}(z)$. Here, $\text{D}_\text{H} = \text{c}/\text{H}$ denotes the angular distance, $\text{D}_\text{H} = \text{c}/\text{H}$ defines the Hubble distance, and $r_d$ represents the sound horizon at the drag epoch. The BAO peak can be precisely identified across various redshifts, allowing us to constrain the cosmological parameters that influence the ratios $\text{D}_\text{H} /r_d$ and $\text{D}_\text{A}/r_d$. By determining a suitable value for $r_d$, we can estimate $H(z)$. The analysis in this study relies on a dataset of $26$ independent data points gathered through proximity BAO measurements \citep{H8,H9,H10,H11,H12,H13,H14,H15,H16,H17,H18,H19}.\\

To represent the posterior probability distribution of the cosmological parameters, we utilized contour plots in Python. The process begins with the implementation of the MCMC algorithm, such as `\textit{emcee}`, to sample the parameter space according to the posterior distribution. Afterward, we assess the density of these samples through Kernel Density Estimation (KDE) or histograms. We then create confidence contours at levels of 68\%, 95\%, and 99\% using libraries like  `\textit{matplotlib}` or `\textit{corner.py}`. Finally, the contours are plotted to show the constraints on the cosmological parameters that emerge from the observational data. This approach serves as a valuable tool for analyzing and comparing various cosmological models.

\section{Conclusion}
\justifying
This chapter looks at how our understanding of the Universe has changed over time, from ancient views to modern discoveries. It also provides a clear foundation for the topics covered in the research. It begins with a discussion of GR and its pivotal role in modern theoretical physics, followed by an overview of the standard cosmological model built upon the FLRW metric, redshift, the Hubble parameter, and the Friedmann equations. The $\Lambda$CDM model, a cornerstone of contemporary cosmology, is introduced along with its key features and associated challenges.

To address contemporary challenges in cosmology, the chapter delves into the framework of modified gravity, focusing on $f(R)$ gravity and its extensions. Linear curvature-matter coupling and linear non-minimal curvature-matter coupling are explored as intermediate models, culminating in the $f(R,L_m)$ gravity framework, which introduces a coupling between curvature and matter through the matter Lagrangian. Then, we discuss the empirical validity of these theories by discussing solar system tests, the geodesic deviation equation, and the Raychaudhuri equation within $f(R,L_m)$ gravity. These tools provide insights into the effects of modified gravity on the structure of spacetime and the motion of particles. Finally, the chapter discusses key Cosmological Observations that provide insights into the nature of the Universe and validate theoretical models.

In the following chapters, we dive deeper into the implications and applications of $f(R,L_m)$ gravity in various cosmological contexts. First, the focus is on understanding the evolution and dynamics of the Universe within the framework of $f(R,L_m)$ gravity, highlighting its impact on cosmology. Next, we explore the late-time cosmic acceleration which is studied through parametrizations of the Hubble parameter within this framework, offering solutions to the challenges posed by observational data and shedding light on the nature of the accelerated expansion. Further, we examine viscous DE, where the EoS is constrained by observational data to investigate its role in cosmic evolution. We then analyze bouncing cosmological models as potential alternatives to the Big Bang singularity, demonstrating the versatility of $f(R,L_m)$ gravity in addressing fundamental cosmological issues. Subsequently, our discussion turns to baryogenesis, examining how $f(R,L_m)$ gravity provides insights into the matter-antimatter asymmetry in the Universe.


%% file: Chapters/Chapter2.tex

\chapter{Cosmology in \texorpdfstring{$f(R,L_m)$}{f(R,Lm)} gravity} 

\label{Chapter2} 

\lhead{Chapter 2. \emph{Cosmology in $f(R,L_m)$ gravity}} 

\vspace{8 cm}
* The following publications cover the work in this chapter: \\
 
\textit{Cosmology in $f(R,L_m)$ gravity}, Physics Letters B, \textbf{831}, 137148 (2022).

\clearpage
In this chapter, we explore the cosmic expansion of the Universe within the context of $f(R, L_m)$ gravity theory. Our focus is on a nonlinear model defined as $f(R,L_m)=\frac{R}{2}+L_m^n + \beta$, where $n$ and $\beta$ are model-specific parameters. We derive the motion equations for the FLRW background and find the exact solution to the corresponding field equations. 
Next, we estimate the optimal ranges for the model parameters utilizing updated datasets of $H(z)$ with 57 data points and the Pantheon dataset containing 1048 points. We further analyze the physical characteristics of the density parameter and the deceleration parameter. Our findings show that the deceleration parameter transitions from a decelerating to an accelerating phase for the Universe. Additionally, we assess the stability of our cosmological model under observational constraints by examining linear perturbations. 
Finally, we study the behavior of the Om diagnostic parameter, which indicates that our model exhibits quintessence-like characteristics. The results from our $f(R, L_m)$ cosmological framework are consistent with recent observational research and effectively account for the late-time acceleration of the Universe.

\section{Introduction}\label{I}
\justifying
Recent observations of SNeIa \citep{Riess,Perlmutter} together with observational studies of the SDSS \citep{sdss}, Wilkinson Microwave Anisotropy Probe (WMAP) \citep{D.N.}, Large Scale Structure (LSS) \citep{T.Koivisto,S.F.}, and CMB \citep{Caldwell} indicate an accelerating behavior of the expansion phase of the Universe. The late-time expansion scenario can be effectively described using $f(R)$ gravity, as discussed in various studies \citep{carr/04}. Researchers have examined the constraints necessary for viable cosmological models in this framework \citep{Cap,amen1/07}. In fact, there are viable models of $f(R)$ gravity that conform to observations of the solar system \citep{noji/03,amen4/08,V.F.,P.J.}. Furthermore, the observational signatures of DE models based on $f(R)$, together with the relevant constraints from solar system observations and the equivalence principle, have been explored from multiple sources \citep{carr/04,tsuj/08,capo3/08,Liu,Alex}. Another model $f(R)$ that unifies the early inflation with DE and passes through local tests has been discussed in \citep{cogn/08,Noj-2,Noj-3}. Moreover, one can check the references \citep{JS,SC,RC} for various cosmological implications of $f(R)$ gravity models.

An extension of $f(R)$ gravity theory that includes an explicit coupling of matter Lagrangian density $L_m$ with a generic function $f(R)$ was proposed in \citep{bert/07}. As a consequence of this matter-geometry coupling, an extra force orthogonal to four velocity vectors appears with the non-geodesic motion of the massive particles. The model was enhanced to accommodate arbitrary couplings in both matter and geometry, as discussed in \citep{hark/08}. The effects of non-minimal couplings between matter and geometry on cosmology and astrophysics have been thoroughly explored in \citep{THK-2,THK-3,SNN,V.F.-2,V.F.-3}. Recently, Harko and Lobo \citep{hark/10} introduced a more advanced version of the curvature-matter coupling theory, called $f(R, L_m)$ gravity. In this framework, $f(R, L_m)$ denotes a general function of the matter Lagrangian density $L_m$ and the Ricci scalar $R$. This $f(R, L_m)$ gravity theory represents the most comprehensive extension of all gravitational theories formulated within the Riemann space. The motion of test particles in $f(R, L_m)$ gravity theory is non-geodesic, and an extra force orthogonal to four velocity vectors arises. Recently, $f(R, L_m)$ gravity theory has led to many fascinating cosmological discoveries, as highlighted in several studies. For more details, refer to the citations \citep{WG,RV-1,RV-2,APP,sol2}.

This chapter is organized as follows. In Sec. \ref{IIb}, we derive the motion equations for the FLRW background. In Sec. \ref{IIIb}, we consider a cosmological $f(R,L_m)$ model, and then we derive the expression for the Hubble parameter and the deceleration parameter. In the next section, Sec. \ref{IVb}, we find the best ranges of the model parameters using the $H(z)$, Pantheon, and the combined $H(z)$ + Pantheon datasets. In addition, we analyze the behavior of cosmological parameters for the values of model parameters constrained by the observational datasets. Additionally, in Sec. \ref{Vb}, we explore the stability of the derived solution in the context of the observational constraint by analyzing a linear perturbation of the Hubble parameter. Further, in Sec. \ref{VIb}, we employ the Om diagnostic test to differentiate our cosmological model from other DE models; finally, in Sec. \ref{VIIb}, we discuss and conclude our results.

\section{Motion equations in \texorpdfstring{$f(R,L_m)$}{f(R,Lm)} gravity}\label{IIb}
\justifying
Taking into account the spatial isotropy and homogeneity of our Universe, we assume the following flat FLRW metric \citep{Ryden} for our analysis
\begin{equation}\label{9b}
ds^2= -dt^2 + a^2(t)[dx^2+dy^2+dz^2].
\end{equation}
Here, $ a(t) $ is the scale factor that measures cosmic expansion at a time $t$. For the line element \eqref{9b}, the non-vanishing components of Christoffel symbols are
\begin{equation}\label{10b}
\Gamma^0_{ij}= a\dot{a}\delta_{ij} , \: \:  \Gamma^k_{0j}= \Gamma^k_{j0}= \frac{\dot{a}}{a} \delta^k_j.
\end{equation}
Here, $i,j,k=1,2,3$. Using Eq. \eqref{8}, we get the non-zero components of the Ricci tensor as
\begin{equation}\label{11b}
R_{00}=-3 \frac{\ddot{a}}{a} \: , \: R_{11}=R_{22}=R_{33}= a \ddot{a}+2 \dot{a}^2.
\end{equation}
Hence, the Ricci scalar is obtained corresponding to the line element Eq. \eqref{9b} is
\begin{equation}\label{12b}
R= 6 \frac{\ddot{a}}{a}+ 6 \bigl( \frac{\dot{a}}{a} \bigr)^2 = 6 ( \dot{H}+2H^2 ).
\end{equation}
Here, $H=\frac{\dot{a}}{a}$ is the Hubble parameter.
The energy-momentum tensor characterizing the Universe filled with perfect fluid type matter content for the line element Eq. \eqref{9b} is given by,
\begin{equation}\label{13b}
T_{\mu\nu}=(\rho+p)u_\mu u_\nu + pg_{\mu\nu}.
\end{equation}
Here, $\rho$ is the density parameter, $p$ is the spatially isotropic pressure, and $u^\mu=(1,0,0,0)$ are components of the four velocities of the cosmic perfect fluid. The Friedmann equations that describes the dynamics of the Universe in $f(R, L_m)$ gravity reads as
\begin{equation}\label{14b}
3H^2 f_R + \frac{1}{2} \left( f-f_R R-f_{L_m}L_m \right) + 3H \dot{f_R}= \frac{1}{2}f_{L_m} \rho ,
\end{equation}
\begin{equation}\label{15b}
\dot{H}f_R + 3H^2 f_R - \ddot{f_R} -3H\dot{f_R} + \frac{1}{2} \left( f_{L_m}L_m - f \right) = \frac{1}{2} f_{L_m}p.
\end{equation}

\section{Cosmological \texorpdfstring{$f(R,L_m)$}{f(R,Lm)} model}\label{IIIb}
\justifying
We consider the following functional form \citep{LB} for our analysis,
\begin{equation}\label{16b} 
f(R,L_m)=\frac{R}{2}+L_m^n + \beta ,
\end{equation}
where $\beta$ and $n$ are free model parameters. Then, for this particular $f(R,L_m)$ model with $L_m=\rho$ \citep{HLR}, the Friedmann  Eqs. \eqref{14b} and \eqref{15b} for the matter-dominated Universe becomes
\begin{equation}\label{17b}
3H^2=(2n-1) \rho^n-\beta,
\end{equation}
\begin{equation}\label{18b}
2\dot{H}+3H^2+\beta=(n-1)\rho^n.
\end{equation}
\justify Moreover, the Eq. \eqref{48} yields the following energy-balance equation corresponding to the model,
\begin{equation}\label{cb}
(2n-1)\dot{\rho}+ 3H(\rho+p)=0.
\end{equation}
In particular, for $n=1$ and $\beta=0$, one can retrieve the usual Friedmann equations for GR. From Eqs. \eqref{17b} and \eqref{18b}, we have
\begin{equation}\label{19b}
\dot{H}+\frac{3n}{2(2n-1)} H^2+\frac{n}{2(2n-1)}\beta=0.
\end{equation}
Then by using $ \frac{1}{H} \frac{d}{dt}= \frac{d}{dln(a)}$, we have the following first order differential equation
\begin{equation}\label{20b}
\frac{dH}{dln(a)}+ \frac{3n}{2(2n-1)}H =-\frac{n\beta}{2(2n-1)}\frac{1}{H}.
\end{equation}
Now, by integrating the above equation, we derive the expression for the Hubble parameter in terms of redshift as follows
\begin{equation}\label{21b}
H(z)=\bigl[ H_0^2 (1+z)^{\frac{3n}{2n-1}}+\frac{\beta}{3} \{ (1+z)^{\frac{3n}{2n-1}}-1 \} \bigr]^\frac{1}{2}.
\end{equation}
Here, $H_0$ is the present value of the Hubble parameter. The deceleration parameter plays a vital role in describing the dynamics of the expansion phase of the Universe, and it is defined as
\begin{equation}\label{22b}
q(z)=-1-\frac{\dot{H}}{H^2}.
\end{equation}
Using Eq.\eqref{21b} in Eq. \eqref{22b}, we have
\begin{equation}\label{23b}
q(z)=-1+\frac{3n(3H_0^2+\beta)}{2(2n-1) \{ 3H_0^2+\beta [ 1-(1+z)^{\frac{3n}{1-2n}} ] \} }.
\end{equation}

\section{Observational constraints}\label{IVb}
\justifying
In this section, we examine the observational aspects of our cosmological model. We used the $H(z)$ and Pantheon datasets to identify the best-fit ranges for the model parameters $n$ and $\beta$. To constrain the model parameters, we employ the standard Bayesian technique and likelihood function along with the MCMC method in \texttt{emcee} python library \citep{mac}. We use the following probability function to maximize the best-fit ranges of the parameters
\begin{equation}\label{24b}
\mathcal{L} \propto exp(-\chi^2/2).
\end{equation} 
Here, $\chi^2$ represents a pseudo-chi-squared function. The $\chi^2$ function used for different datasets is given below.

\subsection{\texorpdfstring{$H(z)$}{H(z)} datasets}
In this section, we have taken an updated set of $57$ data points of $H(z)$ measurements with the corresponding errors $\sigma _{H}$ are tabulated in Table \eqref{Table-1z} with references, in the range of redshift given as $0.07 \leq z \leq 2.41$ \citep{GSS}. In general, there are two well-established techniques for measuring the values of $H(z)$ in a given redshift, namely the line of sight BAO ($26$ points) and the differential age technique ($31$
points) \citep{H1,H2,H3,H4,H5,H6,H7,H8,H9,H10,H11,H12,H13,H14,H15,H16,H17,H18,H19}. Moreover, we have taken $H_0 = 69$ Km/s/Mpc for our analysis \citep{planck_collaboration/2020}. To estimate the mean values of the model parameters $n$ and $\beta$, we define the chi-square function as follows
\begin{equation}\label{25b}
\chi _{H}^{2}(n,\beta)=\sum\limits_{k=1}^{57}
\frac{[H_{th}(z_{k},n,\beta)-H_{obs}(z_{k})]^{2}}{
\sigma _{H(z_{k})}^{2}}.  
\end{equation}
Here, $H_{th}$ denotes the theoretical value of the Hubble parameter obtained by our model, whereas $H_{obs}$ represents its observed value and $\sigma_{H(z_{k})}$ represents the standard deviation. 

\begin{table}[H]
\begin{center}
\begin{tabular}{|c|c|c|c|c|c|c|c|}\hline
\multicolumn{8}{|c|}{57 points of $H(z)$ datasets} \\ \hline
\multicolumn{8}{|c|}{31 points from DA method}  \\ \hline
$z$ & $H(z)$ & $\sigma _{H}$ & Ref. & $z$ & $H(z)$ & $\sigma _{H}$ & Ref. \\ \hline
$0.070$ & $69$ & $19.6$ & \cite{H1} & $0.4783$ & $80$ & $99$ & \cite{H5} \\ \hline
$0.90$ & $69$ & $12$ & \cite{H2} & $0.480$ & $97$ & $62$ & \cite{H1} \\ \hline
$0.120$ & $68.6$ & $26.2$ & \cite{H1} & $0.593$ & $104$ & $13$ & \cite{H3} \\ \hline
$0.170$ & $83$ & $8$ & \cite{H2} & $0.6797$ & $92$ & $8$ & \cite{H3} \\ \hline
$0.1791$ & $75$ & $4$ & \cite{H3} & $0.7812$ & $105$ & $12$ & \cite{H3} \\ \hline
$0.1993$ & $75$ & $5$ & \cite{H3} & $0.8754$ & $125$ & $17$ & \cite{H3} \\ \hline
$0.200$ & $72.9$ & $29.6$ & \cite{H4} & $0.880$ & $90$ & $40$ & \cite{H1} \\ \hline
$0.270$ & $77$ & $14$ & \cite{H2} & $0.900$ & $117$ & $23$ & \cite{H2} \\ \hline 
$0.280$ & $88.8$ & $36.6$ & \cite{H4} & $1.037$ & $154$ & $20$ & \cite{H3} \\ \hline 
$0.3519$ & $83$ & $14$ & \cite{H3} & $1.300$ & $168$ & $17$ & \cite{H2} \\ \hline 
$0.3802$ & $83$ & $13.5$ & \cite{H5} & $1.363$ & $160$ & $33.6$ & \cite{H7} \\ \hline 
$0.400$ & $95$ & $17$ & \cite{H2} & $1.430$ & $177$ & $18$ & \cite{H2} \\ \hline 
$0.4004$ & $77$ & $10.2$ & \cite{H5} & $1.530$ & $140$ & $14$ & \cite{H2} \\ \hline
$0.4247$ & $87.1$ & $11.2$ & \cite{H5} & $1.750$ & $202$ & $40$ & \cite{H2} \\ \hline
$0.4497$ & $92.8$ & $12.9$ & \cite{H5} & $1.965$ & $186.5$ & $50.4$ & \cite{H7}  \\ \hline
$0.470$ & $89$ & $34$ & \cite{H6} &  &  &  &   \\ \hline
\multicolumn{8}{|c|}{26 points from BAO \& other method} \\ \hline
$z$ & $H(z)$ & $\sigma _{H}$ & Ref. & $z$ & $H(z)$ & $\sigma _{H}$ & Ref. \\ \hline
$0.24$ & $79.69$ & $2.99$ & \cite{H8} & $0.52$ & $94.35$ & $2.64$ & \cite{H10} \\ \hline
$0.30$& $81.7$ & $6.22$ & \cite{H9} & $0.56$ & $93.34$ & $2.3$ & \cite{H10} \\ \hline
$0.31$ & $78.18$ & $4.74$ & \cite{H10} & $0.57$ & $87.6$ & $7.8$ & \cite{H14} \\ \hline
$0.34$ & $83.8$ & $3.66$ & \cite{H8} & $0.57$ & $96.8$ & $3.4$ & \cite{H15} \\ \hline
$0.35$ & $82.7$ & $9.1$ & \cite{H11} & $0.59$ & $98.48$ & $3.18$ & \cite{H10} \\ \hline
$0.36$ & $79.94$ & $3.38$ & \cite{H10} & $0.60$ & $87.9$ & $6.1$ & \cite{H13} \\ \hline
$0.38$ & $81.5$ & $1.9$ & \cite{H12} & $0.61$ & $97.3$ & $2.1$ & \cite{H12} \\ \hline
$ 0.40$ & $82.04$ & $2.03$ & \cite{H10} & $0.64$ & $98.82$ & $2.98$ & \cite{H10}  \\ \hline
$0.43$ & $86.45$ & $3.97$ & \cite{H8} & $0.73$ & $97.3$ & $7.0$ & \cite{H13} \\ \hline
$0.44$ & $82.6$ & $7.8$ & \cite{H13} & $2.30$ & $224$ & $8.6$ & \cite{H16} \\ \hline
$0.44$ & $84.81$ & $1.83$ & \cite{H10} & $2.33$ & $224$ & $8$ & \cite{H17} \\ \hline
$0.48$ & $87.79$ & $2.03$ & \cite{H10} & $2.34$ & $222$ & $8.5$ & \cite{H18} \\ \hline
$0.51$ & $90.4$ & $1.9$ & \cite{H12} & $2.36$ & $226$ & $9.3$ & \cite{H19} \\ \hline
\end{tabular}
\end{center}
\caption{Table shows 57 points of $H(z)$ dataset.}\label{Table-1z}
\end{table}

The likelihood contours $1-\sigma$ and $2-\sigma$ for the model parameters $n$ and $\beta$ using the $H(z)$ datasets are presented below.
\begin{figure*}[htbp]
\centering
\includegraphics[scale=0.85]{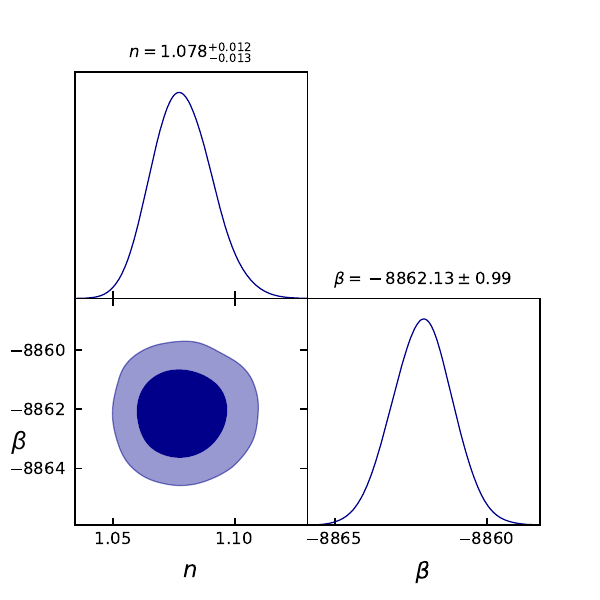}
\caption{Profile of the $1-\sigma$ and $2-\sigma$ contours for the model parameters $n$ and $\beta$ using $H(z)$ datasets}\label{f1a}
\end{figure*}\\
The best-fit ranges obtained for the model parameters are $n=1.078^{+0.012}_{-0.013}$ and $\beta=-8862.13 \pm 0.99$.

\subsection{Pantheon datasets}
Pantheon SNeIa datasets consisting of $1048$ data points have recently been released. The PanSTARSS1 Medium, SDSS, SNLS, Deep Survey, numerous low redshift surveys, and HST surveys contribute to it. Scolnic et al. \citep{Scolnic/2018} put together Pantheon SNeIa samples $1048$ in the redshift range $z \in [0.01,2.3]$. For a spatially flat Universe \citep{planck_collaboration/2020}, the luminosity distance reads as
\begin{equation}\label{26b}
D_{L}(z)= (1+z) \int_{0}^{z} \frac{c dz'}{H(z')}.
\end{equation}
Here, $c$ is the speed of light. For statistical analysis, the $\chi^{2}$ function for supernovae samples is obtained by correlating the theoretical distance modulus 
\begin{equation}\label{27b}
\mu(z)= 5log_{10}D_{L}(z)+\mu_{0}, 
\end{equation}
with 
\begin{equation}\label{28b}
\mu_{0} = 5log(1/H_{0}Mpc) + 25,
\end{equation}
such that
\begin{equation}\label{29b}
\chi^2_{SNeIa}(p_1,....)=\sum_{i,j=1}^{1048}\bigtriangledown\mu_{i}\left(C^{-1}_{SNeIa}\right)_{ij}\bigtriangledown\mu_{j}.
\end{equation}
Here, $p_j$ denotes the free model parameters and $C_{SNeIa}$ represents the covariance metric \citep{Scolnic/2018}, and
  $\quad \bigtriangledown\mu_{i}=\mu^{th}(z_i,p_1,...)-\mu_i^{obs}$,
where $\mu_{th}$ is the theoretical value of the distance modulus, while $\mu_{obs}$ is its observed value.
We have obtained our model's best-fit ranges for parameters $n$ and $\beta$ by minimizing the chi-square function for the supernovae samples. The likelihood contours $1-\sigma$ and $2-\sigma$ for the model parameters $n$ and $\beta$ using the Pantheon data sample are presented below. 
\begin{figure*}[htbp]
\centering
\includegraphics[scale=0.85]{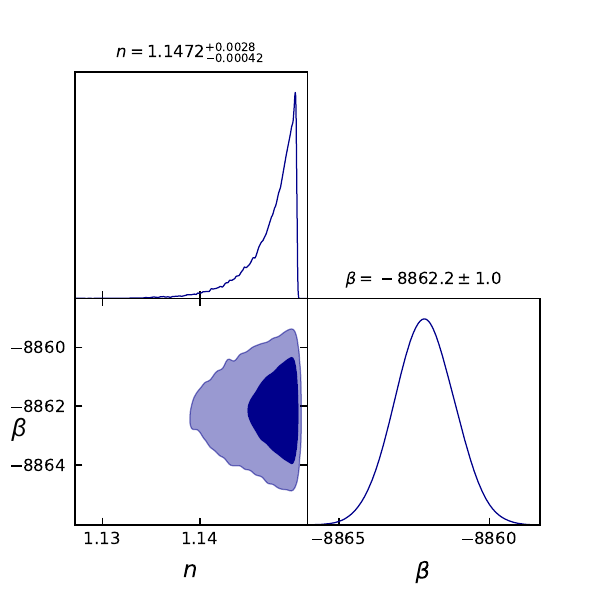}
\caption{Profile of the $1-\sigma$ and $2-\sigma$ contours for the model parameters $n$ and $\beta$ using Pantheon datasets}\label{f2a}
\end{figure*}\\
The best-fit ranges obtained for the model parameters are $n=1.1472^{+0.0028}_{-0.00042}$ and $\beta=-8862.2 \pm 1.0 $.

\subsection{\texorpdfstring{$H(z)$}{H(z)} + Pantheon datasets}
The $\chi^{2}$ function for the $H(z)$ + Pantheon datasets is given as 
\begin{equation}\label{cmb}
\chi^{2}_{total}= \chi^{2}_H + \chi^2 _{SNeIa}.
\end{equation}
The contours $1-\sigma$ and $2-\sigma$ for the model parameters $n$ and $\beta$ using $H(z)$ + Pantheon datasets are presented below.
\begin{figure*}[htbp]
\centering
\includegraphics[scale=0.85]{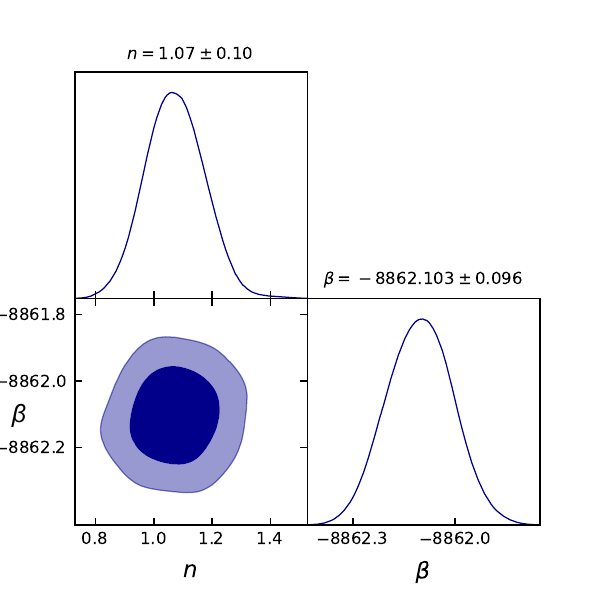}
\caption{Profile of the $1-\sigma$ and $2-\sigma$ contours for the model parameters $n$ and $\beta$ using $H(z)$ + Pantheon datasets}\label{f3a}
\end{figure*}\\
The best-fit ranges obtained for the model parameters are $n=1.07 \pm 0.10$ and $\beta=-8862.103 \pm 0.096 $.

The evolution profile of the density parameter and the deceleration parameter corresponding to the constrained values of the model parameters are presented below.
\begin{figure*}[htbp]
\centering
\includegraphics[scale=0.55]{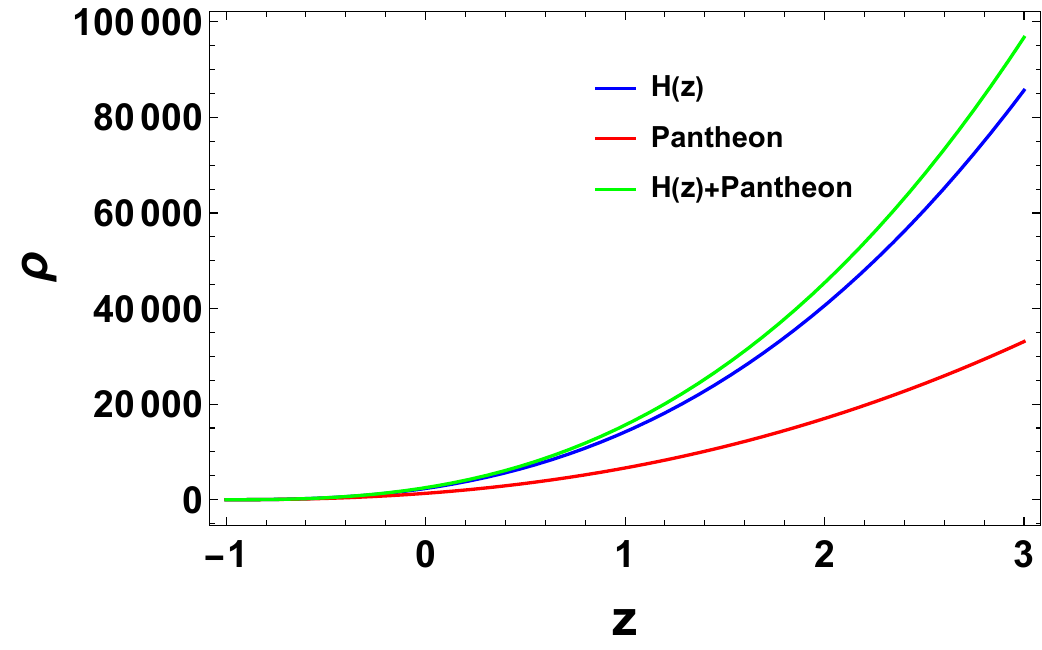}
\caption{Profile of the density parameter vs redshift.}\label{f4a}
\end{figure*}
\begin{figure*}[htbp]
\centering
\includegraphics[scale=0.55]{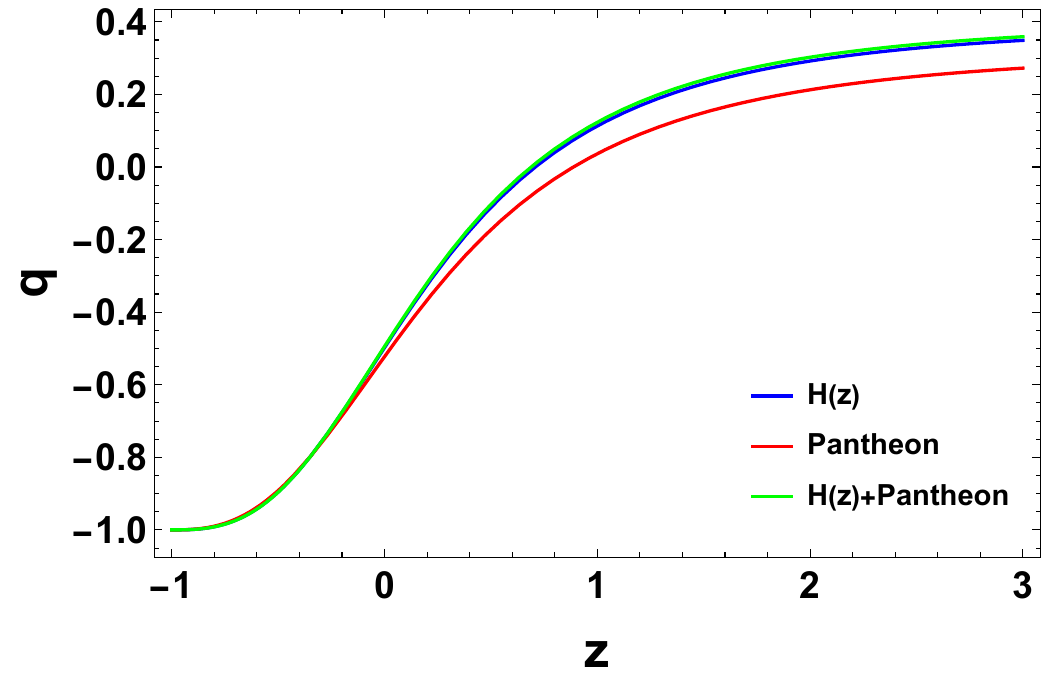}
\caption{Profile of the deceleration parameter vs redshift.}\label{f5a}
\end{figure*}\\
Fig. \ref{f4a} indicates \footnote{In Mathematica, graphs are plotted using functions like \textit{Plot}, \textit{Plot3D}, and \textit{ParametricPlot}, depending on the type of data.} that the density parameter of the cosmic fluid shows positive behavior and vanishes in the far future. The evolution of the deceleration parameter, illustrated in Fig. \ref{f5a}, indicates that our Universe has shifted from a decelerating phase to an accelerating one in more recent times. The transition redshift values derived from the model parameters, as determined by the $H(z)$, Pantheon, and the combined $H(z)$ + Pantheon datasets, are $z_t=0.708^{+0.029}_{-0.031}$, $z_t=0.887^{+0.0075}_{-0.0009}$, and $z_t=0.688^{+0.262}_{-0.224}$, respectively. Furthermore, the current value of the deceleration parameter is noted as $q_0=-0.497^{+0.005}_{-0.004}$ for the $H(z)$ datasets, $q_0=-0.5223^{+0.00003}_{-0.0008}$ for the Pantheon datasets and $q_0=-0.494^{+0.05}_{-0.035}$ for the combined $H(z)$ + Pantheon datasets.

\section{Perturbation analysis of the Hubble parameter}\label{Vb}
\justifying
In this section, we investigate the stability of the solution obtained from our proposed $f(R,L_m)$ model under observational constraints. We have considered a linear perturbation of the Hubble parameter $H(z)$ as
\begin{equation}\label{30b}
H^\ast(z)= H(z)(1+\delta(z)).
\end{equation}
Here, $H^\ast(z)$ represents the perturbed Hubble parameter, and $\delta(z)$ represents the perturbation term. Now, by using Eqs. \eqref{21b} and \eqref{30b} in the matter conservation Eq. \eqref{cb}, we obtained the following expression
\begin{multline}\label{31b}
\left(\frac{(z+1)^{\frac{3 n}{2 n-1}} \left(\beta +H_0^2 (6 \delta (z)+3)-2 \beta  \left((z+1)^{\frac{3 n}{1-2 n}}-1\right) \delta (z)\right)}{2 n-1}\right)^{1/n} \\ 
\times \left(\frac{(z+1) \left(2 \delta '(z) \left(3 H_0^2 (z+1)^{\frac{3 n}{2 n-1}}+\beta  \left((z+1)^{\frac{3 n}{2 n-1}}-1\right)\right)+\frac{3 n \left(\beta +3 H_0^2\right) (z+1)^{\frac{n+1}{2 n-1}} (2 \delta (z)+1)}{2 n-1}\right)}{\beta +(2 \delta (z)+1) \left(3 H_0^2 (z+1)^{\frac{3 n}{2 n-1}}+\beta  \left((z+1)^{\frac{3 n}{2 n-1}}-1\right)\right)}-3\right) =0.
\end{multline}
We solve the Eq. \eqref{31b} numerically \footnote{In Mathematica, numerical solutions for first-order differential equations are obtained using the fourth-order Runge-Kutta method, which ensures stability and higher accuracy.} since it is highly nonlinear, and we present the behavior of the perturbation term $\delta(z)$ corresponding to the values of the model parameters constrained by observational datasets. 
\begin{figure*}[htbp]
\centering
\includegraphics[scale=0.65]{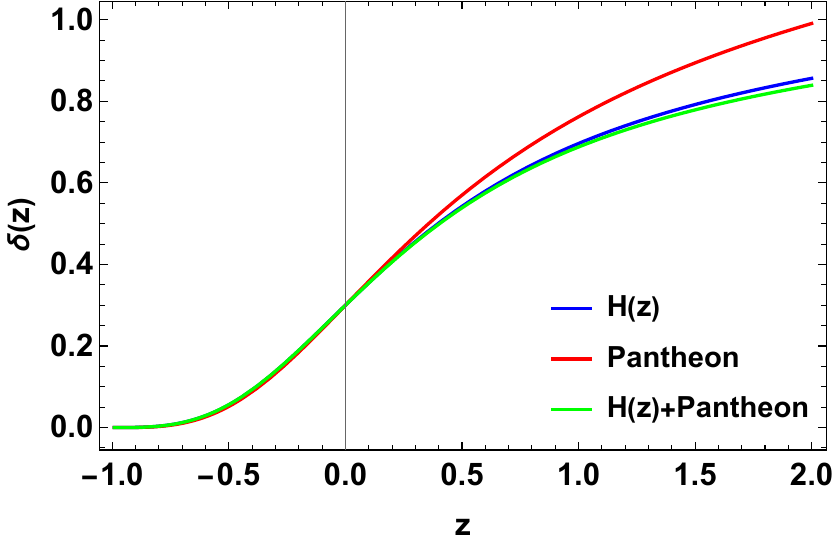}
\caption{Profile of the perturbation term $\delta(z)$ corresponding to the values of model parameters constrained by $H(z)$, Pantheon, and the combined $H(z)$ + Pantheon datasets.}\label{f6a}
\end{figure*}\\
From Fig. \ref{f6a}, it is clear that the perturbation term $\delta(z)$ decays rapidly at late times for the constrained values of the model parameters. Therefore, the solution of our cosmological $f(R, L_m)$ model shows a stable behavior.

\section{\texorpdfstring{$Om(z)$}{Om(z)} diagnostics}\label{VIb}
\justifying Om diagnostic serves as a valuable method for categorizing various cosmological DE models \citep{Om}. It is known for its simplicity, relying solely on the first derivative of the cosmic scale factor. In the context of a spatially flat Universe, it is defined as follows
\begin{equation}\label{omb}
Om(z)= \frac{\big(\frac{H(z)}{H_0}\big)^2-1}{(1+z)^3-1}.
\end{equation}
The negative slope of $Om(z)$ corresponds to quintessence-type behavior, while the positive slope corresponds to phantom behavior. The constant nature of $Om(z)$ represents the $\Lambda$CDM model. 
\begin{figure*}[htbp]
\centering
\includegraphics[scale=0.55]{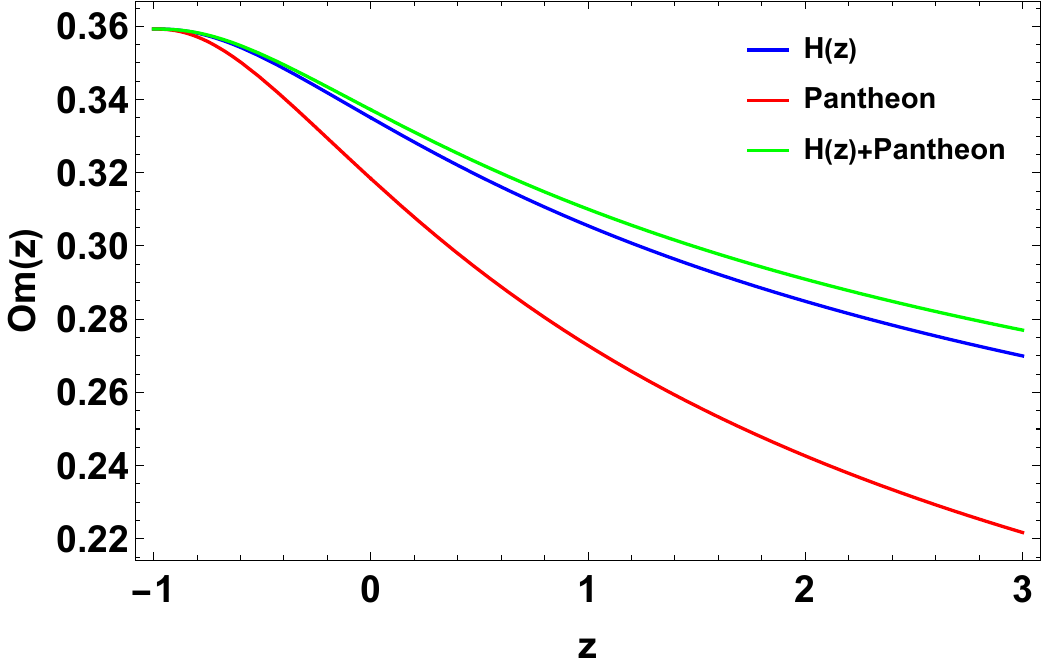}
\caption{Profile of the Om diagnostic parameter corresponding to the values of model parameters constrained by $H(z)$, Pantheon, and the combined $H(z)$ + Pantheon datasets.}\label{f7a}
\end{figure*}
From Fig. \ref{f7a}, we observed that the Om diagnostic parameter for the set of constrained values of the model parameters has a negative slope in the entire domain. Thus, we can conclude from the Om diagnostic test that our cosmological $f(R,L_m)$ model represents quintessence-type behavior.

\section{Conclusions}\label{VIIb}
\justifying
In this chapter, we explored the late-time expansion of the Universe using the framework of $f(R,L_m)$ gravity theory, focusing on a non-linear model. We derived the equations of motion applicable to the FLRW background and provided the analytical solution for our cosmological $f(R,L_m)$ model in Eq. \eqref{21}. Furthermore, we obtained optimal values for the model parameters by analyzing datasets from $H(z)$, the recently released Pantheon dataset, and the combined $H(z)$ + Pantheon datasets.

In addition, we investigated the behavior of the density parameter and the deceleration parameter for the constrained values of the model parameters. The evolution profile of the deceleration parameter in Fig. \ref{f5a} indicated a recent transition of the Universe from a decelerated to an accelerated phase. In contrast, the density parameter in Fig. \ref{f4a} exhibited a positive behavior, as expected. Furthermore, we examined the stability of the obtained solution under observational constraints by considering a linear perturbation of the Hubble parameter. From Fig. \ref{f6a}, we concluded that, for the set of constrained model parameters, the solution obtained from our cosmological $f(R,L_m)$ model demonstrated stable behavior. Finally, the evolution profile of the Om diagnostic parameter, presented in Fig. \ref{f7a}, indicated that our cosmological $f(R,L_m)$ model followed the quintessence scenario. Our findings show that our cosmological $f(R,L_m)$ model was consistent with the constraint $\frac{f_{Lm}(R, L_m)}{f_R(R, L_m)} > 0$ derived in \citep{WG}, which was equivalent to the condition $n>0$ for the model we considered.

Building on our exploration of cosmic expansion within the $f(R,L_m)$ gravity framework, we now extend our analysis to investigate rapid late-time acceleration. By introducing a new parameterization of the Hubble parameter, independent of any specific model, we further examine the transition from a decelerating to an accelerating phase, supported by observational constraints.


%% file: Chapters/Chapter3.tex

\chapter{Late time cosmic acceleration through parametrization of Hubble parameter in \texorpdfstring{$f(R,L_m)$}{f(R,Lm)} gravity} 

\label{Chapter3} 

\lhead{Chapter 3. \emph{Late time cosmic acceleration through parametrization of Hubble parameter in $f(R,L_m)$ gravity}} 


\vspace{10 cm}
* The following publications cover the work in this chapter:\\

\textit{Late time cosmic acceleration through parameterization of the Hubble parameter in $f(R,L_m)$ gravity}, Physics of the Dark Universe, \textbf{46}, 101639 (2024).

\clearpage
This chapter investigates the rapid expansion of the Universe during its later stages, analyzed through the framework of $f(R,L_m)$ gravity theory. We introduce a new parameterization of the Hubble parameter that is independent of any particular model, which allows us to effectively examine the Friedmann equations in the context of the FLRW background. To derive the model parameters, we utilize an MCMC approach, analyzing comprehensive datasets that include CC measurements, independent BAO assessments, and Pantheon+SH0ES datasets. The path traced by the deceleration parameter illustrates the transition from a decelerating to an accelerating phase in the evolution of the Universe. Furthermore, we explore the dynamics of key cosmological variables for two distinct nonlinear $f(R,L_m)$ models, examining aspects such as the density parameter, pressure, the EoS parameter and the associated energy conditions.

\section{Introduction}\label{sec1}
\justifying
Representing late-time cosmic acceleration using a contemporary approach involves the use of parameterization \citep{pac1,pac2}. This method employs particular mathematical functions or parameters, such as the deceleration parameter, the Hubble parameter, and the jerk parameter, to characterize the behavior of cosmic expansion. These values accurately reflect the observed accelerated expansion of the Universe. Flexible modeling is enabled by this method, and it establishes a foundation for comprehending the dynamics and traits of the expanding Universe. In this chapter, we investigate a simple new way to express the Hubble parameter to establish a framework for the Universe that is accelerating \citep{pac3}. Parameterization enables deviations from the standard $\Lambda$CDM model at low and high redshifts. We explored the FLRW background within the $f(R,L_m)$ gravity framework. The best-fit values of the model parameters were found using modern observational datasets. We integrated the CC, BAO, and Pantheon+SH0ES datasets to limit the cosmological model in our study. Additionally, we employed the \textit{emcee} library to facilitate an MCMC technique for parameter estimation \citep{mac}.

The section is organized as follows. Sec. \ref{sec3} discusses the cosmological parameters derived from a novel parameterization of the Hubble parameter, which allows for the calculation of exact solutions to the field equations. In Sec. \ref{sec4}, we use the combined CC, BAO, and Pantheon+SH0ES datasets to constrain the model parameters $H_0$, $\zeta$, and $\eta$. We also examine how the deceleration parameter changes in relation to these parameters. Sec. \ref{sec5} delves into two alternative $f(R,L_m)$ cosmological models and discusses the behavior of various cosmological parameters, such as the density parameter, pressure, EoS parameter, and energy conditions. Finally, we consolidate and report our results in Sec. \ref{sec6}.

\section{Parametrization of the Hubble parameter}\label{sec3}
\justifying
The system of field equations mentioned earlier generally consists of two independent equations, which involve four unknowns: $\rho$, $p$, $f(R,L_m)$, and $H$. Additional constraints are required for a complete solution from a mathematical perspective to analyze the temporal development of the Universe and various cosmological parameters. Various physical motivations for selecting these constraints exist in the literature, the most renowned being the model-independent method to explore DE dynamics \citep{sha}. This method includes the parameterization of cosmological factors, including the Hubble parameter, the deceleration parameter, and the EoS parameter. The parameterization of the EoS parameter and the deceleration parameter has also been extensively investigated \citep{esc,ban,cun}. Numerous other approaches have been proposed to parameterize various cosmological parameters and the slowing parameter. These methods have been extensively explored in the literature to tackle several cosmological problems, including the Hubble tension, the horizon problem, the perpetual deceleration problem, and the initial singularity problem. See \citep{pac1,pac2} for a detailed discussion of these cosmological parametrization techniques. However, several earlier parameterization approaches \citep{sib1,sib2} have shown inconsistencies in accurately representing the current deceleration parameter and the redshift value of the transition, as predicted by $\Lambda$CDM. Therefore, we consider the parameterization for the Hubble function based on these results as \citep{pac3} that led us to better describe these cosmological epochs in alignment with observational data.
\begin{equation}\label{3af}
H(z)=H_0 \bigg[(1-\zeta)+(1+z)(\zeta+\eta z)\bigg]^\frac{1}{2}.
\end{equation}
The Hubble constant at redshift $z=0$ is represented by $H_0$. Parameters $\zeta$ and $\eta$ can be determined from observational data. A key cosmological parameter that describes the expansion dynamics of the Universe is the deceleration parameter. According to cosmological models, the Universe has shifted from an early stage of deceleration $(q > 0)$ to its present stage of acceleration $(q < 0)$, characterized by specific values of the transition redshift $z_t$. Additionally, the deceleration parameter is constrained within the range of $ -1 \leq q \leq 0$, suggesting that the Universe has recently begun accelerating based on the observational evidence discussed here. The definition of this parameter in relation to the Hubble parameter is as follows
\begin{equation}\label{3bf}
q(z)=-1-\frac{\dot H }{H^2}.
\end{equation}
Additionally, by leveraging the connection between redshift and the scale factor of the Universe, which is expressed as $ a(t) = \frac{1}{1+z} $, we can establish how cosmic time is related to redshift as
\begin{equation}\label{3cf}
\frac{\text{d}}{\text{d}t}=\frac{\text{d}z}{\text{d}t}\frac{\text{d}}{\text{d}z}=-(1+z)H(z)\frac{\text{d}}{\text{d}z}.
\end{equation}
Based on the equation we have discussed above, the Hubble parameter can be expressed as follows
\begin{equation}\label{3df}
\frac{\text{d}H}{\text{d}t}=-(1+z)H(z)\frac{\text{d}H}{\text{d}z}.
\end{equation}
Now, by applying Eqs. \eqref{3af} and \eqref{3bf}, and with the assistance of Eq. \eqref{3df}, our model defines the deceleration parameter $q(z)$ in the following way
\begin{equation}\label{3ef}
q(z)=\frac{\zeta(1-z)+\eta(1+z)+2}{2(\eta z^2 +(\zeta+\eta)z+1)}.
\end{equation}
To better understand how these cosmological parameters behave, we will determine the optimal values for the model parameters $\zeta$ and $\eta$. This will be done by analyzing the combined data, including the CC, BAO, and Pantheon+SH0ES datasets.

\section{Observational data and methodology}\label{sec4}
\justifying
This section compares the projected results of the theoretical model through observational data using statistical evaluation. The primary objective is to define the limits for the independent variables $H_0$, $\zeta$, and $\eta$ in the model. The study uses the CC, BAO, and Pantheon+SH0ES datasets. Using an MCMC sample distribution methodology through the \texttt{emcee} Python library \citep{mac} and the likelihood function, Bayesian statistical methods are used to determine a posterior probability.

We utilize the $\chi^2$ function associated with the CC method for conducting an MCMC analysis, which is defined as follows
\begin{equation}\label{4af}
\chi _{CC}^{2}(\Phi)=\sum\limits_{k=1}^{31}
\frac{[H_{th}(z_{k},\Phi)-H_{obs}(z_{k})]^{2}}{\sigma _{H}^2(z_{k})}.  
\end{equation}
In this context, $H_{th}$ signifies the theoretical value of the Hubble parameter based on a specific model defined by the parameters $\Phi$. On the other hand, $H_{obs}$ denotes the observed value of the Hubble parameter. The uncertainty related to the observed value of $H$ in each redshift $z_k$ is represented by the term $\sigma _{H}$.

The $\chi ^2$ function is computed to incorporate the BAO data into the study, much like the CC approach does, which is as follows
\begin{equation}\label{4bf}
\chi _{BAO}^{2}(\Phi)=\sum\limits_{k=1}^{26}
\frac{[H_{th}(z_{k},\Phi)-H_{obs} ^{BAO}(z_{k})]^{2}}{
\sigma _{H}^2(z_{k})}. 
\end{equation}
In this context, $H_{th}$ refers to the theoretical values of the Hubble parameter associated with a specific model characterized by parameters $\Phi$. On the other hand, $\sigma_{H}$ indicates the uncertainty related to the observed values of $H^{BAO}$, while $H^{BAO} _{obs}$ denotes the Hubble parameter that has been observed by the BAO method. 

The dataset known as Pantheon+SH0ES includes distance moduli obtained from the light curves of $1701$ SNeIa, covering a redshift range from $0.001$ to $2.2613$. These data were compiled from $18$ distinct surveys \citep{Brout}. Among these light curves, $77$ corresponds to galaxies that contain Cepheid variable stars. A key advantage of the Pantheon+SH0ES dataset is its ability to place constraints on other model parameters, as well as the Hubble constant $H_0$. To align the model parameters with the Pantheon+SH0ES samples, we utilize a minimization approach on the $\chi^2$ function,
\begin{equation}\label{4df}
\chi^2 _{Pan+SH0ES}=\Delta \mu^T (C^{-1} _{stat+sys})\Delta \mu.
\end{equation}
The distance residual is denoted as $\Delta \mu$. The covariance matrix for the Pantheon+SH0ES sample is represented as $C_{stat+sys}$, constructed by accounting for both systematic and statistical uncertainties,
\begin{equation}\label{4ef}
\Delta \mu_i=\mu_i-\mu_{th}(z_i).
\end{equation}
The observed distance modulus is represented as $\mu_i$. It is crucial to understand that $\mu_i=m_{Bi}-M$, where $m_{Bi}$ denotes the apparent magnitude and $M$ represents the fiducial magnitude. The theoretical distance modulus, indicated as $\mu_{th}$, can be determined using the following expression
\begin{equation}\label{4ff}
\mu_{th} =5\text{log}_{10} \frac{d_L(z,\Phi)}{1Mpc} +25.
\end{equation}
The model-based luminosity distance, denoted as $d_L$, is given in megaparsecs (Mpc)
\begin{equation}\label{4gf}
d_L (z,\Phi)=\frac{c(1+z)}{H_0}\int_0 ^z \frac{\text{d}\gamma}{E(\gamma)}.
\end{equation}
When light travels at the speed of light, denoted $c$, the relationship can be expressed as $E(z) = \frac{H(z)}{H_0}$. In the context of analyzing the SNeIa data, there is a degeneracy between the parameters $H_0$ and $M$. To address this issue, the SNeIa distance residuals are presented in Eq. \eqref{4df} have been adjusted based on findings from previous research \citep{Brout,bro1,sco,per}. In particular, the adjusted residuals, referred to as $\Delta\Bar{\mu}$, are defined as follows
\begin{equation}\label{4hf}
\Delta\Bar{\mu}=
\begin{dcases}
\mu_i -\mu_i ^{\text{Ceph}},& \ $if$ \ i \in \ $Cephied hosts$.\\
\mu_i -\mu_{th}(z_i), &  $otherwise$. \\
\end{dcases}
\end{equation}
In our study, we refer to the Cepheid host of the $i^{th}$ SNeIa as $\mu_i^{\text{Ceph}}$, with data sourced from SH0ES. It is important to understand that the difference between the values $\mu_i$ and $\mu_i^{\text{Ceph}}$ is affected by both the absolute magnitude $M$ and the Hubble constant $H_0$. For our initial analysis, we used an absolute magnitude of $M = -19.253$, which was derived from the distances to the Cepheid hosts provided in the SH0ES dataset \citep{rie}.

The error bar plots for the datasets CC, BAO, and Pantheon+SH0ES are presented in the following figures Fig. \ref{f1f}, Fig. \ref{f2f}, and Fig. \ref{f3f}, respectively.
\begin{figure*}[htbp]
\includegraphics[scale=0.65]{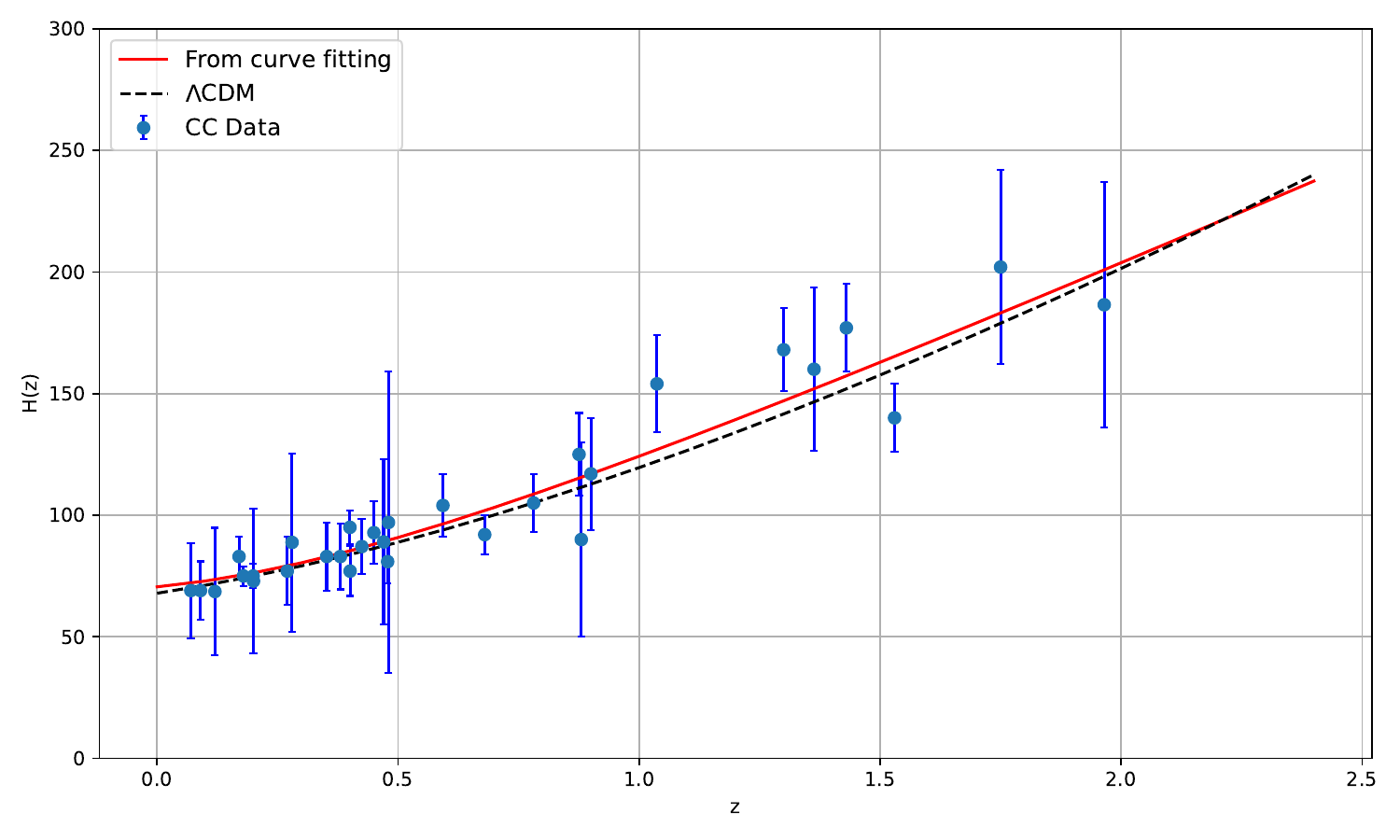}
\caption{Profile of the fitting of the $H(z)$ versus redshift $z$ for our proposed model (red line) in comparison to the standard  $\Lambda$CDM model (black dashed line). An error bar plot that represents the $31$ CC dataset points taken into account for the analysis is also included.}\label{f1f}
\end{figure*}
\begin{figure*}[htbp]
\includegraphics[scale=0.65]{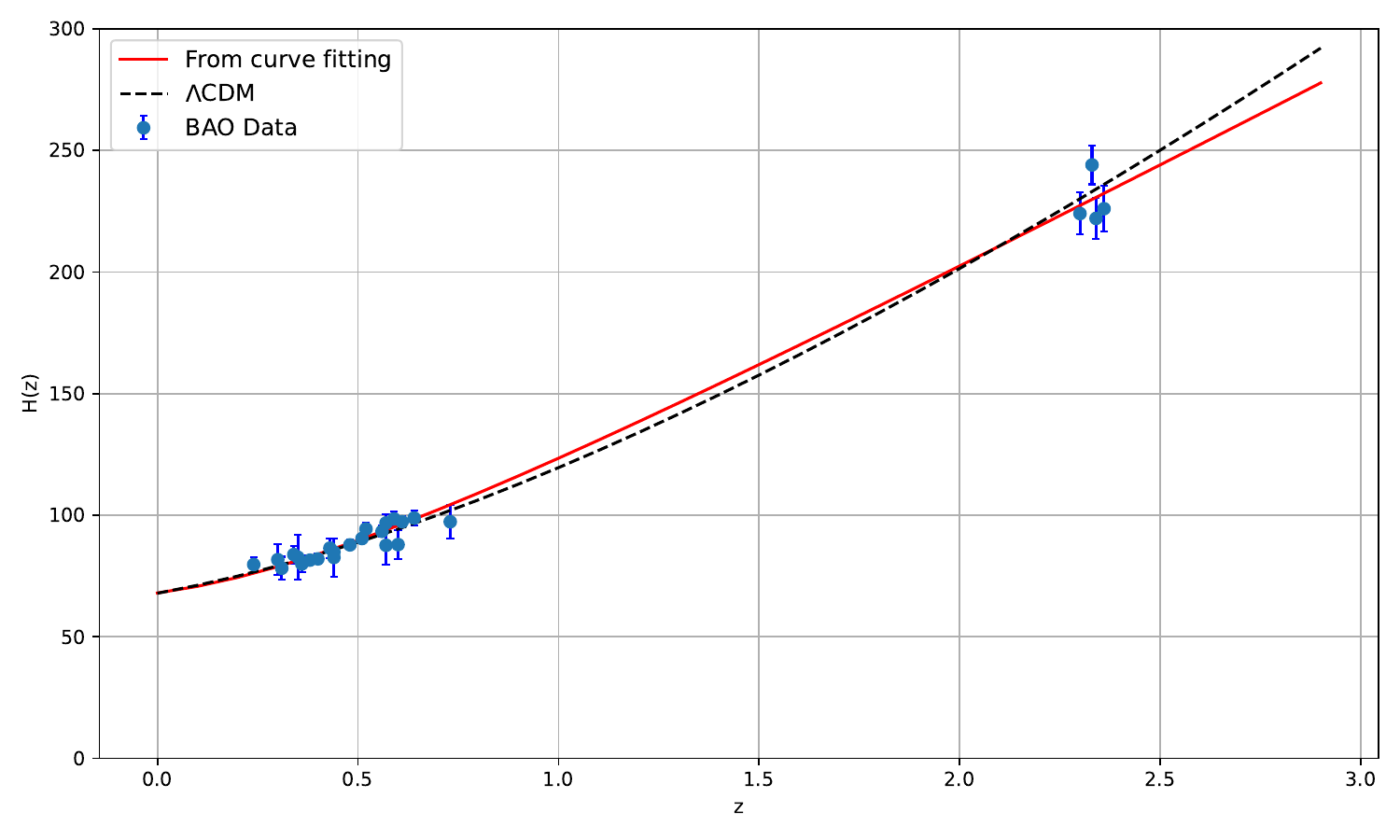}
\caption{Profile of the fitting of the $H(z)$ versus redshift $z$ for our proposed model (red line) in comparison to the standard $\Lambda$CDM model (black dashed line). An error bar plot that represents the $26$ BAO dataset points taken into account for the analysis is also included.}\label{f2f}
\end{figure*}
\begin{figure*}[htbp]
\includegraphics[scale=0.65]{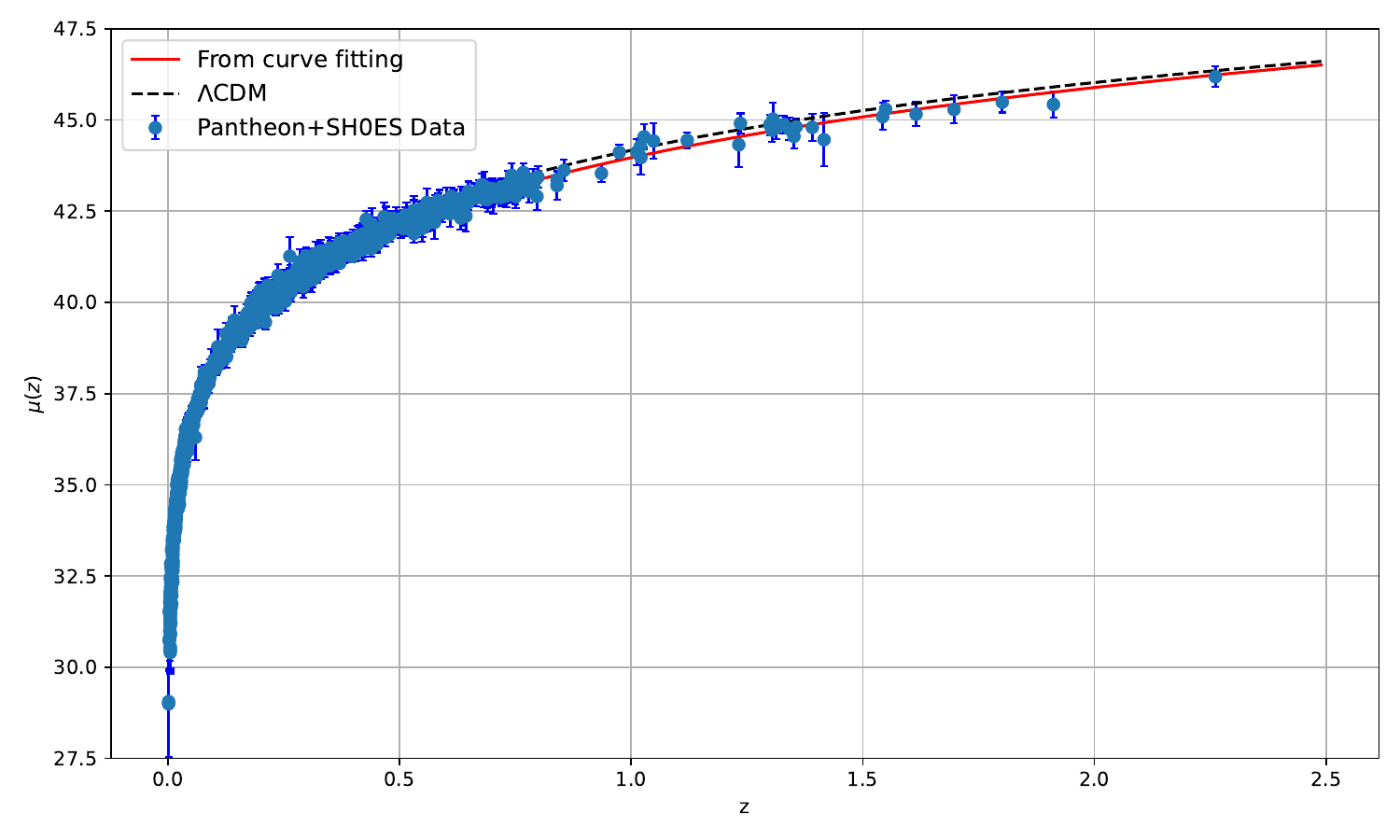}
\caption{Profile of the fitting of the $\mu(z)$ versus redshift $z$ for our proposed model (red line) in comparison to the standard $\Lambda$CDM model (black dashed line). An error bar plot that represents the $1701$ points of the Pantheon+SH0ES dataset taken into account for the analysis is also included.}\label{f3f}
\end{figure*}

To develop the complete $\chi^2_{\text{tot}}$ function for a thorough analysis that includes the CC, BAO, and Pantheon+SH0ES datasets, we combine the individual $\chi^2_{\text{SNeIa}}$ functions as outlined
\begin{equation}\label{13f}
\chi^2 _{tot}=\chi^2 _{CC}+\chi^2 _{BAO}+\chi^2 _{SNeIa}.
\end{equation}

The contour plots depict the combined constraints on the model parameters and showcase our analytical findings. These graphs, representing various confidence levels, highlight areas in the parameter space that align with the observational data. We emphasize regions where the model closely matches the observed data, with contours extending to a confidence level of $3-\sigma$ (99.7\%).
\begin{figure*}[htbp]
\centering
\includegraphics[scale=0.65]{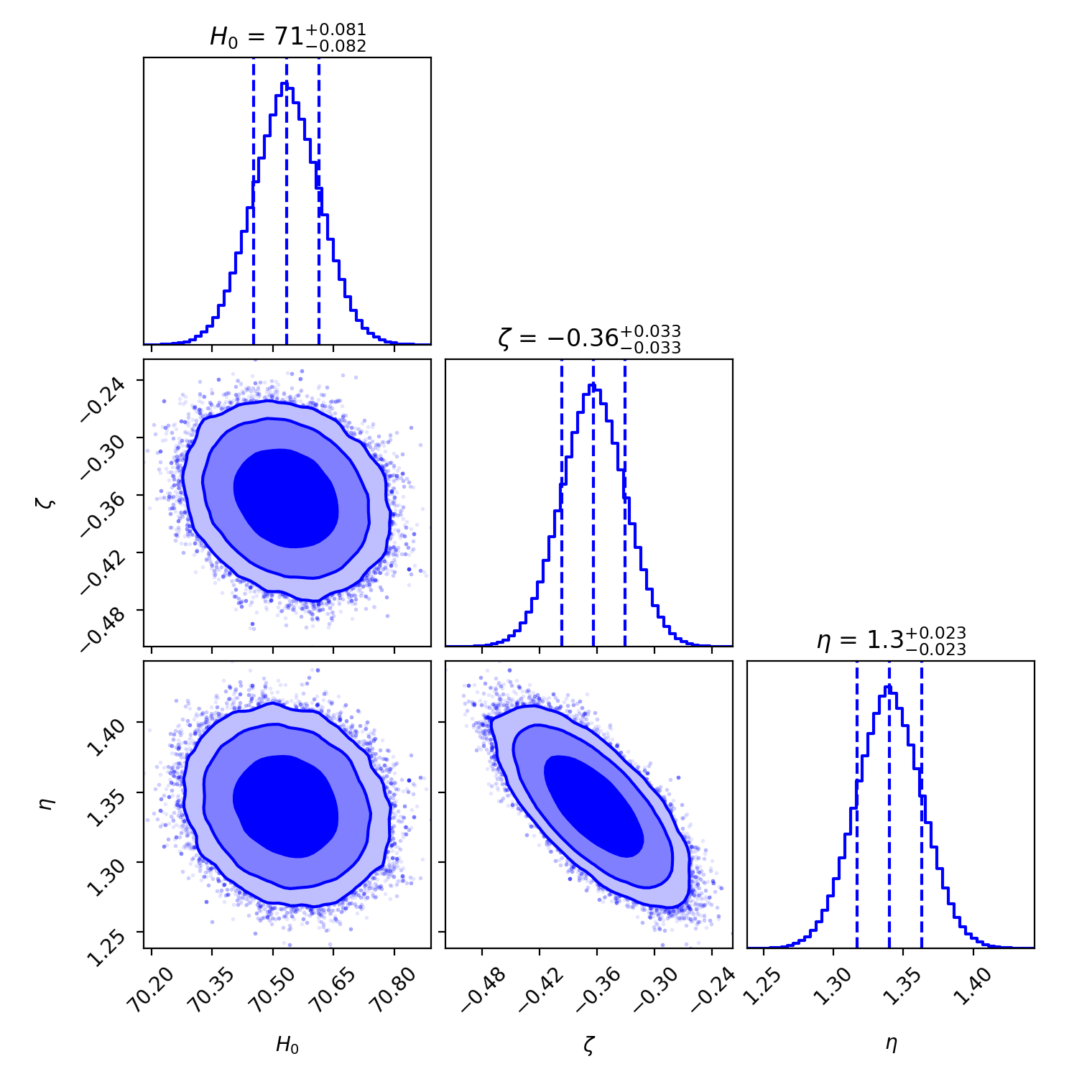}
\caption{Profile of the 2D contour plot of the model parameters $H_0$, $\zeta$, and $\eta$, based on a combined examination of the CC, BAO, and Pantheon+SH0ES datasets, displaying the most likely values and the confidence areas up to 3$-\sigma$.}\label{f4f}
\end{figure*}

We have considered the Gaussian priors for the free parameters as $H_0 \in [50,100]$, $\zeta \in [-2,0]$, and $\eta \in [0,5]$, along with the $150$ walkers and each of them having $2000$ iterations. Using the combined data sets of CC, BAO and Pantheon+SH0ES, we have determined the best fit values for the model parameters with the previously specified setup: $H_0=71^{+0.081}_{-0.082}$, $\zeta=-0.36 \pm  0.033$ and $\eta=1.3 \pm 0.023$. Moreover, we obtained the minimum value of $\chi^2 _{tot}$ as $\chi^2 _{min}=1650.36$. Fig. \ref{f4f} shows the results as 2D contour plots with uncertainties of up to $3-\sigma$. 

\subsection{Model comparison}\label{sbsec4}
It is crucial to perform a statistical analysis using the Akaike Information Criterion (AIC) and the Bayesian Information Criterion (BIC) to gauge the robustness of an MCMC study \citep{lid}. Here is how the first criterion, $AIC$, is defined as
\begin{equation}\label{4if}
AIC=\chi^2 _{min}+2d.
\end{equation}
In this case, $d$ stands for the number of parameters in the given model. To facilitate comparison with the standard $\Lambda$CDM model, we define $\Delta AIC=\mid AIC_{Model}-AIC_{\Lambda CDM}\mid$. A result in $4<\Delta AIC\leq7$ suggests moderate approval of the theoretical model, while a value less than $2$ indicates considerable support. Furthermore, there is no evidence for the suggested model if the value of $\Delta AIC$ is greater than $10$. The following is how the second criterion, $BIC$, is expressed as
\begin{equation}\label{4jf}
BIC=\chi^2 _{min}+dln(N).
\end{equation}
The number of data samples used in the MCMC analysis is denoted by $N$ in this case. Similarly, a range of $2<\Delta BIC\leq6$ indicates moderate support, and a $\Delta BIC$ of less than $2$ indicates strong support for the suggested theoretical model. We derive $AIC_{Model} = 1656.36$ and $BIC_{Model} = 1672.79$ using the minimum value $\chi^2$. This led to $\Delta AIC=1.83$ and $\Delta BIC=3.64$. The $AIC_{\Lambda CDM} = 1658.2$ and $BIC_{\Lambda CDM} = 1669.15$ are the equivalent values for the $\Lambda$CDM model. The proposed $f(R,L_m)$ model has substantial support, as demonstrated by the $\Delta AIC$ value and the $\Delta BIC$ value.

\subsection{Evolution of the \texorpdfstring{$q(z)$}{q(z)} and phase transition}\label{sbsec5}
\begin{figure*}[htbp]
\centering
\includegraphics[scale=0.65]{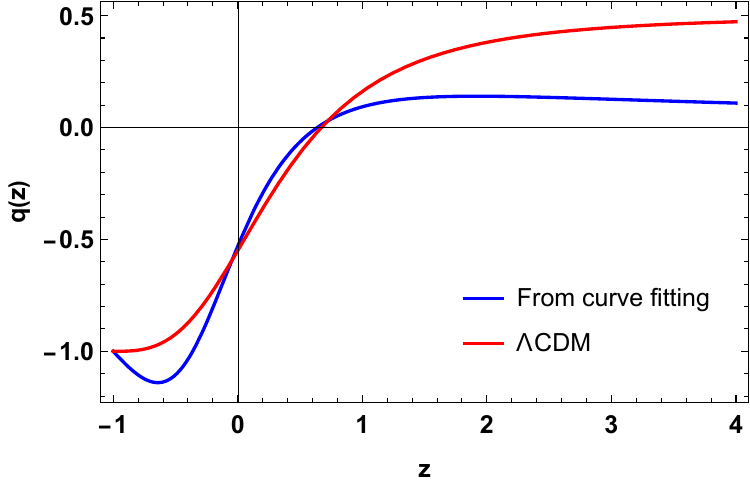}
\caption{Profile of the deceleration parameter, in comparison to the standard $\Lambda$CDM model, with the constraint values from the combined CC, BAO, and Pantheon+SH0ES datasets.}\label{f5f}
\end{figure*}

The evolution of $q(z)$, the deceleration parameter as a function of redshift, plays a crucial role in understanding cosmic expansion and phase transitions in the universe. Initially, in the radiation-dominated and matter-dominated eras, $q(z)$ was positive, indicating a decelerating expansion due to gravitational attraction. However, as DE began to dominate, a phase transition occurred, leading to a negative $q(z)$, signifying an accelerated expansion. This transition marks the shift from a decelerating to an accelerating Universe, profoundly impacting cosmological models and our understanding of the large-scale structure of the cosmos.\\
Fig. \ref{f5f} illustrates the change in the deceleration parameter based on the values of the model parameters. It demonstrates the transition of the cosmological model from a decelerating phase to an accelerating phase. Taking into account the constraints on the model parameters from the combined datasets CC, BAO, and Pantheon+SH0ES, the redshift at which this transition occurs is found to be $z_t = 0.6459$. Furthermore, the current value of the deceleration parameter is $q_0 = -0.5249$. Note that this present value of the deceleration parameter and the redshift of the transition is consistent with the value predicted by the standard $\Lambda$CDM model.
With the theoretical formulations and numerical values of the model parameters in hand, we are ready to delve into the analysis of the model's physical dynamics. In the following section, we examine the physical dynamics of other important cosmological parameters.

\section{Cosmological \texorpdfstring{$f(R,L_m)$}{f(R,Lm)} models}\label{sec5}
\justifying
The focus is on the examination of two non-linear $f(R,L_m)$ models, with the inclusion of free parameters $\alpha$ and $\lambda$

\subsection{Model I : \texorpdfstring{$f(R,L_m)=\frac{R}{2}+L_m ^\alpha$}{f(R,Lm)} }\label{sbsec1}
\begin{figure*}[htbp]
\centering
\subfigure{\includegraphics[width=7.5cm,height=5cm]{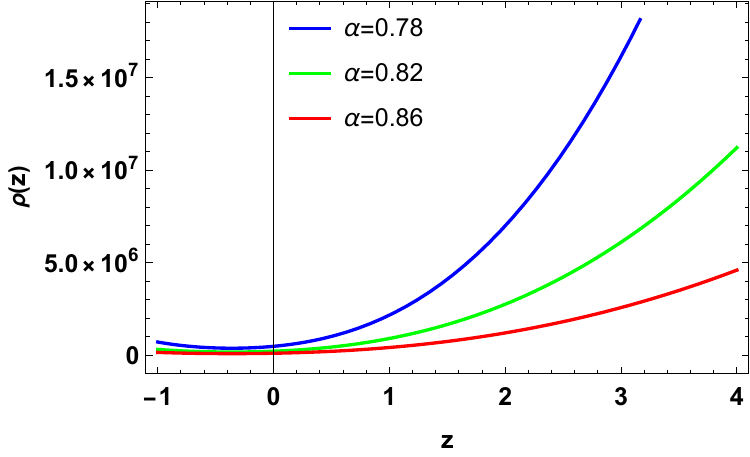}}\,\,\,\,\,\,\,\,\,\,\,\,
\subfigure{\includegraphics[width=7.5cm,height=5cm]{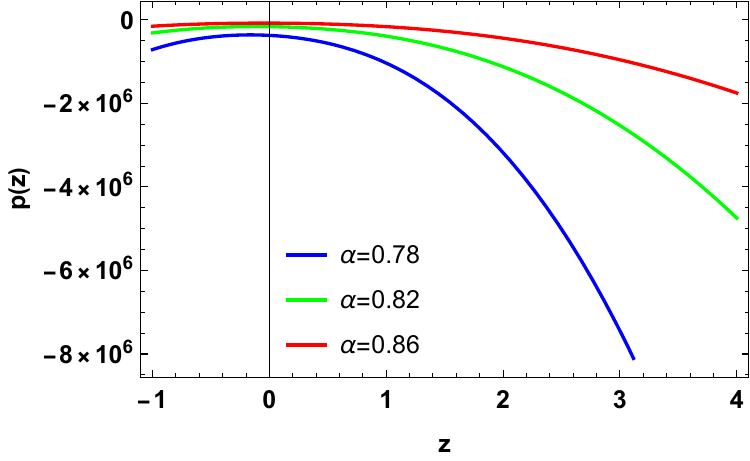}}\,\,\,\,\,\,\,\,\,\,\,\,
\subfigure{\includegraphics[width=7.5cm,height=5cm]{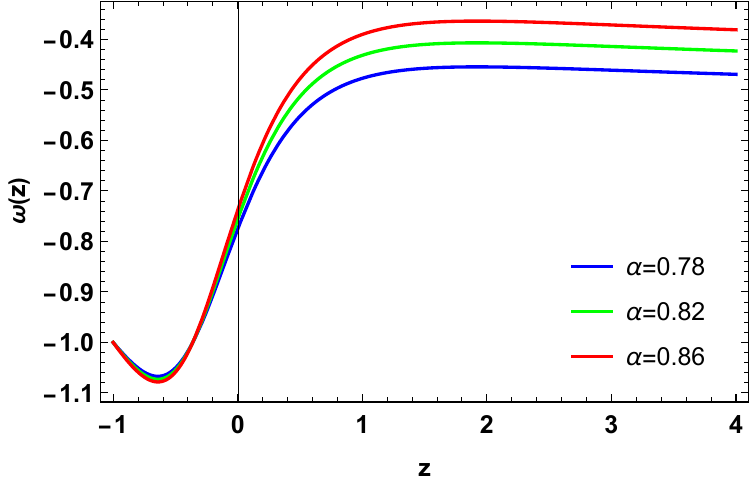}}
\caption{Profile of the density parameter, pressure, and EoS parameter as functions of redshift $z$, shown for $H_0=71$, $\zeta=-0.36$, and $\eta=1.3$, while the model parameter $\alpha$ is varied.}\label{f6f}
\end{figure*}

In this section, we investigate the cosmological implications of the $f(R,L_m)$ gravity by studying the following $f(R,L_m)$ model \citep{sol} 
\begin{equation}\label{5af}
 f(R, L_m) =\frac{R}{2} + L_m ^\alpha .  
\end{equation}
\justify The parameter $\alpha$ represents the free model parameter in this case. It should be noted that the standard Friedmann equations of GR are obtained when $\alpha=1$. In this specific $f(R,L_m)$ model, where $L_m = \rho$ \citep{HLR}, we can represent the Friedmann Eqs. \eqref{14b} and \eqref{15b} as follows
\begin{equation}\label{5bf}
3H^2 = (2\alpha-1) \rho^{\alpha},    
\end{equation}
\begin{equation}\label{5cf}
-2\dot{H}-3H^2 =\left( \alpha(p-\rho)+\rho\right) \rho^{\alpha-1}.
\end{equation} 
Moreover, Eq. \eqref{48} yields the energy balance equation corresponding to the model I, 
\begin{equation}\label{3id}
(2\alpha-1)\dot{\rho}+ 3H(\rho+p)=0.
\end{equation}
The Hubble parameter and its time derivative can be used to express the density parameter, pressure, and EoS parameter according to Eqs. \eqref{5bf} and \eqref{5cf} as
\begin{equation}\label{5df}
\rho=\left(\frac{3H^2}{2\alpha-1}\right)^\frac{1}{\alpha} , 
\end{equation} 
\begin{equation}\label{5ef}
p=-\frac{\big (\frac{3H^2}{2\alpha-1}\big)^\frac{1}{\alpha}\big(3\alpha H^2 +(4\alpha-2)\dot{H}\big)}{3\alpha H^2},
\end{equation} 
\begin{equation}\label{5ff}
\omega=-1 +\frac{(2-4\alpha)\dot{H}}{3\alpha H^2}.
\end{equation} 
Fig. \ref{f6f} shows the density parameter, pressure, and EoS parameter plotted against the redshift $z$, using model parameters derived from the combined CC, BAO, and Pantheon+SH0ES datasets. Eqs. \eqref{3af}, \eqref{3df}, \eqref{5df}, \eqref{5ef}, and \eqref{5ff} were utilized to calculate these values. Observing Fig. \ref{f6f}, it becomes apparent how the values of $\rho$ and $p$ change as the redshift $z$ changes. As a galaxy undergoes evolution, the density parameter increases with redshift and remains positive. In contrast, the pressure tends towards negative values, indicating rapid cosmic expansion and aligning with the overall expansion of the Universe. These results suggest that the logarithmic adjustment applied to the $\Lambda$CDM model within the $f(R,L_m)$ gravity framework can partially elucidate the DE phenomenon. 

It is clear that the equivalent value of the EoS parameter is $\omega=-1$ when $z=-1$. The model illustrates the quintessential DE behavior, as illustrated by the plot of the EoS parameter in Fig. \ref{f6f}.

Energy conditions are sets of requirements placed on the energy-momentum tensor to ensure that the energy remains positive. These conditions originate from the foundational work of Raychaudhuri, commonly known as Raychaudhuri's equation, and are documented as \citep{EC}. Understanding the geodesics of the Universe depends heavily on energy conditions \citep{EC,Ehle,Noji}. When considering a perfect fluid matter distribution within the framework of modified gravity, after plugging in the Eqs. \eqref{5df} and \eqref{5ef} into the energy conditions with the help of Eqs. \eqref{3af} and \eqref{3df}, we get these expressions
\begin{equation}\label{5gf}
\rho+p=  \frac{3^{\frac{1}{\alpha }-1} (\zeta (-2 \alpha +4 \alpha  z+z+1)+(2 \alpha -1) \eta (z+1) (2 z+1)) \left(\frac{H_0^2 (z (\zeta+\eta z+\eta)+1)}{2 \alpha -1}\right)^{1/\alpha }}{\alpha  z (\zeta+\eta z+\eta)+\alpha }\geq0 , 
\end{equation}
\begin{equation}\label{5hf}
\rho-p= \frac{3^{\frac{1}{\alpha }-1} (\zeta (-2 \alpha +4 \alpha  z+z+1)+6 \alpha +\eta (z+1) (2 \alpha  (z-1)+2 z+1)) \left(\frac{H_0^2 (z (\zeta+\eta z+\eta)+1)}{2 \alpha -1}\right)^{1/\alpha }}{\alpha  z (\zeta+\eta z+\eta)+\alpha }\geq0 ,
\end{equation}
\begin{equation}\label{5if}
\rho+3p=\frac{3^{1/\alpha } \left(2 \alpha  \left(\zeta+\eta (z+1)^2-1\right)-(z+1) (\zeta+2 \eta z+\eta)\right) \left(\frac{H_0^2 (z (\zeta+\eta z+\eta)+1)}{2 \alpha -1}\right)^{1/\alpha }}{\alpha  z (\zeta+\eta z+\eta)+\alpha }\geq0 .
\end{equation}
\begin{figure*}[htbp]
\centering
\subfigure{\includegraphics[width=7.5cm,height=5cm]{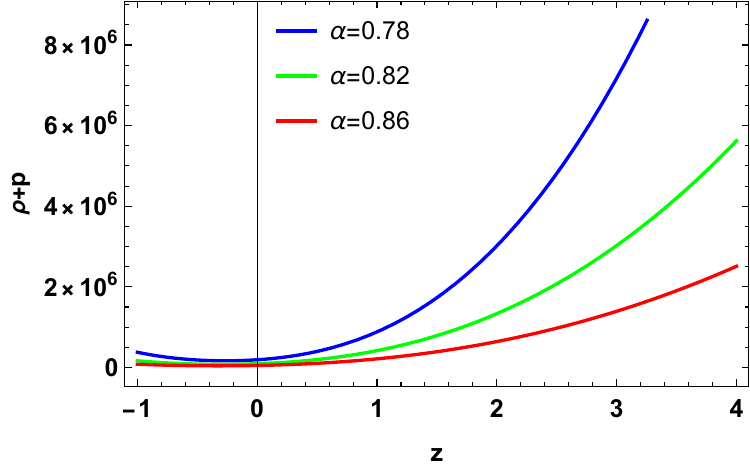}}\,\,\,\,\,\,\,\,\,\,\,\,
\subfigure{\includegraphics[width=7.5cm,height=5cm]{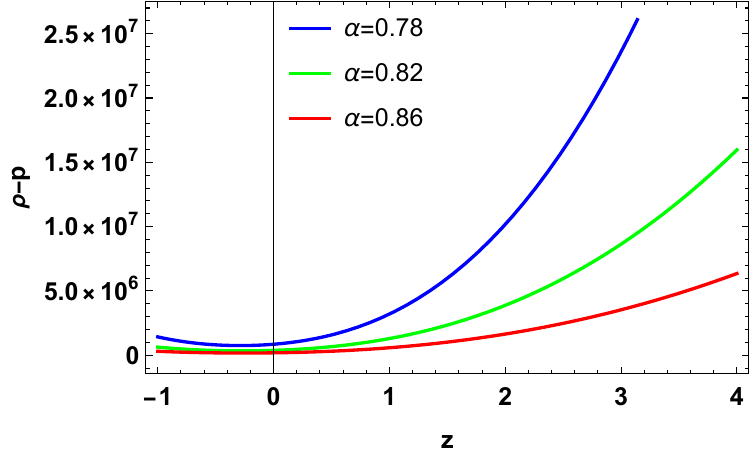}}\,\,\,\,\,\,\,\,\,\,\,\,
\subfigure{\includegraphics[width=7.5cm,height=5cm]{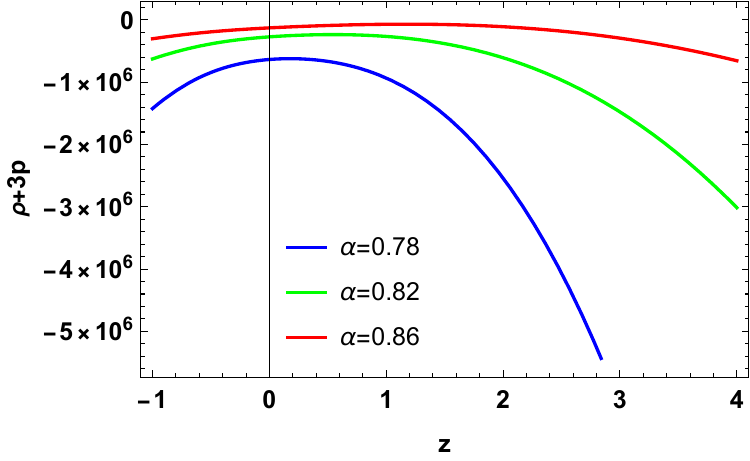}}
\caption{Profile of the NEC, DEC, and SEC for $H_0=71$, $\zeta=-0.36$, and $\eta=1.3$ while varying model parameter $\alpha$ with respect to redshift $z$.}\label{f7f}
\end{figure*}

The graph above illustrates the behavior of various energy conditions, crucial for understanding the expansion of the Universe \citep{bar,mor3,vis}. Examples of these energy conditions include the NEC, WEC, DEC, and SEC. The violation of the SEC is particularly relevant in the context of the observed rapid expansion of the Universe. Cosmological theories suggest that the SEC is likely to be violated during both the current and inflationary phases of the evolution of the Universe. Moreover, the data illustrated in the graph reflect a steady negative trend, which points to a violation of the SEC in relation to changes in the model parameters. The model parameters derived from the combined CC, BAO, and Pantheon+SH0ES datasets demonstrate positive NEC and DEC behavior. Fig. \ref{f7f} verifies the fulfillment of these conditions. It can be inferred that the values obtained from the three datasets also meet the WEC, which includes the NEC and the requirement for the positive density parameter. 

\subsection{Model II : \texorpdfstring{$f(R,L_m) =\frac{R}{2}+(1+\lambda R)L_m$}{f(R,Lm)}}\label{sbsec2}
In accordance with citations \citep{RV-2}, we are considering a specific non-minimal $f(R,L_m)$ function for the second model. This choice is based on extensive research that has explored similar formulations in existing literature,
\begin{equation}\label{5jf}
 f(R, L_m) =\frac{R}{2} + (1 + \lambda R)L_m,  
\end{equation}
where $\lambda$ represents a free model parameter.
The coupling constant has substantially greater values than the weak field limit. A similar pattern is found in scalar-tensor theories, where this parameter fluctuates in response to the mass of the scalar field, a phenomenon known as the chameleon mechanism \citep{khou/04}. Exploring how this coupling parameter operates in different cosmological circumstances is an interesting topic of inquiry.
In particular, when $\lambda=0$, we derive the standard Friedmann equations of GR. For example, consider the $f(R,L_m)$ model where $L_m = \rho$ \citep{HLR}. The Friedmann Eqs. \eqref{14b} and \eqref{15b} can be expressed as follows \citep{myr}
\begin{equation}\label{5kf}
3H^2 (6\lambda \rho-1)+12\lambda\dot{H}\rho+\rho=0 ,
\end{equation}
\begin{equation}\label{5lf}
3H^2 (-2\lambda \rho+4\lambda p+1)+\dot{H}(-2\lambda \rho+6\lambda p+2)+p=0. 
\end{equation}
Moreover, Eq. \eqref{48} yields the following energy-balance equation corresponding to the model II,
\begin{equation}\label{cc1f}
 \dot{\rho}+ 3H(\rho+p)+12\rho\frac{\lambda(\Ddot{H}+4\dot{H}H)}{1+6\lambda(\dot{H}+2H^2)}=0.   
\end{equation}
The Hubble parameter and its time derivative can be used to express the density parameter, pressure, and the EoS parameter in Eqs. \eqref{5kf} and \eqref{5lf} as
\begin{equation}\label{5mf}
\rho=\frac{ 3H^2}{ 18\lambda H^2+12\lambda\dot{H}+1},
\end{equation}
\begin{equation}\label{5nf}
p=-\frac{6\lambda(2H^2 +\dot{H})(3H^2+4\dot{H})+3H^2+2\dot{H}}{(18\lambda H^2+12\lambda\dot{H}+1)(6\lambda(2H^2 +\dot{H})+1)},
\end{equation}
\begin{equation}\label{5of}
\omega=-1+\frac{2\dot{H}}{3H^2}\bigg(\frac{1}{6\lambda(2H^2 +\dot{H})+1}-2\bigg).
\end{equation}
\begin{figure*}[htbp]
\centering
\subfigure{\includegraphics[width=7.5cm,height=5cm]{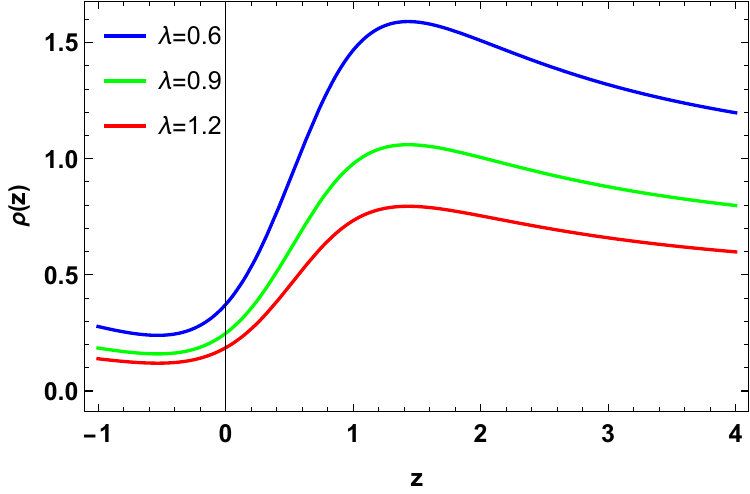}}\,\,\,\,\,\,\,\,\,\,\,\,
\subfigure{\includegraphics[width=7.5cm,height=5cm]{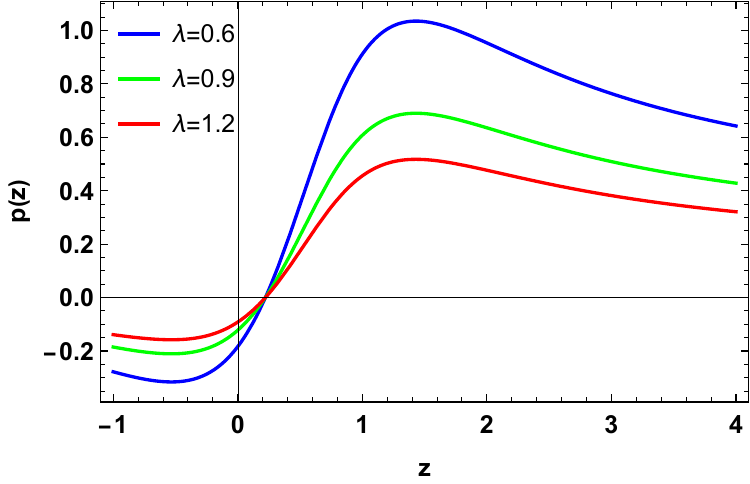}}\,\,\,\,\,\,\,\,\,\,\,\,
\subfigure{\includegraphics[width=7.5cm,height=5cm]{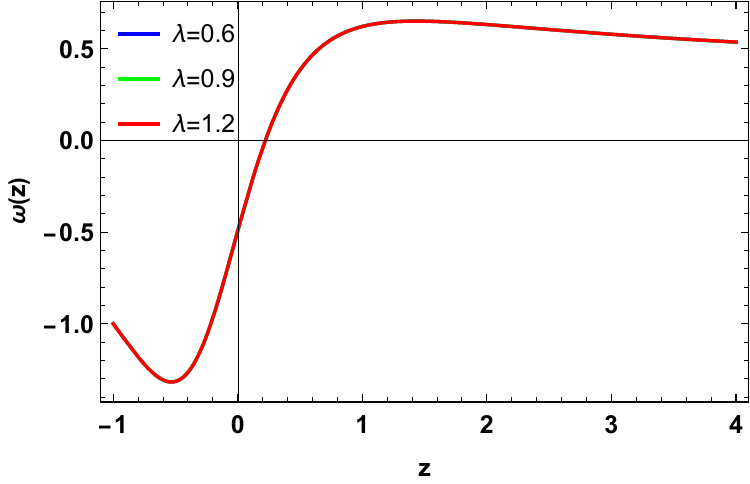}}
\caption{Profile of the density parameter, pressure, and EoS parameter as functions of redshift $z$, shown for $H_0=71$, $\zeta=-0.36$, and $\eta=1.3$, while the model parameter $\lambda$ is varied.}\label{f8f}
\end{figure*}

Upon examining Fig. \ref{f8f}, one can see the shifts in the values of $\rho$ and $p$ as the redshift $z$ changes, with the density parameter remaining positive for all values of the limited model parameters. The pressure undergoes a shift from positive in the past to negative in the present and future. Eqs. \eqref{5mf}, \eqref{5nf}, \eqref{5of}, \eqref{3af}, and \eqref{3df} were utilized to calculate these values. When analyzing the parameters of the fixed model, the graph illustrates the EoS parameter as a function of the redshift $z$. This shows a shift from negative to positive values over the course of cosmic evolution. This shift indicates a previous era of the Universe characterized by positive pressure and deceleration, which was conducive to the development of structures. The EoS parameter indicates negative pressure during the current accelerating phase, and the current EoS parameter values meet the criteria for the acceleration phase.

Using Eqs. \eqref{5mf} and \eqref{5nf} in the energy conditions with Eqs. \eqref{3af} and \eqref{3df} yields the following expressions
\begin{equation}\label{5pf}
\rho+p=\frac{H_0^2 (z+1) (\zeta+2 \eta z+\eta) \left(6 \lambda  H_0^2 \left(\zeta (3 z-1)+\eta \left(2 z^2+z-1\right)+4\right)+1\right)}{\left(6 \lambda  H_0^2 \left(\zeta (2 z-1)+\eta \left(z^2-1\right)+3\right)+1\right) \left(3 \lambda  H_0^2 \left(\zeta (3 z-1)+\eta \left(2 z^2+z-1\right)+4\right)+1\right)}\geq0,
\end{equation}
\begin{multline}\label{5qf}
\rho-p=\frac{6 \lambda  H_0^4 \left(\zeta (2 z-1)+\eta \left(z^2-1\right)+3\right) \left(\zeta (3 z-1)+\eta \left(2 z^2+z-1\right)+4\right)}{\left(6 \lambda  H_0^2 \left(\zeta (2 z-1)+\eta \left(z^2-1\right)+3\right)+1\right) \left(3 \lambda  H_0^2 \left(\zeta (3 z-1)+\eta \left(2 z^2+z-1\right)+4\right)+1\right)}\\
+\frac{H_0^2 \left(5 \zeta z-\zeta+4 \eta z^2+3 \eta z-\eta+6\right)}{\left(6 \lambda  H_0^2 \left(\zeta (2 z-1)+\eta \left(z^2-1\right)+3\right)+1\right) \left(3 \lambda  H_0^2 \left(\zeta (3 z-1)+\eta \left(2 z^2+z-1\right)+4\right)+1\right)}\geq0,
\end{multline}
\begin{equation}\label{5rf}
\rho+3p=\frac{3 H_0^2 \left(6 \lambda  H_0^2 \left(\zeta+\eta (z+1)^2-1\right) \left(\zeta (3 z-1)+\eta \left(2 z^2+z-1\right)+4\right)-\zeta z+\zeta+\eta z+\eta-2\right)}{\left(6 \lambda  H_0^2 \left(\zeta (2 z-1)+\eta \left(z^2-1\right)+3\right)+1\right) \left(3 \lambda  H_0^2 \left(\zeta (3 z-1)+\eta \left(2 z^2+z-1\right)+4\right)+1\right)}\geq0.
\end{equation}
\begin{figure*}[htbp]
\centering
\subfigure{\includegraphics[width=7.5cm,height=5cm]{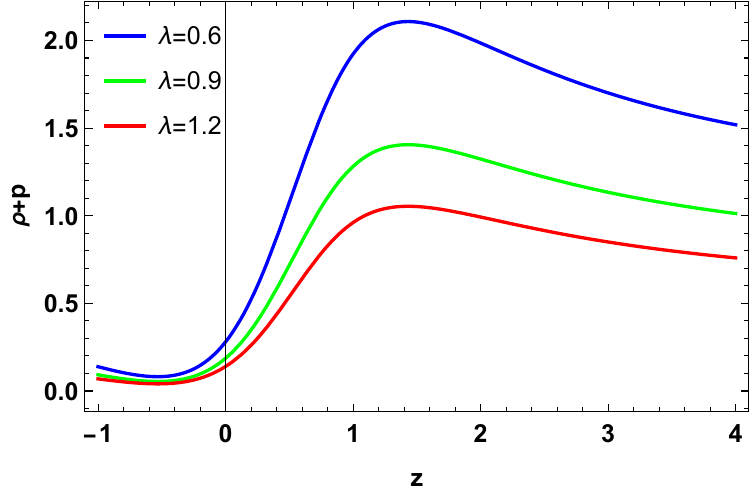}}\,\,\,\,\,\,\,\,\,\,\,\,
\subfigure{\includegraphics[width=7.5cm,height=5cm]{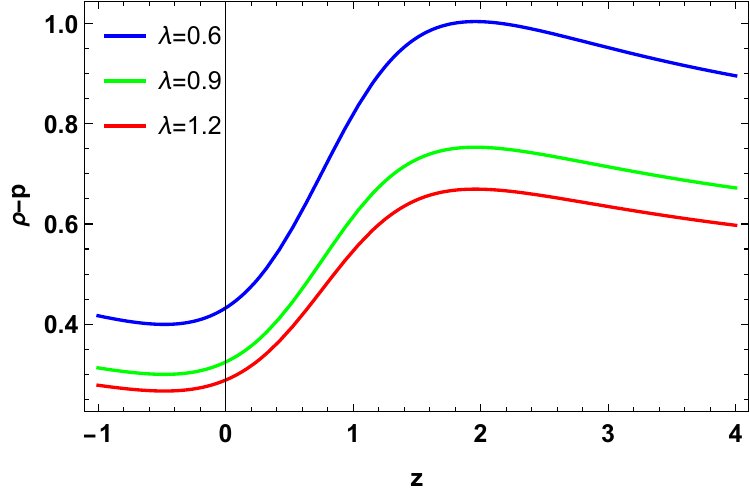}}\,\,\,\,\,\,\,\,\,\,\,\,
\subfigure{\includegraphics[width=7.5cm,height=5cm]{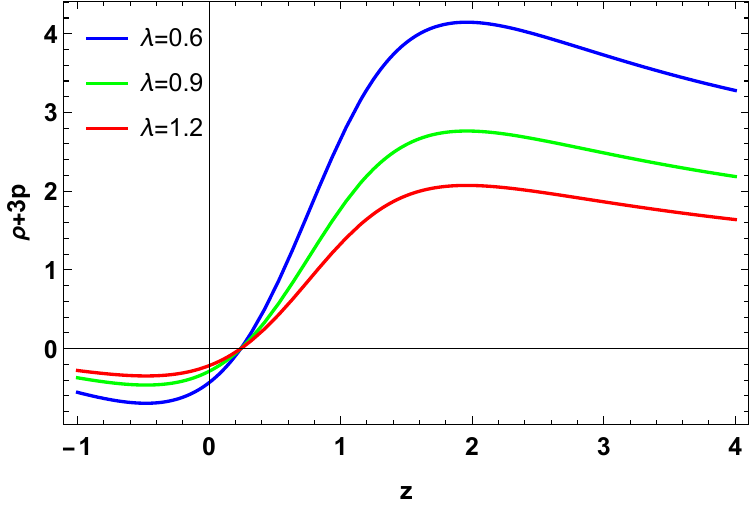}}
\caption{Profile of the NEC, DEC, and SEC for $H_0=71$, $\zeta=-0.36$, and $\eta=1.3$ while varying model parameter $\lambda$ with respect to redshift $z$.}\label{f9f}
\end{figure*}

In Fig. \ref{f9f}, a similar evaluation of the energy conditions was carried out. The model parameters obtained from the combined datasets of CC, BAO, and Pantheon + SH0ES indicate that both the NEC and DEC demonstrate positive behavior, indicating compliance with these conditions. With these parameters, the WEC is also met. In addition, a violation of the SEC occurred in the recent past, which supports the current transition phase from deceleration to acceleration.

\section{Conclusions}\label{sec6}
\justifying
In this chapter, we utilized a parameterization of the Hubble parameter to obtain accurate solutions to the field equations within the FLRW background. The parameterization was expressed as $H(z)=H_0 \big[(1-\zeta)+(1+z)(1+\eta)\big]^\frac{1}{2}$ where $\zeta$ and $\eta$ were free parameters. Furthermore, we identified the optimal values for these model parameters by analyzing the combined datasets from the CC, BAO, and Pantheon+SH0ES observations, resulting in $H_0=71^{+0.081}_{-0.082}$, $\zeta=-0.36 \pm  0.033$ and $\eta=1.3 \pm 0.023$. The updated parameter constraints obtained for the assumed Hubble function differed considerably from the previous work done in \citep{pac3}. We also investigated the behavior of the deceleration parameter. The graph in Fig. \ref{f5f} depicted its evolution, demonstrating a shift from a decelerated phase to an accelerated phase within our cosmological model. The transition redshift, corresponding to the model parameters constrained by the combined CC, BAO, and Pantheon+SH0ES datasets, was $z_t = 0.6459$. Furthermore, the current value of the deceleration parameter was $q_0 = -0.5249$. The present value obtained from the deceleration parameter was more consistent than the previously obtained value $q_0 \approx -0.3$ \citep{pac3}.

Hence, the updated constraints aligned well with the standard $\Lambda$CDM. Two nonlinear functional forms of this theory were analyzed to understand their effects on cosmic evolution. As anticipated, we observed that the density parameter for both models considered in Figs. \ref{f6f} and \ref{f8f} consistently maintained positive values throughout cosmic evolution. The fluctuation in isotropic pressure is depicted in these figures. Our interpretation of the figures indicated that the isotropic pressure was negative both presently and in the future. From the figures above for the EoS parameter, we saw that both models exhibited quintessence behavior at the present epoch and then converged to the $\Lambda$CDM EoS in the distant future, crossing into the phantom region. Note that the crossing of the phantom divide was suggested through several observational studies, such as WMAP + CMB, which represented the value $\omega = -1.079^{+0.090}_{-0.089}$ \citep{G11}, while the Planck 2018 results depicted the value $\omega = -1.03 \pm 0.03$ \citep{G12}. For more on phantom crossing EoS, one can check reference \citep{G13}.

In the end, the validity of the derived solutions was assessed by evaluating the energy conditions for both models. Naturally, we expressed the field equation of modified gravity in a covariant form, i.e., the left-hand side with only Einstein's tensor and everything remaining on the right-hand side. Note that $T_{\mu\nu}$ for the matter component was still bounded by the energy condition constraints. Thus, we could utilize these to find the suitable ranges of the model parameters as in \citep{AVV}. As seen in Figs. \ref{f7f} and \ref{f9f}, it was discovered that all energy conditions showed positive features except the SEC. However, as depicted in Fig. \ref{f9f}, the SEC violation strongly indicated that the cosmic expansion was accelerating, signifying a transition from a decelerated to an accelerated phase in the second model. On the other hand, in Fig. \ref{f7f}, the SEC showed a consistently negative trend, suggesting that the SEC was violated in the range of model parameter variations. Thus, one could conclude that the assumed model of non-minimal coupling more efficiently described the recent epochs of the Universe compared to that of minimal ones. Hence, for future work, the $f(R,L_m)$ models exhibiting non-minimal couplings were highly encouraged.

Building on our investigation of late-time cosmic acceleration through Hubble parameter parametrization in $f(R,L_m)$ gravity, we now extend our analysis by incorporating bulk viscosity effects. In the next chapter, we explore a non-linear $f(R,L_m)$ model, deriving an exact solution for a bulk viscous-dominated Universe and assessing its compatibility with observational data.


%% file: Chapters/Chapter4.tex

\chapter{Constraining viscous dark energy equation of state in \texorpdfstring{$f(R,L_m)$}{f(R,Lm)} gravity} 

\label{Chapter4} 

\lhead{Chapter 4. \emph{Constraining viscous dark energy equation of state in $f(R,L_m)$) gravity}} 
\vspace{10 cm}
* The following publications cover the work in this chapter:\\
 
\textit{Constraining viscous dark energy equation of state in $f(R,L_m)$) gravity}, Universe, \textbf{9}, 163 (2023).

\clearpage
In this chapter, we explore the phenomenon of cosmic late-time acceleration in the context of $f(R,L_m)$ gravity, incorporating an effective EoS while considering bulk viscosity. We focus on a non-linear form of the $f(R,L_m)$ function. We derive an exact solution for our model dominated by bulk viscous matter. Using the combined datasets from CC + Pantheon+SH0ES, we estimate the optimal values for our model's free parameters. Furthermore, we analyze the behavior of the density parameter, effective pressure, and the EoS parameter that accounts for the viscous fluid. The resulting evolution profile of the effective EoS parameter indicates an accelerating phase in the expansion of the Universe.
In contrast, the influence of viscosity pressure demonstrates a negative effect that can contribute to the accelerating expansion of the Universe. Furthermore, the expected behavior of the density parameter is positive. We also explore the statefinder parameters for the proposed $f(R,L_m)$ model and observe that its evolutionary path falls within the quintessence region. Furthermore, we apply the Om diagnostic test, which reveals that our model behaves as quintessence. Lastly, after analyzing the energy condition criteria, we found evidence of past violations of the SEC, whereas the NEC and DEC met the positivity requirements. Overall, our $f(R,L_m)$ cosmological model, incorporating the effects of bulk viscosity, aligns well with recent observational data and effectively describes the cosmic expansion scenario.

\section{Introduction}\label{sec1C}
\justifying
This chapter will delve into the cosmological model defined by $f(R,L_m)$ that incorporates viscosity within the fluid. The concept of introducing a viscosity coefficient into cosmological models has been explored for a long time. From the perspective of hydrodynamics, two primary viscosity coefficients are frequently referenced: the bulk viscosity coefficient $\zeta$ and the shear viscosity coefficient $\gamma$. Under the assumption of spatial isotropy, which is well supported by observations, we can exclude shear viscosity from our considerations. When a system strays from thermal equilibrium, an effective pressure emerges, helping the system return to equilibrium. In this context, the bulk viscosity associated with a cosmological fluid represents this type of effective pressure. Our approach involves integrating the bulk viscosity coefficient $\zeta$ into the framework of the $f(R,L_m)$ gravity model. We assume that $\zeta$ follows a scaling law, which, in turn, simplifies the equations to a form proportional to the Hubble parameter. This scaling law has proved to be quite relevant in our analysis. For further insights, the reader may refer to various intriguing models of viscous fluid cosmology \citep{IB-1,IB-2,IB-3,IB-4,IB-5,JM,AVS,MAT}.

This chapter is structured as follows. In Sec. \ref{sec3c}, we outline the fundamental equations that dictate the dynamics within $f(R,L_m)$ gravity. Sec. \ref{sec4c} involves the assumption of a specific $f(R,L_m)$ function, leading to the derivation of expressions for the Hubble parameter and the EoS parameter, which link the pressure of the bulk viscous matter to its density parameter. In Sec. \ref{sec5c}, we estimate the present values of the Hubble parameter $H_0$ and other model parameters according to observational data by integrating the combined CC + Pantheon+SH0ES datasets. In addition, we analyze the behavior of various parameters, including the density parameter, the effective pressure, and the EoS parameter. Further insights are provided in Sec. \ref{sec6c}, where we examine the trajectory of the $r-s$ parameter within our $f(R,L_m)$ framework to evaluate the DE characteristics implied by the chosen model. Subsequently, Sec. \ref{sec7c} and \ref{sec8c} focus on using the Om diagnostic test and assessing the criteria of the energy condition. The chapter concludes with a summary of our findings in Sec. \ref{sec9c}.

\section{Motion equations in \texorpdfstring{$f(R,L_m)$}{f(R,Lm)} gravity}\label{sec3c}
\justifying
The energy-momentum tensor consists of the density parameter $\rho$ and the pressure $\bar{p}$ of the cosmic fluid, taking into account the effects of viscosity,
\begin{equation}\label{2cc}
T_{\mu\nu}=(\rho+\bar{p})u_\mu u_\nu + \bar{p}g_{\mu\nu}.
\end{equation}
In this context, the density parameter is expressed as $\bar{p} = p-3\zeta H$, where the four-velocity components are given as $u^\mu=(1, 0, 0, 0)$. Here, $p$ represents the conventional pressure and $\zeta > 0$ denotes the coefficient of bulk viscosity.

The relationship between the density parameter and the conventional pressure is described in \citep{J}
\begin{equation}\label{2dc}
p=(\gamma-1)\rho,
\end{equation}
where $\gamma$ is a constant such that $0 \leq \gamma \leq 2$. Therefore, the effective EoS that describes the bulk viscous cosmic fluid is given by \citep{brevik/2005,gron/1990,C.E./1940}
\begin{equation}\label{2ec}
\bar{p}= (\gamma-1)\rho -3\zeta H.
\end{equation}
In a homogeneous and spatially isotropic context, a cosmic fluid that includes viscosity exhibits dissipative effects. The presence of viscosity can decrease the properties of the ideal fluid and adversely affect the overall pressure within the fluid. This concept can be explored further in various references \citep{odintsov/2020,fabris/2006,meng/2009}. 

The Friedmann equations that describe a bulk viscous matter-dominated Universe in $f(R,L_m)$ gravity are formulated as follows
\begin{equation}\label{2fc}
3H^2 f_R + \frac{1}{2} \left( f-f_R R-f_{L_m}L_m \right) + 3H \dot{f_R}= \frac{1}{2}f_{L_m} \rho,
\end{equation}
\begin{equation}\label{2gc}
\dot{H}f_R + 3H^2 f_R - \ddot{f_R} -3H\dot{f_R} + \frac{1}{2} \left( f_{L_m}L_m - f \right) = \frac{1}{2} f_{L_m}\bar{p}.
\end{equation}

\section{Cosmological \texorpdfstring{$f(R,L_m)$}{f(R,Lm)} model }\label{sec4c}
\justifying
We have chosen the following function $f(R,L_m)$ to examine the dynamics of the Universe characterized by viscosity,
\begin{equation}\label{3ac} 
f(R,L_m)=\frac{R}{2}+L_m^\alpha .
\end{equation}
In this context, the variable $\alpha$ denotes a free model parameter. Our investigation of the minimal coupling scenario is influenced by the significant research conducted by Bose et al. \citep{Bose} within the $f(R,T)$ gravity framework. For this particular functional form, where $L_m=\rho$ \citep{HLR}, the Friedmann equations \eqref{2fc} and \eqref{2gc} that describe the Universe governed by bulk viscous matter can be expressed as follows
\begin{equation}\label{3bc}
3H^2=(2\alpha-1) \rho^\alpha,
\end{equation}
\begin{equation}\label{3cc}
2\dot{H}+3H^2=  \left\{ (\alpha-1)\rho-\alpha \bar{p} \right\} \rho^{\alpha-1} .
\end{equation}
By employing Equation \eqref{48}, we formulated the matter conservation equation for our bulk viscous cosmological model, 
\begin{equation}\label{3dc}
(2\alpha-1)\dot{\rho}+ 3H(\gamma \rho-3\zeta H) = 0.
\end{equation}
From Eqs. \eqref{3bc} and \eqref{3cc}, one can have
\begin{equation}\label{3ec}
\dot{H}+\frac{3\alpha\gamma}{2(2\alpha-1)} H^2 = \frac{3}{2} \left( \frac{3}{2\alpha-1} \right)^{\frac{\alpha-1}{\alpha}}\alpha\zeta H^{\frac{3\alpha-2}{\alpha}}.
\end{equation}
We substitute $ \frac{1}{H} \frac{d}{dt}= \frac{d}{dln(a)}$ so that Eq. \eqref{3ec} becomes
\begin{equation}\label{3fc}
\frac{dH}{dln(a)}+\frac{3\alpha\gamma}{2(2\alpha-1)} H = \frac{3}{2} \left( \frac{3}{2\alpha-1} \right)^{\frac{\alpha-1}{\alpha}}\alpha\zeta H^{\frac{2(\alpha-1)}{\alpha}}.
\end{equation}
On integrating the Eq. \eqref{3fc}, we obtained the expression for the Hubble parameter as follows
\begin{equation}\label{3gc}
H(z)=\big\{ H_0^{\frac{2-\alpha}{\alpha}} (1+z)^{\frac{3\gamma(2-\alpha)}{2(2\alpha-1)}} + \frac{3\zeta}{\gamma} \left( \frac{2\alpha-1}{3} \right)^{\frac{1}{\alpha}} [ 1-(1+z)^{\frac{3\gamma(2-\alpha)}{2(2\alpha-1)}} ] \big\}^\frac{\alpha}{2-\alpha}.
\end{equation}
Notably, when $\alpha=1$, $\gamma=1$, and $\zeta=0$, the solution simplifies to $H(z) = H_0(1+z)^{\frac{3}{2}}$, which indicates that the Universe is predominantly composed of ordinary matter.

The effective EoS parameter for our bulk viscous cosmological model is given by
\begin{equation}\label{3hc}
\omega_{eff} = \frac{p_{eff}}{\rho} = \gamma -1 - \frac{3\zeta H}{\rho}.
\end{equation}
Using Eqs. \eqref{3bc} and \eqref{3gc}, one can acquired
\begin{equation}\label{3ic}
\omega_{eff} = \gamma -1 - 3\zeta \left( \frac{2\alpha-1}{3} \right)^{\frac{1}{\alpha}} \big\{ H_0^{\frac{2-\alpha}{\alpha}} (1+z)^{\frac{3\gamma(2-\alpha)}{2(2\alpha-1)}} + \frac{3\zeta}{\gamma} \left( \frac{2\alpha-1}{3} \right)^{\frac{1}{\alpha}} [ 1-(1+z)^{\frac{3\gamma(2-\alpha)}{2(2\alpha-1)}} ] \big\}^{-1}.
\end{equation}

\section{Data, methodology, and physical interpretation}\label{sec5c}
\justifying
In this section, we determine the best-fit values for our model parameter to characterize different cosmic epochs utilizing the CC and Pantheon+SH0ES datasets. To obtain the optimal values for $H_0$ and the model parameters $\alpha$, $\gamma$, and $\zeta$, we use the CC datasets in conjunction with the Pantheon+SH0ES samples. 

Now, we take the $\chi^{2}$ function for the CC + Pantheon+SH0ES datasets as
\begin{equation}\label{15c}
\chi^2 _{tot}=\chi^2 _{CC}+\chi^2 _{SNeIa}.
\end{equation}
We present the likelihood contours $1-\sigma$ and $2-\sigma$ for the model parameters $\alpha$, $\gamma$, $\zeta$ and $H_0$ using the combined CC + Pantheon+SH0ES datasets below in Fig. \ref{f1b}.
\begin{figure*}[htbp]
\centering
\includegraphics[scale=0.85]{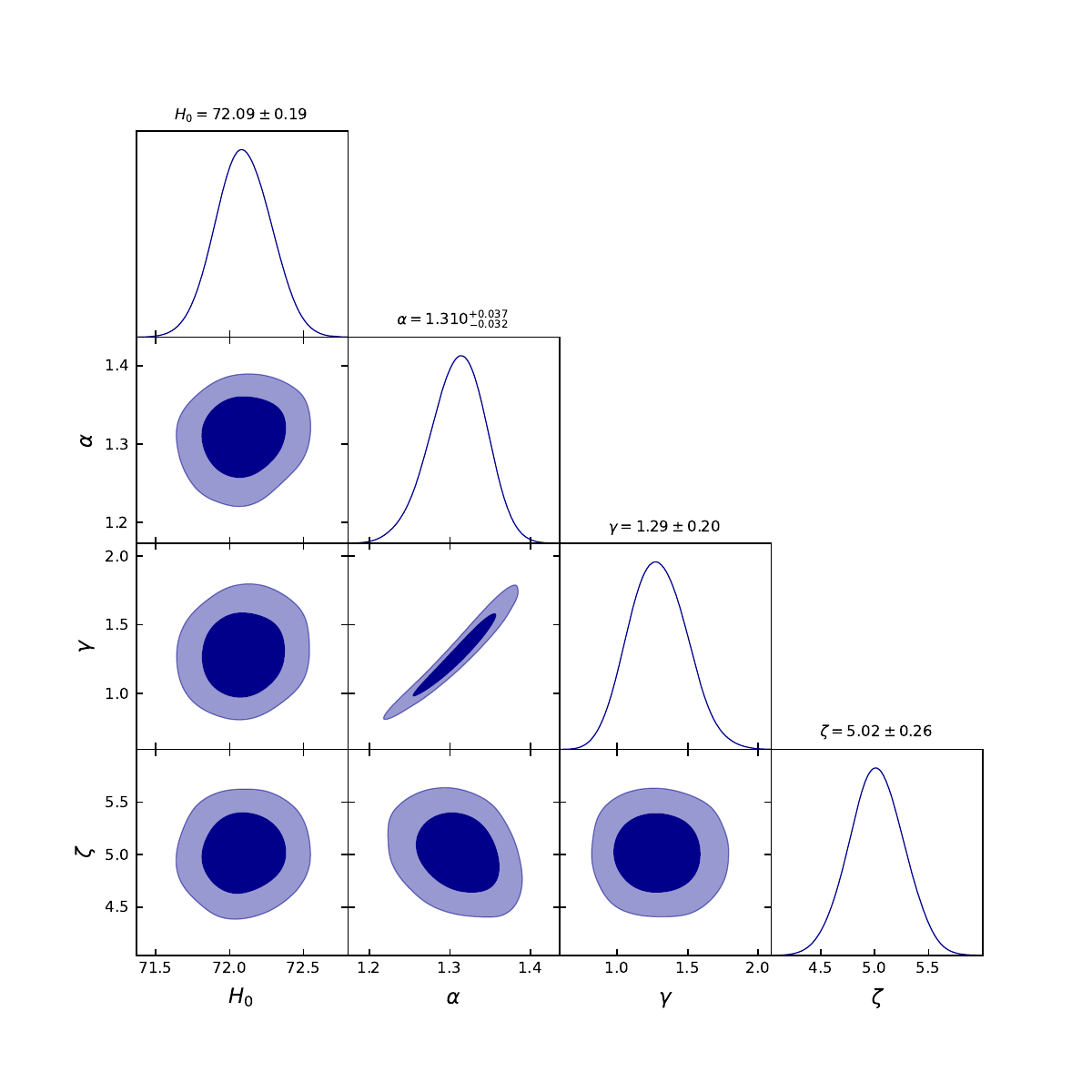}
\caption{Profile of the $1-\sigma$ and $2-\sigma$ contours for the model parameters $H_0$, $\alpha$, $\gamma$, and $\zeta$  using combined CC + Pantheon+SH0ES datasets}\label{f1b}
\end{figure*}\\
The best-fit values obtained are: $\alpha=1.310^{+0.037}_{-0.032}$, $\gamma = 1.29 \pm 0.20$, $\zeta= 5.02 \pm 0.26$, and $H_0= 72.09 \pm 0.19$.

Now, we will explore the cosmological implications derived from the observational constraints we have obtained. We will examine the dynamics of the density parameter, the pressure component that includes viscosity, and the effective EoS parameter using the mean values of $H_0$ and the model parameters $\alpha$, $\gamma$, and $\zeta$ as constrained by the CC + Pantheon+SH0ES datasets.

\begin{figure*}[htbp]
\centering
\includegraphics[scale=0.3]{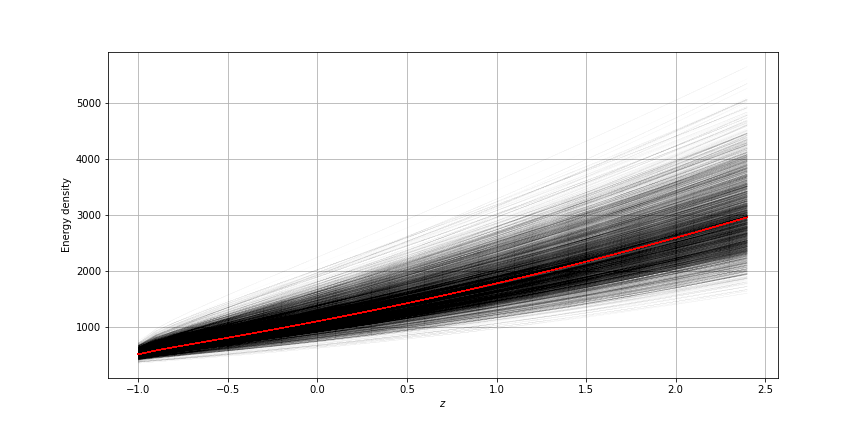}
\caption{Profile of the reconstruction of the density parameter is illustrated as a function of redshift for our model, based on a sample of $7500$ instances. These samples are generated through re-sampling the chains using \textit{emcee}. We display all the resulting curves along with the curve that represents the best-fit parameters (shown in red).}\label{f2b}
\end{figure*}
\begin{figure*}[htbp]
\centering
\includegraphics[scale=0.3]{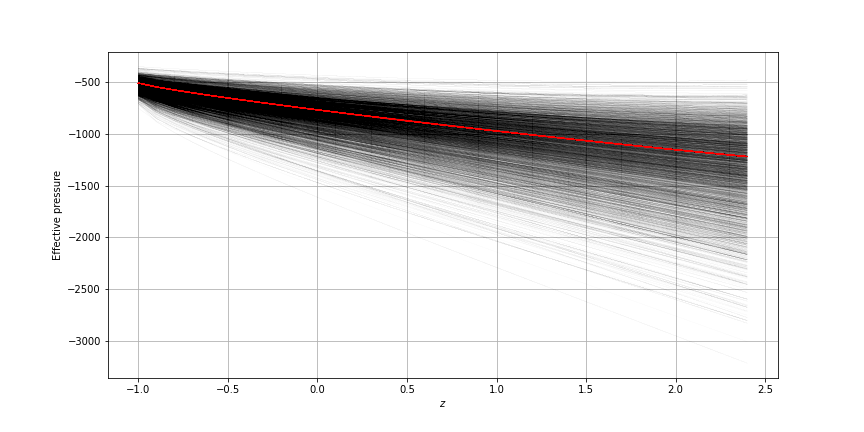}
\caption{Profile of the reconstruction of the effective pressure is illustrated as a function of redshift for our model, based on a sample of $7500$ instances. These samples are generated through re-sampling the chains using \textit{emcee}. We display all the resulting curves along with the curve that represents the best-fit parameters (shown in red).}\label{f3b}
\end{figure*}
\begin{figure*}[htbp]
\centering
\includegraphics[scale=0.3]{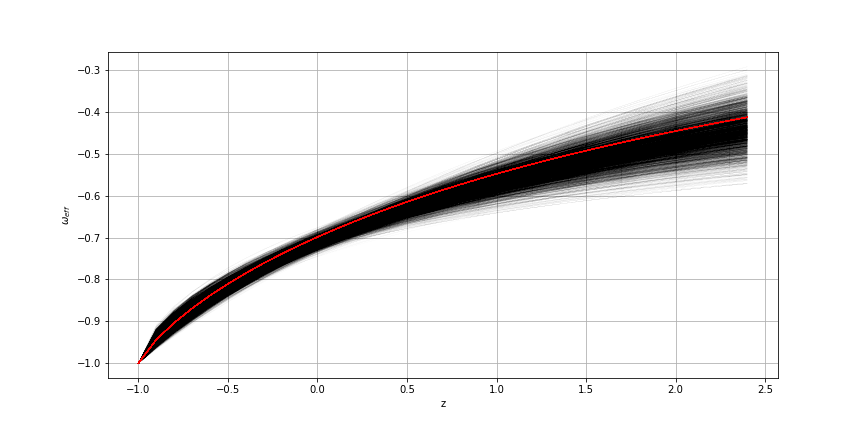}
\caption{Profile of the reconstruction of the effective EoS parameter is illustrated as a function of redshift for our model, based on a sample of $7500$ instances. These samples are generated through re-sampling the chains using \textit{emcee}. We display all the resulting curves along with the curve that represents the best-fit parameters (shown in red).}\label{f4b}
\end{figure*}

\justify We analyzed the density parameter, effective pressure and effective EoS parameter in relation to redshift, as illustrated in Figs. \ref{f2b}, \ref{f3b}, and \ref{f4b}. This analysis is based on $7500$ samples generated by resampling chains using the \textit{ emcee}. According to Fig. \ref{f2b}, the density parameter demonstrates a consistent positive trend and is projected to decrease as the Universe continues to expand into the distant future. In Fig. \ref{f3b}, the effective pressure component shows a negative trend, indicating a potential for an accelerating expansion of the Universe. Furthermore, the current value of the effective EoS parameter is estimated at $\omega_0 \approx -0.71$. This is supported by the behavior of the effective EoS parameter presented in Fig. \ref{f4b}, which further supports the idea that the expansion of the Universe is accelerating.

\section{Statefinder diagnostic}\label{sec6c}
\begin{figure*}[htbp]
\centering
\includegraphics[scale=0.6]{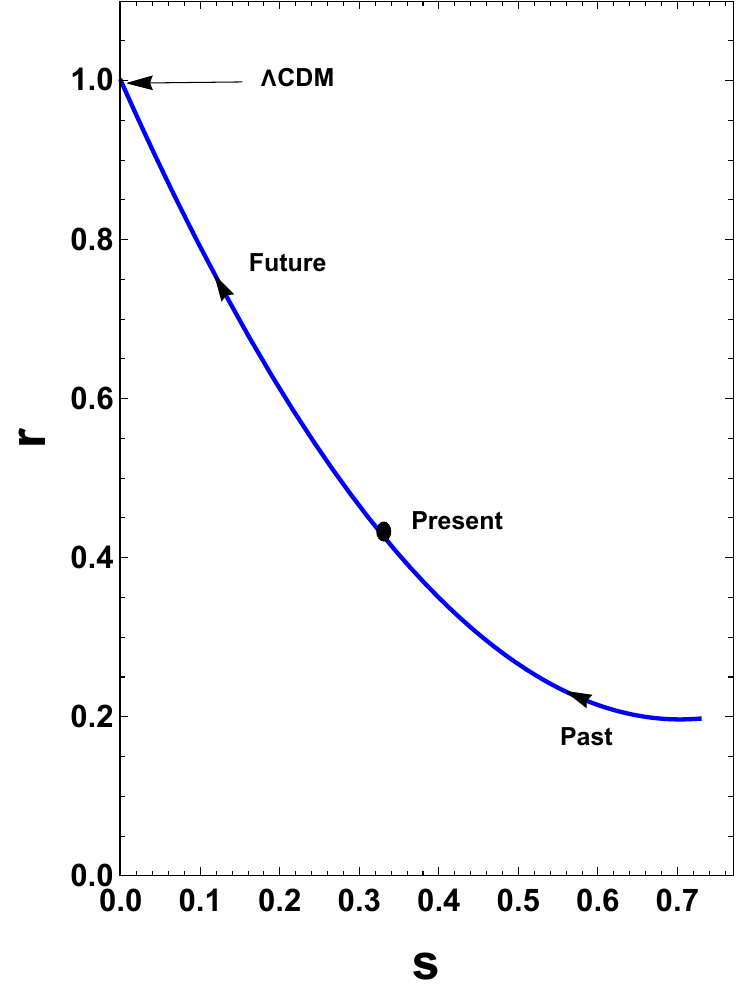}
\caption{Profile of the evolution trajectory of a given model in the $r-s$ plane with the agreement of obtained observational constraints.}\label{f5b}
\end{figure*}

\justifying
The role of DE in the expansion of the Universe is widely recognized. In recent decades, there has been a significant increase in research focused on understanding the origin and fundamental properties of DE. This surge has led to the emergence of numerous DE models, necessitating a way to distinguish between them, either quantitatively or qualitatively. To address this, Sahni et al. \citep{V.S.} proposed a statefinder diagnostic method, which uses a pair of geometric parameters known as statefinder parameters, denoted as $(r,s)$, to differentiate among various DE models. This method is defined as follows
\begin{equation}\label{16c}
 r=\frac{\dddot{a}}{aH^3} ,
\end{equation}
\begin{equation}\label{17c}
s=\frac{(r-1)}{3(q-\frac{1}{2})}.
\end{equation}
We analyze the statefinder parameters $(r,s)$ within our cosmological $f(R,L_m)$ model. The evolutionary path of this model, aligned with the observational constraints we obtained, is illustrated in Fig. \ref{f5b}. The distinction between the trajectory of our model and that of the $\Lambda$CDM model provides the necessary differentiation. Specifically, the point $(r=1, s=0)$ corresponds to the $\Lambda$CDM framework, while the values where $r>1$ and $s<0$ indicate a Chaplygin gas model and $r<1$ with $s>0$ signify a quintessence model. Currently, our model yields statefinder parameters close to $(r,s)=(0.43,0.33)$. As seen in Fig. \ref{f5b}, it is clear that the dark component influenced by the modified geometry and the effects of bulk viscosity exhibits the characteristic behavior of quintessence.

\section{\texorpdfstring{$Om(z)$}{Om(z)} diagnostics}\label{sec7c}
\justifying
Om diagnostic is a newly introduced method that can effectively differentiate between various DE models \citep{Om}. It is simpler than statefinder analysis, as it provides a formulation that relies solely on the Hubble parameter. For a spatially flat constraint, it is expressed in Eq. \eqref{omb}.
\begin{figure*}[htbp]
\centering
\includegraphics[scale=0.6]{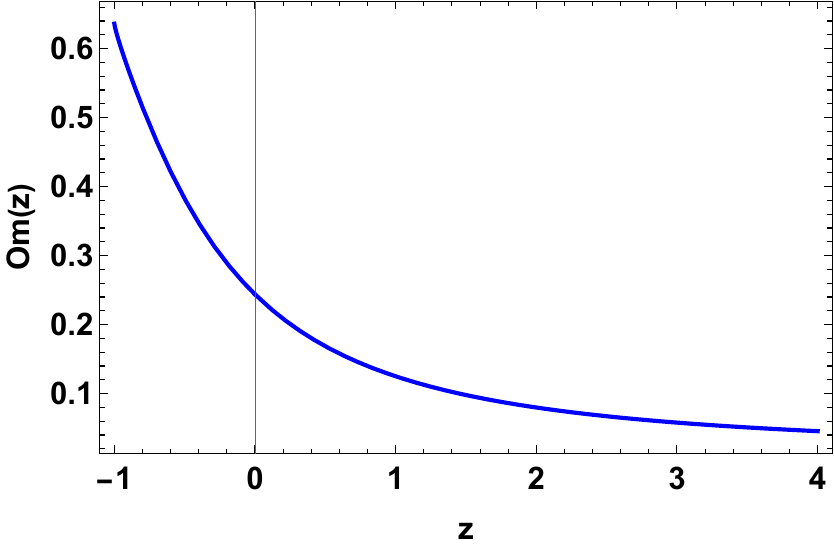}
\caption{Profile of the Om diagnostic parameter with the agreement of obtained observational constraints.}\label{f6b}
\end{figure*}\\
Fig. \ref{f6b} shows that the Om diagnostic parameter exhibits a negative slope throughout the entire domain. Consequently, this suggests that our bulk viscous matter-dominated $f(R,L_m)$ model aligns with the quintessence scenario.

\section{Energy conditions}\label{sec8c}
\justifying
We evaluate the feasibility of the solution obtained for the proposed $f(R,L_m)$ model using energy conditions as a benchmark \citep{EC}. 

\begin{figure*}[htbp]
\centering
\includegraphics[scale=0.6]{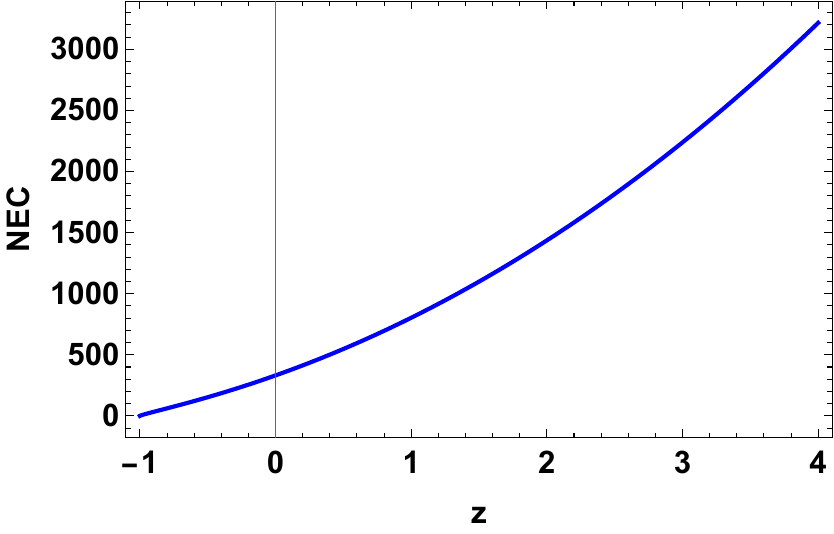}
\caption{Profile of the NEC vs redshift.}\label{f7b}
\end{figure*}
\begin{figure*}[htbp]
\centering
\includegraphics[scale=0.6]{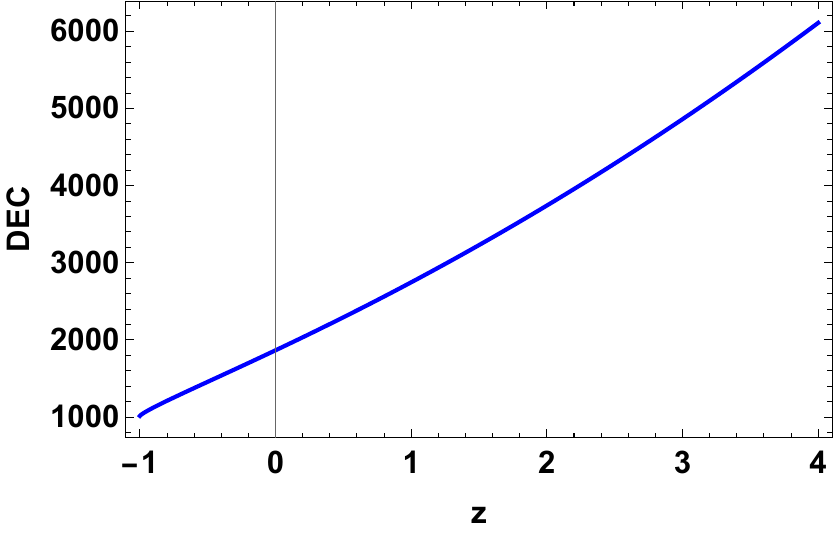}
\caption{Profile of the DEC vs redshift.}\label{f8b}
\end{figure*}
\begin{figure*}[htbp]
\centering
\includegraphics[scale=0.6]{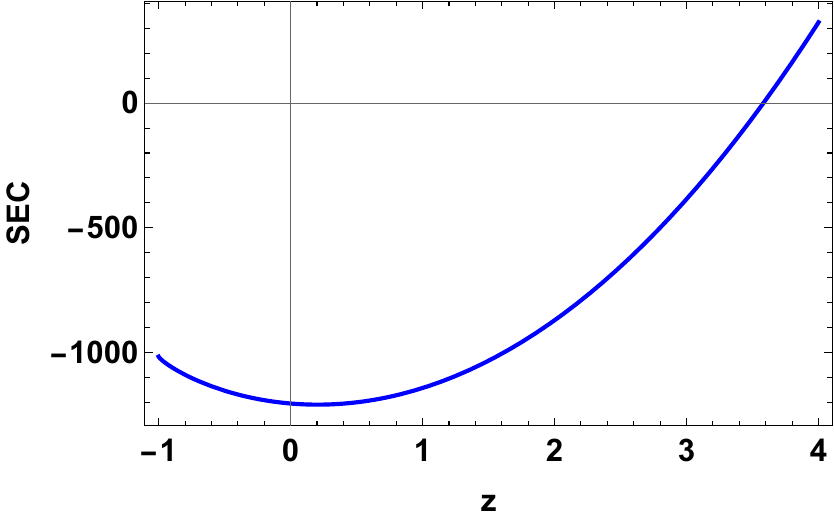}
\caption{Profile of the SEC vs redshift.}\label{f9b}
\end{figure*}

Based on Figs. \ref{f7b} and \ref{f8b}, it is evident that both NEC and DEC meet the positivity conditions across the full redshift range, which aligns with the parameter values estimated from observational data. Additionally, the WEC holds because it includes both the density parameters and the NEC. Finally, Fig. \ref{f9b} illustrates that the SEC was violated in the recent past, suggesting that this violation supports the idea of cosmic acceleration.

\section{Conclusions}\label{sec9c}
\justifying
In this chapter, we adopted the effective EoS expressed in Eq. \ref{2ec}, which was drawn from the conventional value found in Einstein theory, complemented by a proportionality constant $\zeta$ that had been commonly referenced in the literature \citep{IB-1}. We successfully derived the exact solution for our model, which focused on bulk viscous matter within the context of the $f(R,L_m)$ framework. Subsequently, we used a combination of observational datasets, including CC and Pantheon+SH0ES, to refine estimates for the current value of the Hubble parameter $H_0$, as well as the model parameters. The results yielded best-fit values of
$\alpha=1.310^{+0.037}_{-0.032}$, $\gamma = 1.29 \pm 0.20$, $\zeta= 5.02 \pm 0.26$, and $H_0= 72.09 \pm 0.19$.

We investigated the behavior of the density parameter, the pressure component that included viscosity, and the effective EoS parameter in relation to the redshift, as illustrated in Figs. \ref{f2b}, \ref{f3b}, and \ref{f4b}. This analysis was based on 7,500 samples obtained by resampling the chains using \textit{emcee}. From Fig. \ref{f2b}, it was evident that the cosmic density parameter demonstrated the expected positive trend. In contrast, the effective pressure component, shown in Fig. \ref{f3b}, revealed a negative trend, which could have contributed to the accelerating expansion of the Universe. Furthermore, we determined that the current value of the effective EoS parameter was approximately $\omega_0 \approx -0.71$. The trajectory of the EoS parameter, depicted in Fig. \ref{f4b}, further supported the notion of an accelerating expansion phase of the Universe. Following this, we evaluated the $(r,s)$ parameters for our proposed $f(R,L_m)$ model, obtaining a value of the statefinder parameter of nearly $(r,s)=(0.43,0.33)$. As shown in Fig. \ref{f5b}, we observed that the evolutionary path of our $f(R,L_m)$ model resided within the quintessence region. Finally, the Om diagnostic, illustrated in Fig. \ref{f6b}, suggested that the proposed $f(R,L_m)$ model aligned with the quintessence scenario.

Furthermore, the energy conditions shown in Figs. \ref{f7b}, \ref{f8b}, and \ref{f9b} met the positivity criteria throughout the redshift range for the NEC and the DEC. However, a violation of the SEC was observed. This breach of the SEC, particularly in the recent past, supported the notion of current acceleration of the Universe. Furthermore, the energy conditions shown in Figs. \ref{f7b}, \ref{f8b}, and \ref{f9b} met the positivity criteria throughout the redshift range for the NEC and the DEC. However, a violation of the SEC was observed. This breach of the SEC, particularly in the recent past, supported the notion of current acceleration of the Universe. In conclusion, our cosmological $f(R,L_m)$ model, which included the effects of bulk viscosity in the fluid, provided a robust framework for understanding late-time cosmic behavior according to observational data.

Having explored the constraints on viscous DE and its EoS in $f(R,L_m)$ gravity, we now turn our attention to an alternative approach to addressing cosmic singularities. In the next chapter, we investigate the concept of matter bounce nonsingular cosmology within the same modified gravity framework, analyzing specific nonlinear functional forms and their implications for cosmic evolution.


%% file: Chapters/Chapter5.tex

\chapter{Bouncing cosmological models in \texorpdfstring{$f(R,L_m)$}{f(R,Lm)} gravity} 

\label{Chapter5} 

\lhead{Chapter 5. \emph{Bouncing cosmological models in $f(R,L_m)$ gravity}} 

\vspace{10 cm}
* The following publications cover the work in this chapter:\\
 
\textit{Bouncing cosmological models in $f(R,L_m)$ gravity}, Physica Scripta, \textbf{99}, 065031 (2024).
\clearpage
This chapter investigates the concept of matter bounce non-singular cosmology within the framework of $f(R,L_m)$ gravity. We analyze two specific non-linear functional forms of $f(R,L_m)$. We derive the associated Friedmann equations for both models in the FLRW background. The impact of model parameters and the bouncing-scale factor is discussed concerning the EoS parameter as well as the pressure and density parameters. We also delve into the dynamical characteristics of cosmographic parameters, including jerk, lerk, and snap. Furthermore, analysis reveals that violations of the NEC and the SEC indicate non-singular accelerating solutions for both examined nonlinear $f(R,L_m)$ functions. Lastly, we assess the behavior of the adiabatic speed of sound to evaluate the feasibility of the proposed cosmological bouncing scenario.

\section{Introduction}\label{sec1e}
\justifying
It is well known that the inflationary scenario proposes that the Universe underwent rapid expansion at its beginning. However, the singularity occurred before inflation began, so the scenario needs to be completed to explain the entire history of the Universe. An alternate way to resolve this issue is to consider the matter bounce scenario \citep{MB-1,Haro,Saho}, suggesting that the Universe first contracted before expanding without encountering a singularity. This theory proposes an initial matter-dominated contraction epoch followed by a bounce that further causes a generation of casual fluctuations. 

However, it should be noted that in a flat Universe, the non-singular bounce may violate the NEC. Such a nonsingular cosmology with violation of the NEC is supported in generalized Galileon theories \citep{MB-3}. The idea of a big bounce scenario in the context of modified gravity replacing the Big Bang singularity is an intriguing topic for research \citep{Mand2,Zuba,Bhat,Yous}. There are plenty of interesting results of bouncing scenarios in the context of $f(R)$ gravity that have appeared in the literature, such as Carlos et al. \citep{Barr} have studied bouncing cosmologies in Palatini $f(R)$ gravity formalism, Niladri et al. \citep{Paul} studied cosmological bounce in spatially flat FLRW background in metric $f(R)$ gravity, Odintsov and  Oikonomou investigated matter bounce loop quantum cosmology from $f(R)$ gravity \citep{Odin}, Myrzakulov and Sebastian studied bouncing solutions with viscous fluid \citep{Myrz}, bouncing cosmology with future singularity from modified gravity has been investigated in \citep{Odin1}, and others \citep{Ilya,Aman,Odin2,sing,sing1,bane,caru}.

The present section investigates a matter bounce non-singular cosmological scenario in the $f(R,L_m)$ gravity background. In Sec. \ref{sec3e}, we present the idea of matter bounce. We specify a scale factor parameterization and investigate the intricate dynamical development of a Universe undergoing a non-singular bounce. In this section, we discuss the cosmography of the bouncing model in terms of cosmic time. In Sec. \ref{sec5e} we investigate the matter bounce scenario in the $f(R,L_m)$ gravity framework using two nonlinear functions of $f(R,L_m)$. Further in Sec. \ref{sec5e}, we derive the Friedmann equations and the corresponding dynamical parameters $\rho$, $p$, and $\omega$ corresponding to the assumed nonlinear functional forms $f(R,L_m)$. Then, we have presented the profiles of different energy conditions. Moreover, we investigate the stability of the bouncing scenarios considered. Finally, in Sec. \ref{sec6e}, we presented the outcomes of our findings.

\section{The bouncing model}\label{sec3e}
\begin{figure*}[htbp]
\centering
\subfigure[]{\includegraphics[width=7.5cm,height=5cm]{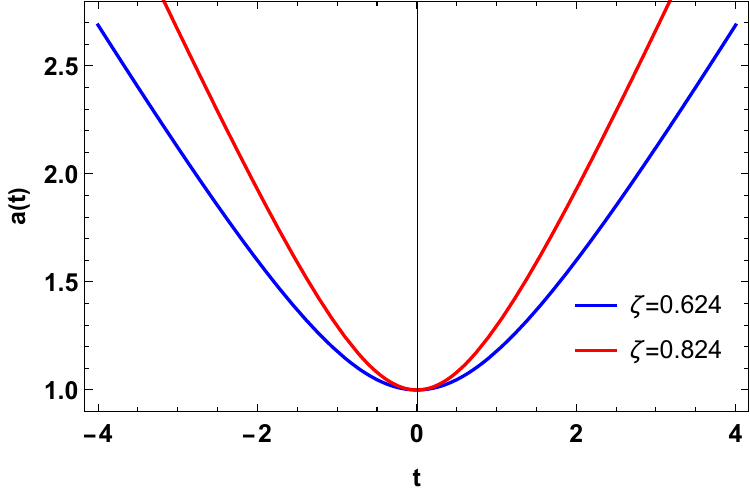}}\,\,\,\,\,\,\,\,\,\,\,\,
\subfigure[]{\includegraphics[width=7.5cm,height=5cm]{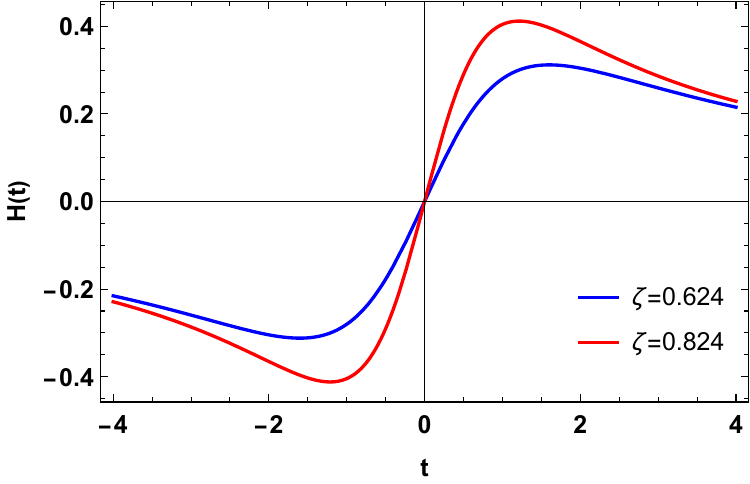}}\,\,\,\,\,\,\,\,\,\,\,\,
\subfigure[]{\includegraphics[width=7.5cm,height=5cm]{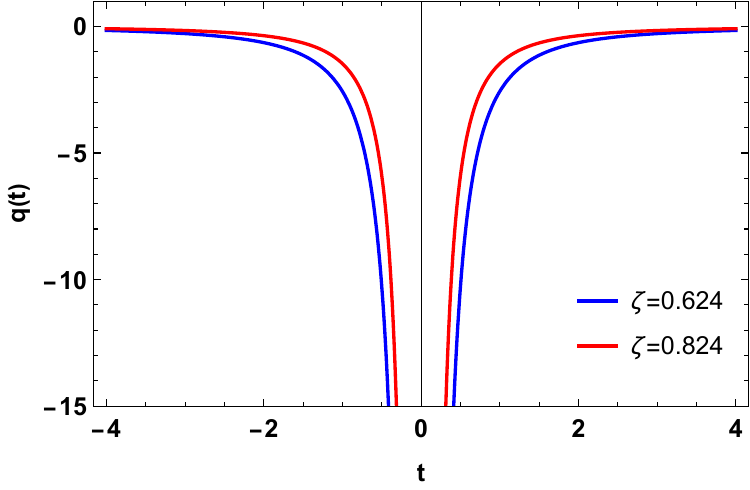}}
\caption{Profile of the scale factor, Hubble parameter, and deceleration parameter for $a_0=1$, with different values of $\zeta$ against cosmic time.}\label{f1e}
\end{figure*}

\justifying
To understand the behavior of the Universe, we must first understand its physical and dynamic entities. The Hubble parameter is used to assert all the physical parameters. It is worth noting that its inflationary scenario does not account for the entire previous history of the cosmos. As a result, the matter bounce has been suggested as a potential solution to this problem. We are also focused on the matter bounce scenario with the respective scale factor \citep{Zuba}
\begin{equation}\label{3ae}
a(t)=(a_0 ^2 +\zeta^2 t^2)^\frac{1}{2}.
\end{equation}
As this study is for the accelerated expansion phase, $\zeta$ is a positive bouncing parameter. The radius of the scale factor at the bouncing point $t=0$ is denoted by $a_0$. Fig. \ref{f1e} (a) depicts the behavior of the scale factor versus cosmic time for various values of $\zeta$. The scale factor expands symmetrically on both sides of the bounce point $t=0$, implying an initial contraction, a bounce, and an expansion. The bouncing parameter $\zeta$ influences the slope of the scale factor. In summary, the bouncing parameter is an essential component that affects the slope of a curve of the scale factor $a$. The corresponding expression for the Hubble parameter of the scale factor \eqref{3ae} is defined as 
\begin{equation}\label{3be}
H(t)=\frac{\zeta^2 t }{a_0 ^2  + \zeta^2 t^2}.
\end{equation}
The Hubble parameter has negative values before the bounce and positive values after the bounce, which disappears at the bounce. Fig. \ref{f1e} (b) depicts the adaptive behavior of the Hubble parameter in the preceding bouncing scenario. The Hubble parameter meets the requirements for bouncing cosmology because it starts negative, passes through zero at t = 0, and then turns positive. The deceleration parameter for the bouncing scenario described above is given by
\begin{equation}\label{3ce}
q(t)=-\frac{a_0 ^2}{\zeta^2 t^2 }.
\end{equation}
\\
As a result, Fig. \ref{f1e} (c) depicts the behavioral patterns of the deceleration parameter, whose negative range indicates an accelerated phase of expansion. The deceleration parameter exhibits a singularity near the bounce due to the selected responses of the bouncing model. With this favorable behavior of the basic parameters, we can proceed with the chosen scale factor to evaluate the cosmological model. Given the positive trends in the fundamental parameters, we can proceed with the selected scale factor to assess the cosmological model.

The traditional cosmographic approach is based on the Taylor series expansion of material objects. Cosmography is an effective approach to bridge the gap between cosmological models and popular methods to understand the dynamics of the Universe \citep{V.S.,Viss,Alam,Capo1,Capo2}. The cosmographic framework includes several cosmological parameters, including the Hubble parameter ($H$), the deceleration parameter ($q$), the jerk parameter ($j$), the snap parameter ($s$), and the lerk parameter ($l$) \citep{Avil,Sale,Mand}. The Hubble parameter and deceleration parameter have already been discussed in relation to the FLRW background, while the remaining cosmographic parameters can be derived as follows 
\begin{equation}\label{5ae}
j(t)= -\frac{3a_0 ^2}{\zeta^2 t^2},
\end{equation}
\begin{equation}\label{5be}
s(t)=-\frac{3 \left(a_0^4 -4 a_0^2 \zeta ^2 t^2\right)}{\zeta ^4 t^4},
\end{equation}
\begin{equation}\label{5ce}
l(t)=\frac{15 \left(3a_0^4-4 a_0^2 \zeta ^2 t^2\right)}{\zeta ^4 t^4}.
\end{equation}
\begin{figure*}[htbp]
\centering
\subfigure[]{\includegraphics[width=7.5cm,height=5cm]{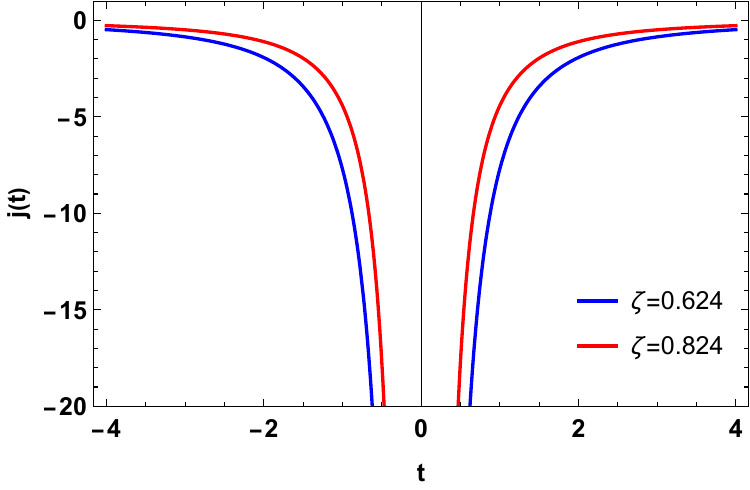}}\,\,\,\,\,\,\,\,\,\,\,\,
\subfigure[]{\includegraphics[width=7.5cm,height=5cm]{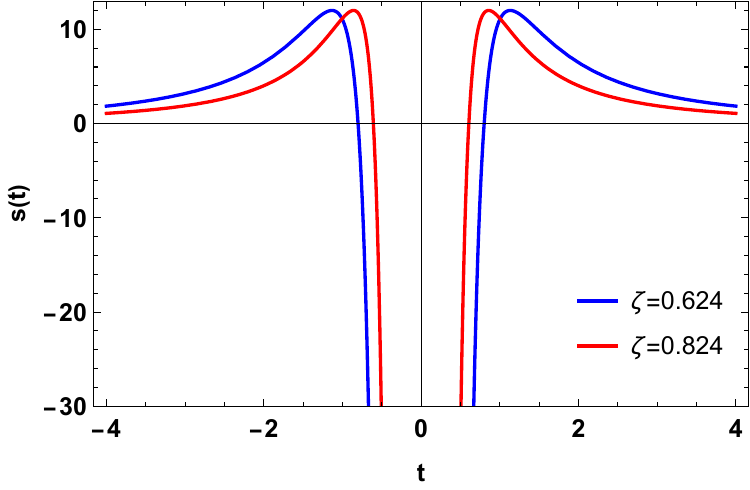}}\,\,\,\,\,\,\,\,\,\,\,\,
\subfigure[]{\includegraphics[width=7.5cm,height=5cm]{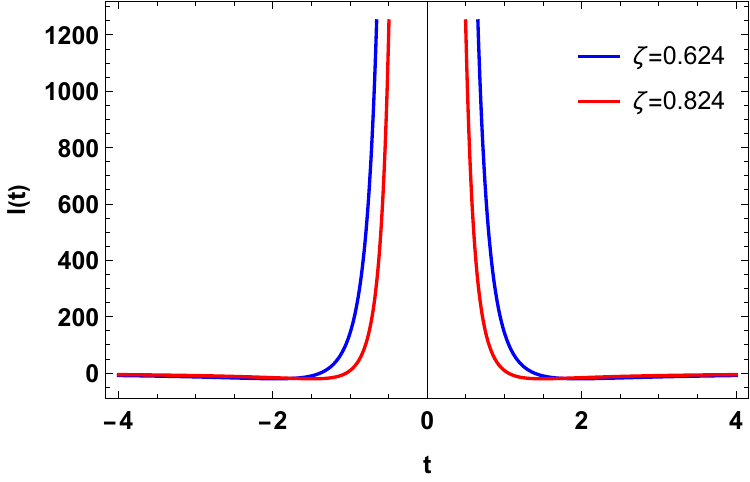}}
\caption{Profile of the jerk parameter, snap parameter, and lerk parameter for $a_0=1$, with different values of $\zeta$ against cosmic time.}\label{f2e}
\end{figure*}\\

Fig. \ref{f2e} (a) depicts the graphical representation of the behavior of the jerk parameter for the chosen parameters, while Fig. \ref{f2e} (b) depicts the behavior of the snap parameter. Regardless of the negative and positive time zones, the behavior of $\Lambda$CDM leads to the jerk parameter, which displays a singular bounce during the bouncing epoch. The snap parameter displays similar patterns of behavior. However, it deviates from the intended $\Lambda$CDM behavior. But for the temporal area and the progression of the Universe is away from a matter-dominated one, the lerk parameter remains positive, shown in Fig. \ref{f2e} (c).

\section{Matter bounce scenario in \texorpdfstring{$f(R,L_m)$}{f(R,Lm)} gravity}\label{sec5e}
\justifying
To probe the implications of the assumed bouncing model under the $f(R,L_m)$ gravity background, we consider the following two non-linear $f(R,L_m)$ functional forms: 

\textbf{Model I :} We consider the model $f(R,L_m) = \frac{R}{2}+ L_m ^{\beta}+\gamma$. The corresponding modified Friedmann equations with $L_m=\rho$ \citep{HLR} and for a perfect fluid distribution reads as
\begin{equation}\label{n4e}
3H^2 = (2\beta-1) \rho^{\beta}-\gamma,
\end{equation}
\begin{equation}\label{n5e}
-2\dot{H}-3H^2 =\big( \beta(p-\rho)+\rho\big) \rho^{\beta-1}+\gamma.
\end{equation} 
Moreover, Eq. \eqref{48} yields the following energy-balance equation corresponding to the model I,
\begin{equation}\label{cc1e}
 (2\beta-1)\dot{\rho}+ 3H(\rho+p)=0.   
\end{equation}

\textbf{Model II :} We consider the model $f(R,L_m) = \frac{R}{2}+ \lambda R^2 + \alpha L_m $. The assumed model $f(R,L_m)$ reduces to the GR case for the values $\lambda=0$ and $\alpha=1$. We choose this specific model because of its significance in contemporary theoretical physics, especially in modified gravity theories and cosmology. This functional form draws inspiration from the Starobinsky model, which has been thoroughly explored and confirmed to be consistent across various cosmological scenarios, including inflation \citep{Chai}, post-inflationary gravitational wave production \citep{Kosh}, and the evolution of star clusters \citep{Manz}. The corresponding modified Friedmann equations with $L_m=\rho$ \citep{HLR} and for a perfect fluid distribution reads as
\begin{equation}\label{2ge}
3H^2 = \frac{\alpha \rho - 36\lambda(2 H \ddot{H}-\dot{H}^2)}{1+72\lambda\dot{H}},
\end{equation}
\begin{equation}\label{2he}
-2\dot{H}-3H^2 =\alpha p+12\lambda(2\dddot{H}+9\dot{H}^2+18\dot{H}H^2+12\ddot{H}H).
\end{equation} 
Moreover, Eq. \eqref{48} yields the following energy-balance equation corresponding to the model II,
\begin{equation}\label{cc2e}
 3\dot{\rho}+ 3H(\rho+p)=0.   
\end{equation}

\subsection{Dynamical parameters}\label{sbsec1e}
\begin{figure*}[htbp]
\centering
\subfigure[]{\includegraphics[width=7.5cm,height=5cm]{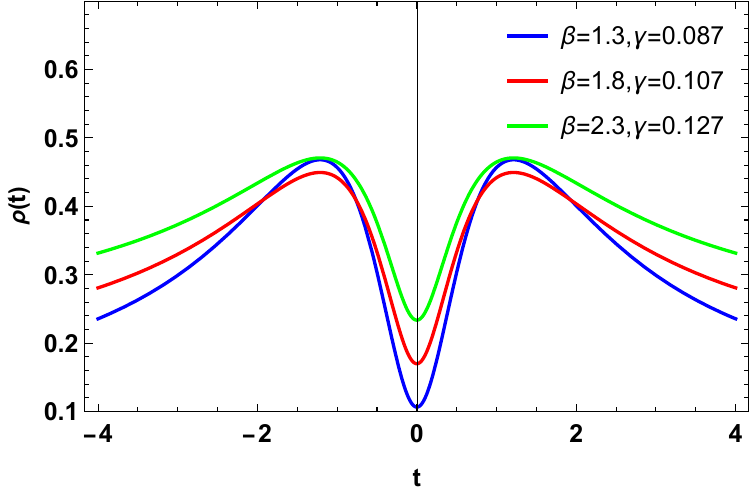}}\,\,\,\,\,\,\,\,\,\,\,\,
\subfigure[]{\includegraphics[width=7.5cm,height=5.15cm]{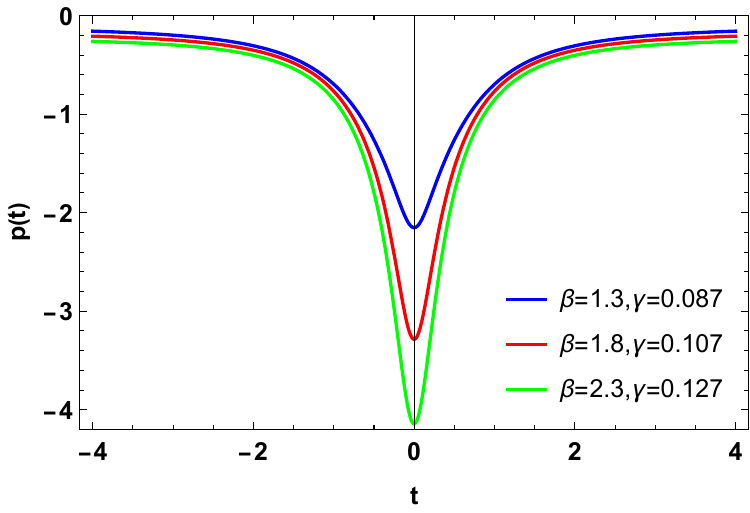}}\,\,\,\,\,\,\,\,\,\,\,\,
\subfigure[]{\includegraphics[width=7.5cm,height=5.2cm]{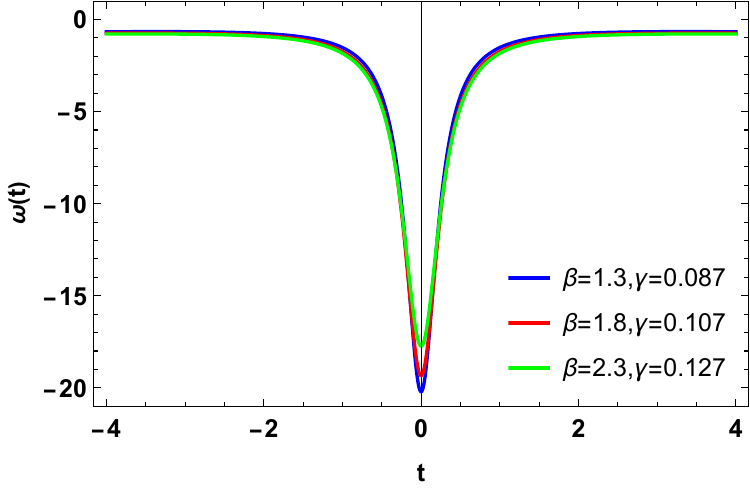}}
\caption{Profile of the density parameter, pressure, and  EoS parameter for $a_0=1$ and  $\zeta=0.824$, with different values of other model parameters $\beta$ and $\gamma$ against cosmic time.}\label{f3e}
\end{figure*}

It is essential to look at the decisive parameters of the model to understand the effective elements of the Universe. The action of decisive parameters indicates whether or not the model is bouncing. The decisive parameters, such as the density parameter, pressure and the EoS parameter for $f(R,L_m)$ model I, were obtained using Eq. \eqref{3be} in \eqref{n4e} and \eqref{n5e} are as follows
\begin{equation}\label{n6e}
\rho(t)=\left(\frac{\gamma+\frac{3\zeta^4 t^2 }{(a_0^2+\zeta^2 t^2)^2}}{(2\beta-1)}\right)^\frac{1}{\beta},
\end{equation}
\begin{equation}\label{n7e}
p(t)=-\frac{\left(a_0^4 \beta  \gamma +2 a_0^2 \zeta ^2 \left(\beta  \left(\gamma  t^2+2\right)-1\right)+\zeta ^4 t^2 \left(\beta  \left(\gamma  t^2-1\right)+2\right)\right)\left(\gamma+\frac{3\zeta^4 t^2 }{(a_0^2+\zeta^2 t^2)^2}\right)^\frac{1}{\beta}}{\beta (2\beta-1)^\frac{1}{\beta}\big(a_0 ^4 \gamma +2a_0^2 \gamma \zeta ^2 t^2 + (3+\gamma t^2)\zeta^4t^2\big)},
\end{equation}
\begin{equation}\label{n8e}
\omega(t)=-\frac{a_0^4 \beta  \gamma +2 a_0^2 \zeta ^2 \left(\beta  \left(\gamma  t^2+2\right)-1\right)+\zeta ^4 t^2 \left(\beta  \left(\gamma  t^2-1\right)+2\right)}{\beta  \left(a_0^4 \gamma +2 a_0^2 \gamma  \zeta ^2 t^2+\zeta ^4 t^2 \left(\gamma  t^2+3\right)\right)}.
\end{equation}
The expressions obtained for the density parameter, the pressure, and the EoS parameter possess the bouncing parameter $\zeta$ and the model parameters $\beta$ and $\gamma$. We presented the profile of the above cosmological parameters against cosmic time for the appropriate values of the model parameters as shown in Fig. \ref{f3e} (a), \ref{f3e} (b) and \ref{f3e}(c), respectively.

Fig. \ref{f3e} (a) presents a profile of the density parameter for the values of the chosen parameters. To satisfy some specific energy conditions, the density parameter must be positive throughout the domain of cosmic time. Therefore, we have considered different values of the parameters $\beta$ and $\gamma$, with particular values of $a_0$ and $\zeta$, to satisfy the positive criteria of the density parameter. The density parameter satisfies the positive condition for the range of $\beta \in (0.5,\infty)$ and $\gamma \in(0.05,\infty )$. As the bounce time approaches, the density parameter increases significantly, peaks, and then decreases at the moment of the bounce. Following the bounce, it climbs briefly before decreasing as cosmic time advances. Fig. \ref{f3e} (b) presents a profile of the pressure component against cosmic time. During cosmic evolution, the pressure was a negative quantity. The pressure component is negative for the range of $\beta \in (0.5,\infty)$ and $\gamma \in(0.015,\infty )$. It exhibits a tiny negative value at the prebounce epoch and then increases to a high negative value at the bounce. In addition, it approaches a tiny negative value in the post-bouncing zone. Fig. \ref{f3e} (c) presents a profile of the EoS parameter for different values of model parameters $\beta$ and $\gamma$ for model I, where the bouncing parameter has a significant impact on the EoS parameter for non-zero constant values of $a_0$ and $\zeta$. The EoS parameter exhibits negativity in the range of $\beta \in (0.49,\infty)$ and $\gamma \in(0,\infty )$. The EoS parameter corresponding to model I crosses the $\Lambda$CDM line and lies in the phantom region near the bouncing epoch. In contrast, it remains in the quintessence phase for the region away from the bouncing epoch. The EoS parameter follows a similar pattern for both the negative and positive time zones.

The cosmological parameters such as density parameter, pressure, and the EoS parameter for $f(R,L_m)$ model II were obtained by using Eq. \eqref{3be} in Eqs. \eqref{2ge} and \eqref{2he} are as follows
\begin{equation}\label{3de}
\rho(t)=\frac{3a_0^2 \zeta ^4 \left(t^2-12 \lambda \right)+3\zeta ^6 t^2 \left(t^2-36 \lambda \right)}{\alpha  \left(a_0^2+\zeta ^2 t^2\right)^3},
\end{equation}
\begin{equation}\label{3ee}
p(t)=-\frac{\zeta ^2 \left(3 a_0^2 \zeta ^2 \left(t^2-12 \lambda \right)+2 a_0^4+\zeta ^4 t^2 \left(36 \lambda +t^2\right)\right)}{\alpha  \left(a_0^2+\zeta ^2 t^2\right)^3},
\end{equation}
\begin{equation}\label{3fe}
\omega(t)= -\frac{\zeta ^2 \left(3 a_0^2 \zeta ^2 \left(t^2-12 \lambda \right)+2 a_0^4+\zeta ^4 t^2 \left(36 \lambda +t^2\right)\right)}{3 \left(a_0^2 \zeta ^4 \left(t^2-12 \lambda \right)+\zeta ^6 t^2 \left(t^2-36 \lambda \right)\right)} .
\end{equation}

\begin{figure*}[htbp]
\centering
\subfigure[]{\includegraphics[width=7.5cm,height=5.16cm]{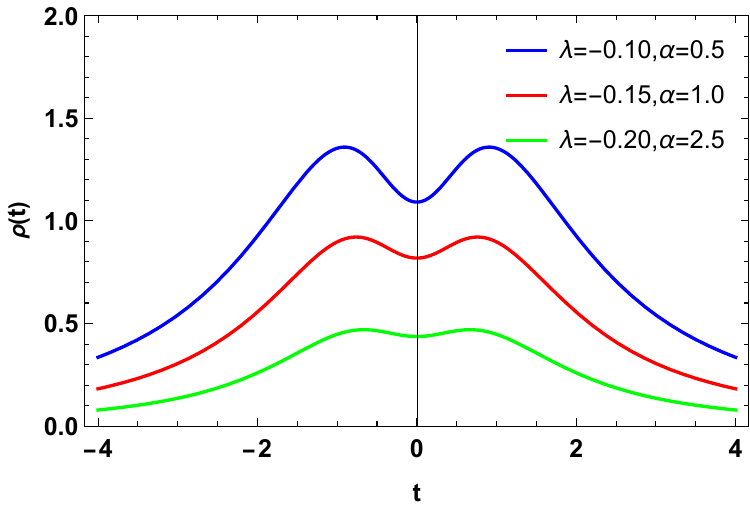}}\,\,\,\,\,\,\,\,\,\,\,\,
\subfigure[]{\includegraphics[width=7.5cm,height=5cm]{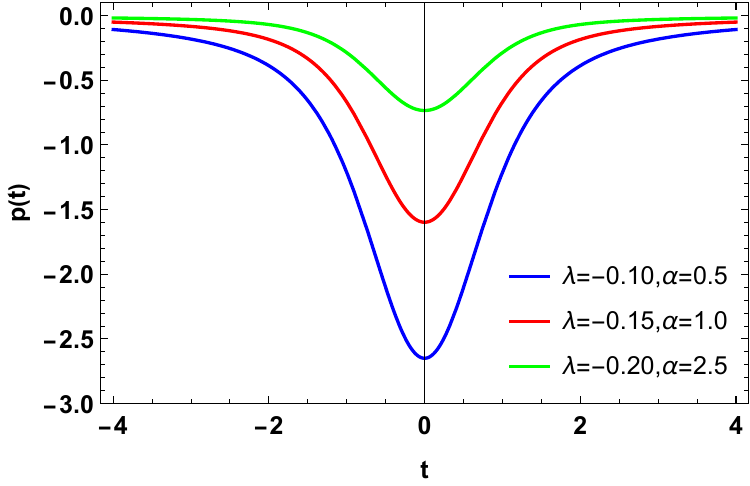 }}\,\,\,\,\,\,\,\,\,\,\,\,
\subfigure[]{\includegraphics[width=7.5cm,height=5.2cm]{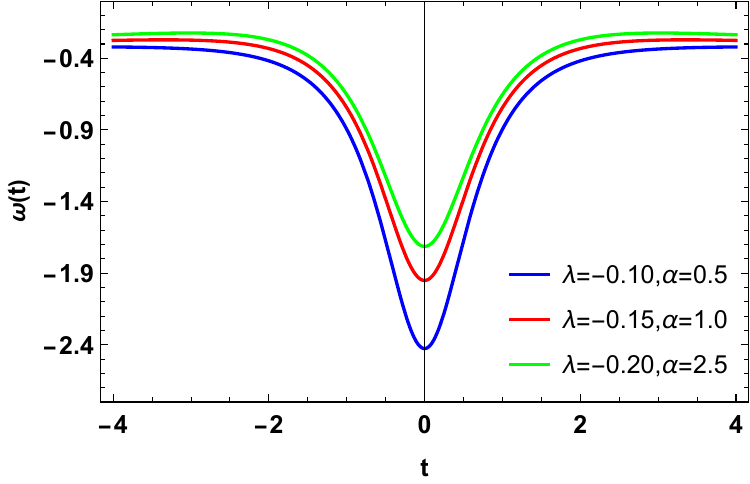}}
\caption{Profile of the density parameter, pressure, and  EoS parameter for $a_0=1$ and  $\zeta=0.624$, with different values of other model parameters $\lambda$ and $\alpha$ against cosmic time.}\label{f4e}
\end{figure*}

The above equations show that the density parameter, pressure, and the EoS parameter are functions of the bounce parameter $\zeta$ and the model parameters $\alpha$ and $\lambda$. In the present section, we investigate appropriate values of these parameters and plot them versus cosmic time, as shown in Fig. \ref{f4e} (a), \ref{f4e} (b) and \ref{f4e} (c), respectively.

Fig. \ref{f4e} (a) depicts the graphical representation of the behavior of the density parameter for the chosen parameters. The density parameter should be positive throughout cosmic history to fulfill specific energy conditions. As a result, we have considered different values of the model parameters $\lambda$ and $\alpha$, with a particular value of $a_0$ and $\zeta$, to guarantee a positive density parameter throughout cosmic development. The density parameter satisfies the positive condition for the range of $\lambda \in (-\infty,-0.02)$ and $\alpha \in(0,\infty )$. As the bounce time approaches, the density parameter experiences a significant increase, peaks, and then declines at the moment of the bounce. Following the bounce, it climbs briefly before decreasing as cosmic time advances. Fig. \ref{f4e} (b) shows a graphical representation of the pressure in cosmic time. During cosmic evolution, the pressure was a negative quantity. The pressure component is negative for the range of $\lambda \in (-0.7,0)$, and $\alpha \in(0,\infty )$. It increases from a tiny negative value in the prebounce zone to a high negative value at the bounce. In contrast, it increases from a high negative value to a tiny negative value in the post-bouncing zone. Fig. \ref{f4e} (c) depicts the dynamical behavior of the EoS parameter for different values of model parameters $\lambda$ and $\alpha$ for Model II, where the bouncing parameter has a significant impact on the EoS parameter for non-zero constant values of $a_0$ and $\zeta$. The EoS parameter exhibits negativity in the range of $\lambda \in (-0.67,0)$ and $\alpha \in (-\infty,\infty)$. The model transitions across the $\Lambda$CDM line and remains in the quintessence phase as the EoS parameter shifts away from the bouncing epoch. During this evolution, the EoS parameter evolves from the phantom region on both sides of the bouncing period. Both the negative and positive time zones had a similar pattern of behavior in the growth of the EoS parameter.

\subsection{Energy conditions}\label{sbsec2e}
In the context of GR, it is believed that the energy conditions remain essentially positive. A set of linear equations called energy conditions predict that matter should have a positive density parameter. The amplification of energy conditions is the cause of gravitational collapse, the Big Bang singularity, and the absence of traversable wormholes. As a result, under the assumption of a perfect fluid matter distribution, the energy conditions \citep{EC} can be determined from the standard GR. The model parameters affect every energy condition in the expression above. Following DEC, matter moves along time-like or null-world trajectories, and the SEC implies that gravity is attractive \citep{Ehle,Noji,Capo3,Mand1}. We obtained the following expressions for NEC, SEC and DEC corresponding to $f (R, L_m)$ model I, obtained by using Eqs. \eqref{n6e} and \eqref{n7e},\\
\begin{equation}\label{n9e}
 \rho+p= \left[\frac{\frac{3 \zeta ^4 t^2}{\left(a_0^2+\zeta ^2 t^2\right)^2}+\gamma }{2 \beta -1}\right]^\frac{1}{\beta} \left[1-\frac{a_0^4 \beta  \gamma +2 a_0^2 \zeta ^2 \left(\beta  \left(\gamma  t^2+2\right)-1\right)+\zeta ^4 t^2 \left(\beta  \left(\gamma  t^2-1\right)+2\right)}{\beta  \left(a_0^4 \gamma +2 a_0^2 \gamma  \zeta ^2 t^2+\zeta ^4 t^2 \left(\gamma  t^2+3\right)\right)}\right]\geq0,
\end{equation}
\begin{equation}\label{n10e}
 \rho+3p=  \left[\frac{\frac{3 \zeta ^4 t^2}{\left(a_0^2+\zeta ^2 t^2\right)^2}+\gamma }{2 \beta -1}\right]^\frac{1}{\beta} \left[1-3\frac{a_0^4 \beta  \gamma +2 a_0^2 \zeta ^2 \left(\beta  \left(\gamma  t^2+2\right)-1\right)+\zeta ^4 t^2 \left(\beta  \left(\gamma  t^2-1\right)+2\right)}{\beta  \left(a_0^4 \gamma +2 a_0^2 \gamma  \zeta ^2 t^2+\zeta ^4 t^2 \left(\gamma  t^2+3\right)\right)}\right]\geq0,
\end{equation}
\begin{equation}\label{n11e}
\rho-p= \left[\frac{\frac{3 \zeta ^4 t^2}{\left(a_0^2+\zeta ^2 t^2\right)^2}+\gamma }{2 \beta -1}\right]^\frac{1}{\beta} \left[1+\frac{a_0^4 \beta  \gamma +2 a_0^2 \zeta ^2 \left(\beta  \left(\gamma  t^2+2\right)-1\right)+\zeta ^4 t^2 \left(\beta  \left(\gamma  t^2-1\right)+2\right)}{\beta  \left(a_0^4 \gamma +2 a_0^2 \gamma  \zeta ^2 t^2+\zeta ^4 t^2 \left(\gamma  t^2+3\right)\right)}\right]\geq0.
\end{equation}
\begin{figure*}[htbp]
\centering
\subfigure[]{\includegraphics[width=7.5cm,height=5cm]{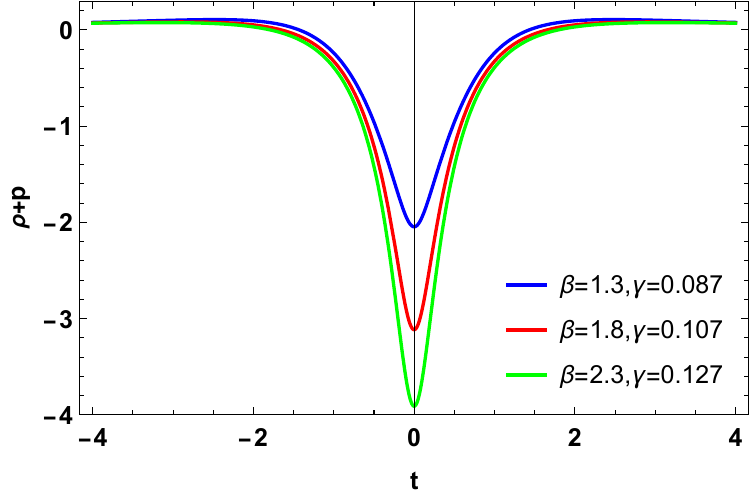}}\,\,\,\,\,\,\,\,\,\,\,\,
\subfigure[]{\includegraphics[width=7.5cm,height=5cm]{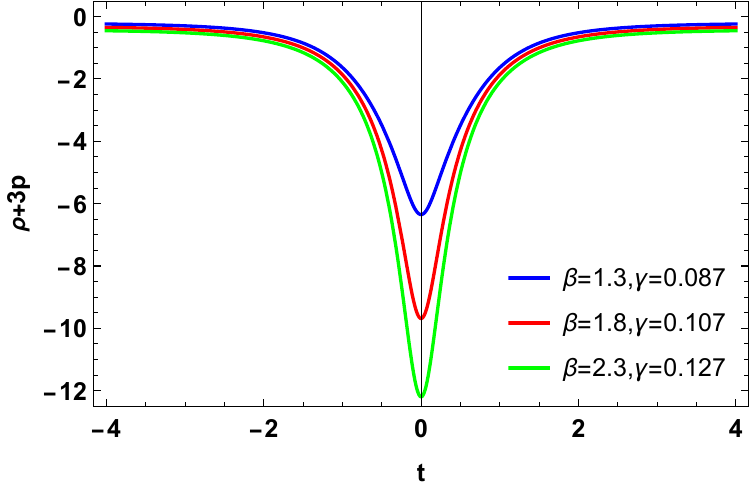}}\,\,\,\,\,\,\,\,\,\,\,\,
\subfigure[]{\includegraphics[width=7.5cm,height=5.2cm]{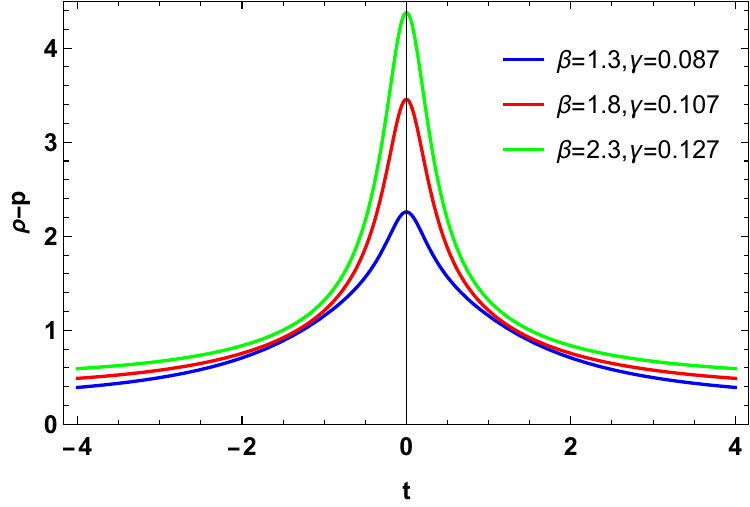}}
\caption{Profile of the NEC, SEC, and DEC for $a_0=1$ and $\zeta=0.824$, with different values of other model parameters $\beta$ and $\gamma$ against cosmic time.}\label{f5e}
\end{figure*}

Fig. \ref{f5e} (a) presents a profile of the NEC, whereas Fig. \ref{f5e} (b) presents a profile of the SEC for the values of the chosen parameters. To complete a nonsingular bounce, the EoS parameter must cross the phantom split $\omega <-1$ and then violate the NEC. The bouncing behavior requires a violation of the NEC during the bouncing phase. Fig. \ref{f5e} (a) indicates the apparent violation of the NEC, while Fig. \ref{f5e} (b) shows how the violation of the NEC leads to the violation of the SEC, which causes the model to evolve in the phantom phase $\omega <-1$. These models also show how the Universe is expanding faster than usual. The figures clearly show that there is no singularity around the bouncing epoch. Fig. \ref{f5e} (c) shows the positive behavior of the DEC, which is expected in the case of a perfect fluid-type matter distribution. In this case, NEC and SEC are violated and DEC satisfies the $\beta \in (0.59,\infty)$ range and $\gamma \in (0,\infty)$.  Moreover, the region close to the bouncing point experiences a uniform shift in energy conditions.

Again, the expressions for the NEC, SEC and DEC corresponding to $f(R,L_m)$ model II, obtained by using Eqs. \eqref{3de} and \eqref{3ee} reads as,
\begin{equation}\label{4ae}
 \rho+p= \frac{2 \zeta ^6 t^2 \left(t^2-72 \lambda \right)-2 a_0^4 \zeta ^2}{\alpha  \left(a_0^2+\zeta ^2 t^2\right)^3}\geq0,
\end{equation}
\begin{equation}\label{4be}
  \rho+3p=-\frac{6 \zeta ^2 \left(a_0^2 \zeta ^2 \left(t^2-12 \lambda \right)+a_0^4+36 \zeta ^4 \lambda  t^2\right)}{\alpha  \left(a_0^2+\zeta ^2 t^2\right)^3}\geq0,
\end{equation}
\begin{equation}\label{4ce}
 \rho-p= \frac{2 \zeta ^2 \left(a_0^2+2 \zeta ^2 \left(t^2-18 \lambda \right)\right)}{\alpha  \left(a_0^2+\zeta ^2 t^2\right)^2}\geq0.
\end{equation}

\begin{figure*}[htbp]
\centering
\subfigure[]{\includegraphics[width=7.5cm,height=5cm]{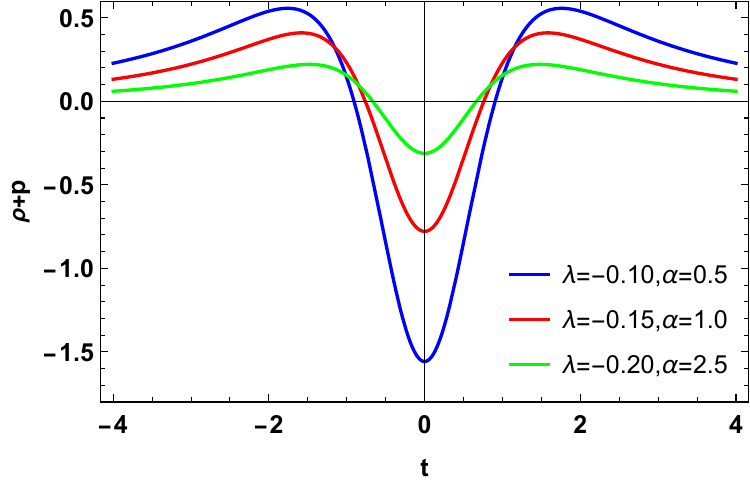}}\,\,\,\,\,\,\,\,\,\,\,\,
\subfigure[]{\includegraphics[width=7.5cm,height=5cm]{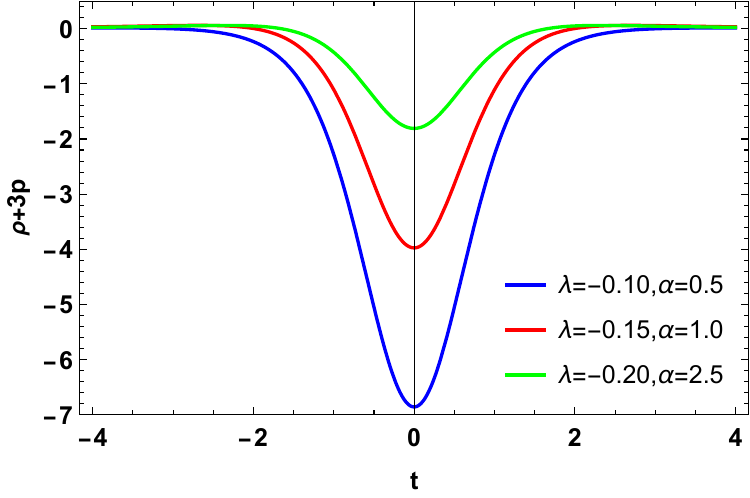}}\,\,\,\,\,\,\,\,\,\,\,\,
\subfigure[]{\includegraphics[width=7.5cm,height=5.2cm]{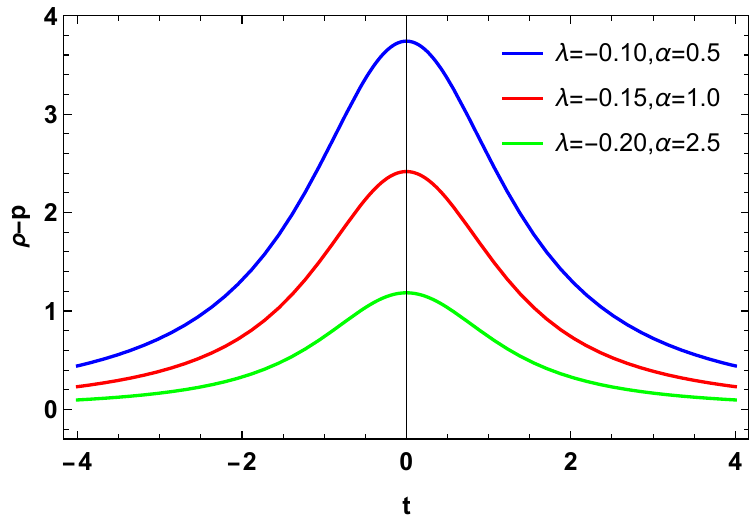}}
\caption{Profile the of NEC, SEC, and DEC for $a_0=1$ and $\zeta=0.624$, with different values of other model parameters $\lambda$ and $\alpha$ against cosmic time.}\label{f6e}
\end{figure*}

 Fig. \ref{f6e} (a) depicts the graphical representation of the behavior of the NEC for the chosen parameters, while Fig. \ref{f6e} (b) depicts the behavior of the SEC. Fig. \ref{f6e} (a) illustrates the clear violation of the null energy constraint. Fig. \ref{f6e} (b) shows how a violation of NEC leads to a violation of SEC, which is less important in the study. In this case, the NEC is violated for the range of $\lambda \in (-0.6,\infty)$ and $\alpha \in (0,\infty)$, whereas the SEC is violated for the range of $\lambda \in (-\infty,0.2)$ and $\alpha \in (0,\infty)$.  The figures clearly show that there is no singularity around the bouncing epoch. However, in the case of a perfect fluid, the DEC is supposed to remain positive, and the same outcome has been seen in Fig. \ref{6} (c). The DEC satisfies for the range of $\lambda \in (-\infty,0.05)$, and $\alpha \in (0,\infty)$.  Instead, the area close to the bouncing point experiences a uniform shift in energy conditions. The energy conditions NEC and SEC turn negative near the bounce across the positive and negative time zone by the predicted matter bounce scenario, and there is a clear indication of violation of the energy condition, which causes the model to evolve in the phantom phase with $\omega <-1$. 

\subsection{Stability analysis}\label{sbsec3}
The squared speed of sound approach \citep{Suda,Char,Fara} is used in this section to address the stability of bouncing models in $f(R,L_m)$ gravity. The symbol for the squared sound speed is $C_s ^2$, and its definition is $C_s ^2=\frac{dp}{d\rho}$. Mechanically stable structures must produce nonnegative outcomes when the square of the speed of sound is applied. Therefore, the bouncing models described above meet the criteria for being stable for positive values of the squared sound speed $C_s ^2$. The squared speed of sound $C_s ^2$ must remain within zero and one for mechanical stability evaluation. The following is the cosmic-time-based equation for the squared speed of sound $C_s ^2$ that can be determined, corresponding to the model I, which we have by using Eqs. \eqref{n6e} and \eqref{n7e} as
\begin{multline}\label{n12e}
 C_s^2 =\frac{3 a_0^6 \beta  (4 \beta -3) \gamma +a_0^4 \zeta ^2 \left(6 \beta  (2 \beta -3)+\beta  (20 \beta -13) \gamma  t^2+6\right)+a_0^2 \zeta ^4 t^2 \left(3 \beta  (4 \beta +3)+\beta  (4 \beta +1) \gamma  t^2-12\right)}{3 \beta  (a_0^2-\zeta ^2 t^2) \left(a_0^4 \gamma +2 a_0^2 \gamma  \zeta ^2 t^2+\zeta ^4 t^2 \left(\gamma  t^2+3\right)\right)}\\ + \frac{ \zeta ^6 t^4 \left(-3 \beta +\beta  (5-4 \beta ) \gamma  t^2+6\right)}{3 \beta  (a_0^2-\zeta ^2 t^2) \left(a_0^4 \gamma +2 a_0^2 \gamma  \zeta ^2 t^2+\zeta ^4 t^2 \left(\gamma  t^2+3\right)\right)}.
\end{multline}
Corresponding to Model II, we have by using Eqs. \eqref{3de} and \eqref{3ee} as
\begin{equation}\label{6ae}
 C_s ^2 =\frac{6 \zeta ^4 t \left(3 a_0^2 \zeta ^2 \left(t^2-12 \lambda \right)+2 a_0^4+\zeta ^4 t^2 \left(36 \lambda +t^2\right)\right) - \zeta ^2 \left(6 a_0^2 \zeta ^2 t+2 \zeta ^4 t \left(36 \lambda +t^2\right)+2 \zeta ^4 t^3\right)}{3 \left(2 a_0^2 \zeta ^4 t+2 \zeta ^6 t \left(t^2-36 \lambda \right)+2 \zeta ^6 t^3\right)-18 \zeta ^2 t \left(a_0^2 \zeta ^4 \left(t^2-12 \lambda \right)+\zeta ^6 t^2 \left(t^2-36 \lambda \right)\right)}.
\end{equation}
\begin{figure*}[htbp]
\centering
\subfigure[]{\includegraphics[width=7.5cm,height=5.2cm]{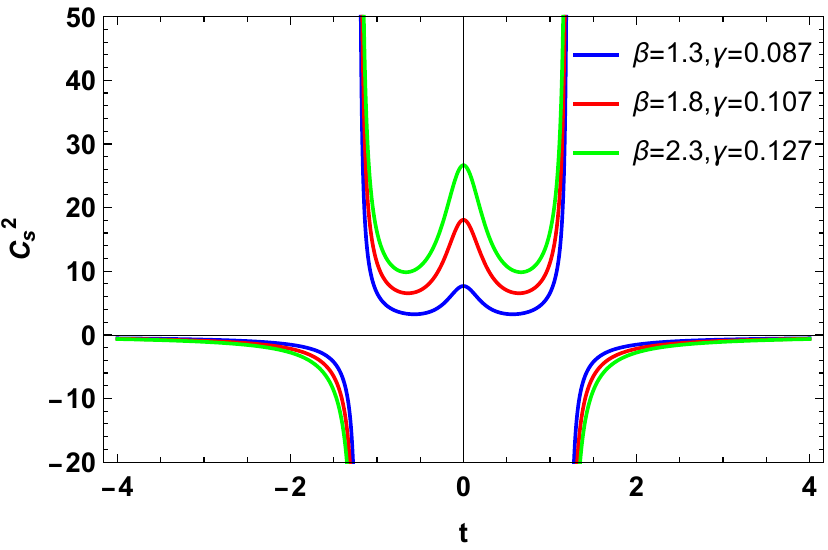}}\,\,\,\,\,\,\,\,\,\,\,\,
\subfigure[]{\includegraphics[width=7.5cm,height=5.2cm]{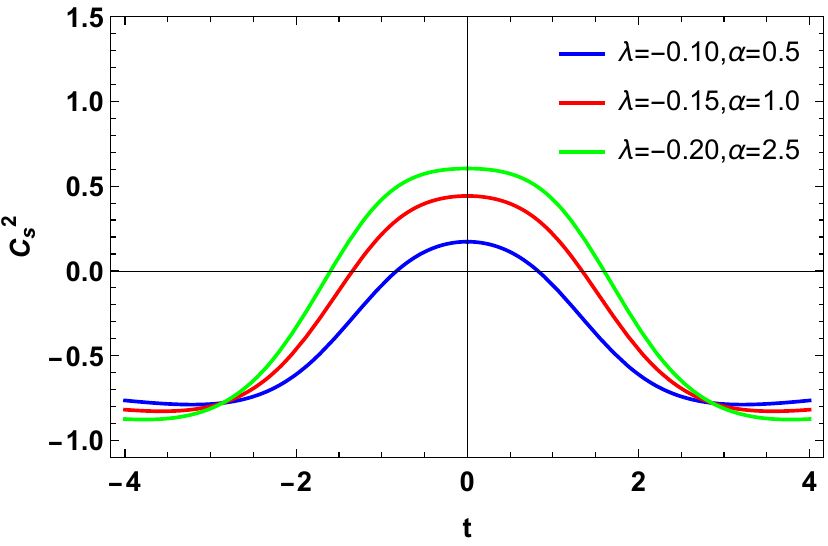 }}
\caption{Profile of stability analysis for $a_0=1$, $\zeta=0.824$ for model I, and $\zeta=0.624$ for model II, with different values of other model parameters $\beta$, $\gamma$ and $\lambda$, $\alpha$ against cosmic time respectively.}\label{f7e}
\end{figure*}

Figs. \ref{f7e} (a) and (b) show the stability analysis for model I and model II. It can be seen from Fig. \ref{f7e} (a) for different values of the model parameters $\beta$ and $\gamma$ with bouncing parameter $\zeta$, the squared speed of sound shows positive values but still greater than one that predicts the unstable behavior of model I. As demonstrated in Fig. \ref{f7e} (b), the squared speed of sound exhibits positive values between zero and one in a specific time range for various deals of the model parameters $\lambda \in (-0.6,0)$, and $\alpha \in (-\infty,\infty) $, which predicts stable behavior for model II. Given that the proposed model displays bouncing behavior at epoch $t = 0$, and considering that the stability range encompasses this bouncing epoch, we conclude that model I is unstable while model II is stable.

\section{Conclusions}\label{sec6e}
\justifying
In this chapter, we investigated the possibility of reproducing specific bouncing models within the $f(R,L_m)$ gravity framework. We observed the bounce at the moment $t=0$ by analyzing the behavior of the dynamical parameters. It was evident that a cosmic bounce occurred when the Hubble parameter was equal to $H=0$. Additionally, for the bounce scenario to hold, a violation of the NEC was necessary, and our analysis confirmed that this violation occurred within the bounce interval. The behavior of the EoS parameter, along with the density parameter, supported the bouncing behavior, showing a clear distinction between the negative and positive regions. The geometrical parameters, including the Hubble parameter and the deceleration parameter, were obtained by constructing a parametric model of the scale factor in the form $a(t)=(a_0^2 +\zeta^2 t^2)^\frac{1}{2}$. Similarly to other cosmographic parameters, such as the jerk, snap, and lerk parameters, the behavior of the deceleration parameter during the bounce exhibited a singularity. An intriguing aspect of the bouncing model was that the lerk parameter reached singularity in its positive profile, while the jerk and snap parameters reached singularity in their negative profiles.

We considered two non-linear $f(R,L_m)$ cosmological models. We derived the expressions for the density parameter, pressure, EoS parameter, and different energy conditions, presenting a descriptive picture of the initial circumstances of the Universe at the bounce for both considered non-linear models. In the pre-bouncing epoch, the density parameter increased, reached its highest point, and then experienced a significant decline. During the post-bouncing epoch, it rose again, peaked once more, and subsequently decreased as cosmic time progressed (refer to the left panels of Fig. \ref{f3e} and \ref{f4e}). Compared to the assumed nonlinear models, the pressure components exhibited high negative values near the bounce. In contrast, they showed a small negative value away from the bouncing epoch (see the right panels of Figs. \ref{f3e} and \ref{f4e}). Due to the non-vanishing nature of the scale factor, we discovered that the initial conditions of the density parameter and pressure in the Universe were finite, eliminating the first singularity problem. For both models, we found that, in the region near the bounce, the EoS parameter crossed the $\Lambda$CDM line ($\omega=-1$)) and was in the phantom phase ($\omega<-1$). In comparison, in regions distant from the bouncing epoch, it existed in the quintessence phase ($\omega > -1$), as shown in the lower panels of Fig. \ref{f3e} and \ref{f4e}. The violation of the NEC and SEC at the bounce region needed to be demonstrated to achieve any non-singular bounce with a standard matter source. The violation of NEC and SEC near the bouncing epoch was presented in Fig. \ref{f5e} for Model I and in Fig. \ref{f6e} for Model II. However, the DEC showed positive behavior for both models. The fact that the SEC was repeatedly violated during cosmic history was interesting because it suggested that the expansion of the Universe accelerated in the late cosmic period. Finally, using the squared speed of sound ($C_s^2$) method, we examined the stability of the bouncing solutions obtained, as presented in Fig. \ref{f7e}. Model I showed unstable behavior, whereas Model II exhibited stable behavior.

Building on our investigation of bouncing cosmological models in $f(R,L_m)$ gravity, we now shift our focus to another intriguing aspect of cosmic evolution, the gravitational baryogenesis epoch. In the next chapter, we explore how $f(R,L_m)$ gravity can influence the generation of baryon asymmetry in the Universe. By considering a Universe filled with DE and a perfect fluid during radiation dominance, we examine the viability of this modified gravity framework in explaining the observed baryon-to-entropy ratio.


%% file: Chapters/Chapter6.tex
\chapter{Baryogenesis in \texorpdfstring{$f(R,L_m)$}{f(R,Lm)} gravity}  

\label{Chapter6} 

\lhead{Chapter 6. \emph{Baryogenesis in $f(R,L_m)$ gravity}} 

\vspace{10 cm}
* The following publications cover the work in this chapter: \\
 
\textit{Baryogenesis in $f(R,L_m)$ gravity}, Physics of the Dark Universe, \textbf{40}, 101223 (2023).
\clearpage
This chapter focuses on the reconstruction of the gravitational baryogenesis epoch within the context of the $f(R,L_m)$ theory of gravity. Assume the Universe is filled with DE and a perfect fluid, along with a nonzero baryon-to-entropy ratio during a period of radiation dominance. We limit our investigation to the gravitational baryogenesis framework, highlighting model parameter values that align with observational data regarding the baryon-to-entropy ratio. Our analysis indicates that $f(R,L_m)$ gravity can contribute significantly and consistently to the process of gravitational baryogenesis.

\section{Introduction}\label{sec1d}
\justifying
Even before cosmology became an independent research division, one of the unanswered questions was the abundance of matter over antimatter in our Universe. Significant observational evidence, such as Big Bang Nucleosynthesis (BBN) \citep{Burl} and CMB \citep{Caldwell}, has strongly indicated that matter predominates over antimatter in the Universe. This superiority is known as baryogenesis. Recent research suggests that the asymmetrical correlation between matter and antimatter originated at the beginning of the cosmos. However, the true origin of baryon asymmetry (BA) remains a mystery that requires further investigation.

Numerous theories have emerged to solve this mystery of BA by evaluating interactions in the primitive Universe that go beyond the standard model, some of which are Affleck-Dine baryogenesis \citep{Stew,Yama,Akit}, spontaneous baryogenesis \citep{Taka,Bran,Simo}, electroweak baryogenesis \citep{Trod,Morr}, grand unified theories \citep{Kolb}, baryogenesis of thermal and black hole evaporation \citep{Dolg}, these multiple baryogenesis contexts discuss how this Universe could have more matter than antimatter during the matter or radiation epoch. The gravitational baryogenesis process employs one of the Sakharov criteria \citep{Sakh} proposed that the baryon asymmetry can be produced by three necessary conditions: (1) processes that violate baryon number, (2) violation of charge (C) and charge parity (CP) and (3) out of thermal equilibrium interactions. The key component is a CP-violating interaction stipulated by coupling between the baryon matter current $J^\mu$ and the derivative of the Ricci scalar curvature $R$, in the form
\begin{equation}\label{1ad}
\frac{1}{M_*^2}\int{\sqrt{-g} J^\mu \partial_\mu (R) d^4x}.
\end{equation}
In Eq. \eqref{1ad}, the parameter $M_*$ represents the cut-off scale of the effective theory, while $R$, $J^\mu$, and $g$ stand for the Ricci scalar, baryonic matter current, and the trace of the metric tensor, respectively. The baryon-to-entropy ratio $\frac{n_B}{s} \propto \dot{R}$, in the case of the FLRW background, where the dot represents the cosmic-time derivative. In the situation of a radiation-dominated era with $\omega=\frac{1}{3}$, the net BA generated by Eq.\eqref{1ad} is zero.

Several authors have studied the mysterious concept of baryogenesis in the context of modified gravity in the past few years. In \citep{Lamb1,Ramo}, $f(R)$ gravity theories are addressed in terms of gravitational baryogenesis, while in \citep{Odin4} Gauss-Bonnet gravity, in \citep{Bent} Gauss-Bonnet braneworld cosmology, in \citep{Oiko} $f (T)$ gravity, in \citep{Bhat4} $f(P)$ gravity, teleparallel gravity \citep{Bhat2}, in \citep{Noza,Baff,Saho4} $f (R,T)$ gravity, in \citep{Sale4} $f(R,T,X)$ gravity, in \citep{Azha} $f(T,B)$ gravity, and in \citep{Bhat1} $f(Q,T)$ gravity, etc. 
This chapter looks into the gravitational baryogenesis framework in the $f(R,L_m)$ gravity theory.

The following is how the present chapter is organized: Sec. \ref{sec3d} will provide some essential baryogenesis factors before investigating gravitational baryogenesis in $f(R,L_m)$ gravity. Then, we explain baryogenesis in $f(R,L_m)$ gravity in detail and imply generating observationally acceptable baryon-to-entropy ratios for the $f(R,L_m)$ gravity model. We will also review the generalized form of baryogenesis that applies to the assuming gravity model. In the final Sec. \ref{sec4d}, we will discuss the conclusions of the present work.

\section{Baryogenesis in \texorpdfstring{$f(R,L_m)$}{f(R,Lm)} gravity}\label{sec3d}
\justifying
In this part, we shall demonstrate how $f(R,L_m)$ gravity addresses the difficulty of gravitational baryogenesis in the cosmos. A crucial parameter to figure out asymmetry is known as BAF, given by
\begin{equation}\label{3ad}
\eta _B =\frac{n_B - \Bar{n}_B}{s},
\end{equation}
where $n_B$ is the baryon number, $\bar{n}_B$ is the antibaryon number, and $s$ is the entropy of the Universe. Observational evidence such as BBN \citep{Burl} and CMB \citep{Caldwell} confirms that the restriction on this BAF is $\frac{n_B}{s}\simeq 9\times10^{-11}$. For the $f(R,L_m)$ gravity, we consider an interaction term that violates CP and is produced by the baryon asymmetry of the Universe of the form
\begin{equation}\label{3bd}
\frac{1}{M_*^2}\int{\sqrt{-g} J^\mu \partial_\mu (R+L_m) d^4x}.
\end{equation}
As a result, for an interaction of Eq. \eqref{3bd} that violates CP, the resulting baryon-to-entropy ratio in $f(R,L_m)$ gravity is as follows
\begin{equation}\label{3cd}
\frac{n_B}{s}\simeq -\frac{15 g_B }{4\pi ^2  g_{*s} }\frac{(\dot{R}+\dot{L_m})}{M^2 _* T_D}.
\end{equation}
In Eq. \eqref{3cd}, $g _B$ is the total number of intrinsic degrees of freedom of baryons, $g_*s$ is the total number of degrees of freedom of massless particles, and the critical temperature is $T_D$ is the temperature of the Universe when all interactions cause baryon asymmetry to begin. We will assume that a thermal equilibrium exists, with the density parameter being proportional to temperature $T$ as
\begin{equation}\label{3dd}
\rho(T)=\frac{\pi^2}{30} g_{*s} T^4.
\end{equation}
Within the framework of Einstein's GR, suppose the matter content is represented as a perfect fluid with a constant EoS parameter and the Ricci scalar, as follows
\begin{equation}\label{3ed}
 R=-8\pi G (1-3\omega)\rho.   
\end{equation}
In a Universe dominated by radiation within the framework of GR, the baryon-to-entropy ratio is zero. However, for other forms of matter, this ratio typically differs from zero. In the context of $f(R,L_m)$ gravity theories, it is possible to generate a net BA during the radiation-dominated era. To explore this phenomenon, we examine a specific $f(R,L_m)$ model that allows us to recover the epoch of baryogenesis. We compute the baryon-to-entropy ratio for this model, considering a Universe comprising DE and a perfect fluid characterized by a constant EoS parameter. 

\subsection{The perfect fluid with \texorpdfstring{$f(R,L_m)$}{f(R,Lm)} gravity}
The recently proposed model we use in this section was presented in chapter-\ref{Chapter2}. That is
\begin{equation}\label{3fd}
  f(R,L_m) = \frac{R}{2} + L_m ^{\alpha} + \zeta . 
\end{equation}
where $\alpha$ and $\zeta$ are model parameters. Then, for this specific $f(R,L_m)$ model with $L_m=\rho$ \citep{HLR}, the Friedmann Eqs. \eqref{14b} and \eqref{15b} are transformed into
\begin{equation}\label{3gd}
3H^2 = (2\alpha-1) \rho ^{\alpha}-\zeta,
\end{equation}
\begin{equation}\label{3hd}
-2\dot{H}-3H^2 =(\alpha p+ (1-\alpha)\rho)\rho^{\alpha-1}+\zeta.
\end{equation} 
Moreover, Eq. \eqref{48} yields the following energy-balance equation corresponding to the model, the same as Eq. \eqref{3id}.

We will proceed assuming that the scale factor follows a power-law form given by $a(t)= B t^{\beta}$, where $\beta = \frac{2}{3(1+\omega)}$ and $B$ is a constant parameter \citep{Odin4}. Consequently, we can derive the expressions for the Hubble parameter and the density parameter in this model as follows
\begin{equation}\label{3jd}
H(t) =  \frac{\beta }{t},
\end{equation}
\begin{equation}\label{3kd}
\rho(t) =  \big[\frac{3\beta ^2 +\zeta t^2}{(2\alpha-1)t^2}  \big]^{\frac{1}{\alpha}}.
\end{equation}
By equating Eqs. \eqref{3kd} and \eqref{3dd}, we derive the decoupling time $t_D$ as a function of the decoupling temperature $T_D$, which can be expressed as
\begin{equation}\label{3ld}
t_D=\big[\frac{3\beta ^2}{(2\alpha-1)(\frac{\pi ^2}{30} g_{*s} T_D ^4)^\alpha -\zeta}\big]^\frac{1}{2}.
\end{equation}
Applying Eq. \eqref{3ld}, we obtain the final expression for the baryon-to-entropy ratio within the context of the current $f(R,L_m)$ model,
\begin{multline}\label{3md}
\frac{n_B}{s}\simeq \left(\frac{-15 g_B}{4\pi ^2 g_{*s}M^2 _*  T_D} \right)\left(\frac{\left[(2\alpha-1)(\frac{\pi ^2}{30} g_{*s}T_D ^4 )^\alpha -\zeta \right]^\frac{3}{2}}{\sqrt{27} \beta ^3}\right) \left(12(\beta-2\beta ^2)-\frac{ 6\beta \left(\frac{\pi ^2}{30} g_{*s}T_D ^4 \right)^{1-\alpha}}{\alpha (2\alpha-1)}\right).
\end{multline}
We know $\beta=\frac{1}{2}$ in the radiation-dominated phase.
Hence, Eq. \eqref{3md} reduces to
\begin{equation}\label{3nd}
\frac{n_B}{s}\simeq \left(\frac{-15 g_B}{4\pi ^2 g_{*s}M^2 _*  T_D} \right)\left(\frac{8\left[(2\alpha-1)(\frac{\pi ^2}{30} g_{*s}T_D ^4 )^\alpha -\zeta \right]^\frac{3}{2}}{\sqrt{27} }\right) \left(-\frac{ 3 \left(\frac{\pi ^2}{30} g_{*s}T_D ^4 \right)^{1-\alpha}}{\alpha (2\alpha-1)}\right).
\end{equation}
As shown in Eq. \eqref{3nd}, the resulting baryon-to-entropy ratio is nonzero. The ratio in Eq. \eqref{3nd} can be adjusted to satisfy the observational constraints depending on the matter content. Still, the most interesting feature of a perfect fluid-dominated Universe for the case of $f(R,L_m)$ gravity baryogenesis is that the ratio is nonzero in the radiation-dominated case.
\begin{figure*}[htbp]
\centering
\includegraphics[scale=1]{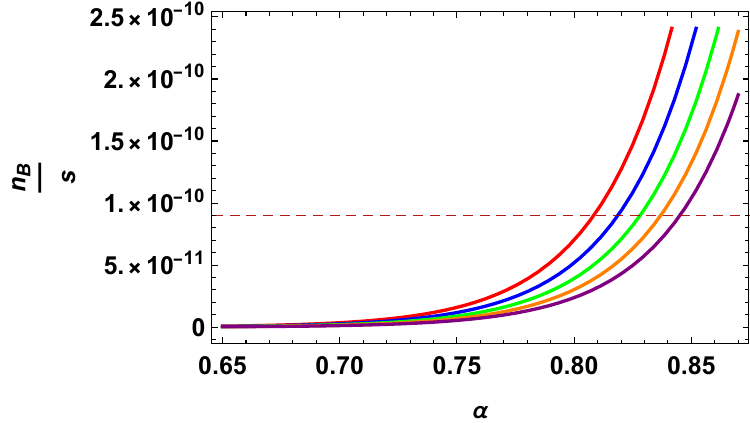}
\caption{Profile of the baryon-to-entropy ratio for the given model. The graphs are plotted for $\alpha$, for the varying values of $\beta$, $\beta=0.55$ (Red), $\beta=0.60$ (Blue), $\beta=0.65$ (Green), $\beta=0.70$ (Orange), $\beta=0.75$ (Purple), and $\zeta=2$. The dashed line represents the observational value.}\label{f1d}
\end{figure*}\\
Substituting $g_{*s} = 106$, $g_B =1$, $T_D = 2\times 10^{12} GeV$ and $M_* = 2 \times 10^{16} GeV$ \citep{Lamb}, with model parameters $\alpha=0.79$ and $\zeta=2$  in Eq. \eqref{3nd} the resultant baryon-to-entropy ratio reads $\frac{n_B}{s}\simeq 7.28749 \times 10^{-11}$ which is in excellent agreement with observations. The intersections of the curves in Fig. \ref{f1d} with the curve that reveals the observational value (dashed line) are in fine contract value of baryon to entropy ratio for particular values of $\alpha$ about including $0.75$ and $0.85$. Interestingly, when the parameter $0.5\leq \alpha \leq 0.69$, every curve tends to zero, consistent with theoretical results.

\subsection{Generalized gravitational baryogenesis}
We will now attempt to investigate the effects of a more comprehensive and generalized CP-violating interaction proportional to $\partial _\mu (f(R, L_m))$ rather than $\partial _\mu (R+L_m)$ in attempting to address the baryon asymmetry of the Universe for the chosen $f(R, L_m)=\frac{R}{2} + L_m ^\alpha + \zeta$  model. In $f(R,L_m)$ gravity, we can express the generalized CP-violating interaction as \citep{Noza}
\begin{equation}\label{4ad}
\frac{1}{M_*^2}\int{\sqrt{-g} J^\mu \partial_\mu (f(R,L_m)) d^4x}.
\end{equation}
The resulting baryon-to-entropy ratio for \eqref{4ad} is
\begin{equation}\label{4bd}
\frac{n_B}{s}\simeq \frac{-15 g_B (\dot{R} f_R+\dot{L_m}f_{L_m})}{4 g_{*s}M^2 _* \pi ^2 T_D}.
\end{equation}
Substituting Eqs. \eqref{3ed}, \eqref{3fd}, \eqref{3jd}, and \eqref{3ld} in Eq. \eqref{4bd}, we get the baryon-to-entropy ratio as
\begin{multline}\label{4cd}
\frac{n_B}{s}\simeq \left(\frac{-15 g_B}{4\pi ^2 g_{*s}M^2 _*  T_D} \right)\left(\frac{\left[(2\alpha-1)(\frac{\pi ^2}{30} g_{*s}T_D ^4 )^\alpha -\zeta \right]^\frac{3}{2}}{\sqrt{27} \beta ^3}\right) \left(6(\beta-2\beta ^2)-\frac{ 6\beta}{(2\alpha-1)}\right).
\end{multline}
We finally obtain, as discussed in the previous section, a radiation-dominated Universe by setting $\beta=\frac{1}{2}$, in Eq. \eqref{4cd}
\begin{equation}\label{4dd}
\frac{n_B}{s}\simeq \left(\frac{-15 g_B}{4\pi ^2 g_{*s}M^2 _*  T_D} \right)\left(\frac{8\left[(2\alpha-1)(\frac{\pi ^2}{30} g_{*s}T_D ^4 )^\alpha -\zeta \right]^\frac{3}{2}}{\sqrt{27}}\right) \left(-\frac{ 3}{4(2\alpha-1)}\right).
\end{equation}
By substituting $M_*$, $g_{*s}$, $g_B$, $T_D$ as before, $\alpha=0.93$, and $\zeta=2$, the baryon-to-entropy ratio obtained is $\frac{n_B}{s}\simeq 7.01\times10^{-11} $, which is also in excellent agreement with observations.
\begin{figure*}[htbp]
\centering
\includegraphics[scale=1]{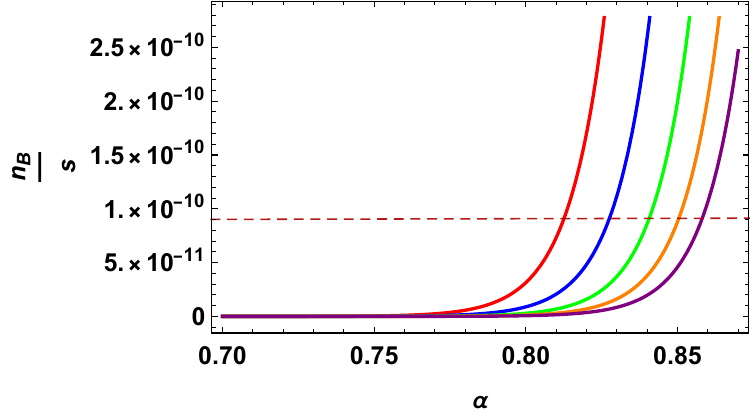}
\caption{Profile of the baryon-to-entropy ratio for the given model. The graphs are plotted for $\alpha$, for the varying values of $\beta$, $\beta=0.018$ (Red), $\beta=0.026$ (Blue), $\beta=0.036$ (Green), $\beta=0.046$ (Orange), $\beta=0.056$ (Purple), and $\zeta=2$. The dashed line represents the observational value.}\label{f2d}
\end{figure*}\\
We show $\frac{n_B}{s}$ for the generalized baryogenesis interaction as a function of $\alpha$ in Fig. \ref{2}.
The points where the curves in Fig. \ref{2} intersect with the observational value curve (dashed line) correspond to an acceptable range for the baryon-to-entropy ratio at specific values of $\alpha$, roughly around $0.79$ and $0.86$. Interestingly, when the parameter $0.6\leq \alpha \leq 0.79$, every curve tends to zero, consistent with theoretical results. Acceptable baryon-to-entropy values were obtained by $f (R, L_m)=\ \frac {R}{2} + L_m ^\alpha + \zeta$, which resulted in physically acceptable baryon-to-entropy ratios. Thus, in $f(R,L_m)$ gravity, the baryogenesis problem can be rectified.

\section{Conclusions}\label{sec4d}
\justifying
In this chapter, we explore the methodology of gravitational baryogenesis through the lens of $f(R,L_m)$ gravity theory. We evaluated the baryon-to-entropy ratio for a specific model given by $f(R, L_m)=\frac{R}{2}+L_m ^\alpha+\zeta$, which is based on the CP-violating interaction that generates BA in the cosmos. We analyze the matter content as a perfect fluid characterized by a constant EoS parameter $\omega$. Our findings indicate that, unlike in GR, the baryon-to-entropy ratio during a radiation-dominated era is nonzero for our model. Specifically, we determine the baryon-to-entropy ratio for a radiation-dominated cosmos as $\beta=\frac{1}{2}$. When the dynamics of the filled perfect fluid is governed by the theory of $f(R,L_m)$ gravity, we establish the baryon-to-entropy ratio to be $\frac{n_B}{s}\simeq 7.28749 \times 10^{-11}$, which aligns well with the observational value of $9\times 10^{-11}$.

We assumed that a scale factor of the form $a(t)= B t^{\beta}$, of $\beta = \frac{2}{3(1+\omega)}$, and $B$ is a constant parameter for this chapter. Because $\beta=\frac{1}{2}$ were the appropriate values required to obtain a viable baryon-to-entropy ratio, we assert that for such a scale factor, the power law part predominated at early times, which is consistent with observations \citep{Riess,Perlmutter}. Finally, we conclude our research by investigating a more complete and generalized baryogenesis interaction proportional to $\partial _\mu f(R, L_m)$. In this kind of interaction, our model produced a theoretical value in leading order of $\frac{n_B}{s}\simeq 7.01\times10^{-11} $, which is close to the observed value and the value obtained in the first case.


%% file: Chapters/Conclusion.tex

\chapter{Concluding Remarks and Future Perspectives} 

\label{Chapter7} 

\lhead{Chapter 7. \emph{Concluding Remarks and Future Perspectives}} 


In this work, we have explored the evolution of the Universe through the lens of $f(R,L_m)$ gravity theory, systematically addressing challenges in cosmology and theoretical physics. The findings demonstrate the versatility and compatibility of $f(R,L_m)$ gravity in describing late-time cosmic acceleration, baryogenesis, bouncing cosmology, and other cosmological phenomena. The study establishes $f(R,L_m)$ gravity as a promising framework to address outstanding cosmological questions by employing diverse observational datasets and advanced parameterization methods. The summary of the discussion, investigation, and the corresponding outcomes of each chapter are discussed below.

\section{Concluding remarks}\label{sec1g}
\justifying
In Chapter \ref{Chapter1}, we discussed the historical evolution of cosmological theories, from ancient perspectives to modern advancements, providing a foundation for research. We examined GR and its role in theoretical physics, together with the key features and challenges of the $\Lambda$CDM model. In addition, we explore the FLRW metric, redshift, the Hubble parameter, and the Friedmann equations. To address modern cosmological challenges, we introduced modified gravity, focusing on $f(R)$ gravity and its extension, $f(R,L_m)$ gravity, which coupled curvature and matter through the matter Lagrangian. Furthermore, we assessed the empirical validity of these theories through solar system tests, the geodesic deviation equation, and the Raychaudhuri equation, highlighting the effects of modified gravity on spacetime and particle motion. Finally, this chapter examined key cosmological observations that offered insights into the nature of the Universe and supported theoretical models.

In Chapter \ref{Chapter2}, we examined the late-time cosmic expansion within the framework of $f(R,L_m)$ gravity, focusing on the non-linear model $f(R, L_m) = \frac{R}{2} + L_m^n + \beta$, where $n$ and $\beta$ were free parameters. We derived the motion equations for the flat FLRW background and constrained the model parameters using the $H(z)$ and Pantheon datasets, as well as their combination. Our analysis determined optimal values for $n \approx 1.07$ and $\beta \approx -8862$, revealing a recent transition from deceleration to acceleration at redshifts $z_t\approx 0.68$. The density parameter and the deceleration parameter exhibited the expected behavior, and the stability analysis confirmed the robustness of the model under observational constraints. Furthermore, the Om diagnostic indicated that the model aligned with the quintessence scenario, consistent with the theoretical stability criterion ($n>0$).

In Chapter \ref{Chapter3}, we parameterize the Hubble function as $H(z)=H_0 \big[(1-\zeta)+(1+z)(1+\eta)\big]^\frac{1}{2}$, where $\zeta$ and $\eta$ were free model parameters. Using CC, BAO, and Pantheon + SH0ES datasets, we constrained the parameters to $H_0=71^{+0.081}_{-0.082}$, $\zeta=-0.36 \pm  0.033$ and $\eta=1.3 \pm 0.023$, which aligned well with the $\Lambda$CDM model. The evolution of the deceleration parameter showed a transition redshift of $z_t=0.6459$ and a present value of $q_0=-0.5249$, indicating consistency with the observed cosmic acceleration. We analyze two models: $f(R,L_m)=\frac{R}{2}+L_m^\alpha $ and $f(R,L_m)=\frac{R}{2}+(1+\lambda R)L_m$. Both models maintained positive density parameters, negative pressure, and quintessence-like behavior of the EoS parameters, transitioning to the $\Lambda$CDM model in the future through a phantom crossing. The SEC violation in both models supported late-time cosmic acceleration, with the non-minimal coupling model more effectively describing recent epochs. These results reinforced the viability of $f(R,L_m)$ gravity in explaining cosmic acceleration and encouraged further exploration of non-minimal coupling models.

In Chapter \ref{Chapter4}, we explore a model $f(R,L_m)$ with bulk viscosity, represented by the equation $f(R,L_m)=\frac{R}{2}+L_m^\alpha$, where $\alpha$ was treated as a free parameter. We adopted the effective EoS as stated in Eq. \ref{2ec}, which corresponded to the value of the Einstein case, utilizing the proportionality constant $\zeta$ seen in Einstein theory \citep{IB-1}, a concept commonly referenced in existing literature. We derived the exact solution for the model dominated by bulk viscous matter and constrained the parameters using the combined $H(z)$ + Pantheon+SH0ES datasets. The results yielded $\alpha=1.310^{+0.037}_{-0.032}$, $\gamma=1.29\pm0.20$, $\zeta=5.02\pm0.26$, and $H_0=72.09\pm 0.19$. The density parameter, pressure, and EoS evolution confirmed an accelerating expansion with $\omega_0\approx-0.71$. The statefinder parameters $(r,s)\approx(0.43,0.33)$ and the Om diagnostics indicated quintessence-like behavior. The energy conditions confirmed the positivity of NEC and DEC, while SEC violations in the recent past supported acceleration. This $f(R,L_m)$ model with bulk viscosity effectively explained late-time cosmic acceleration with observational consistency.

In Chapter \ref{Chapter5}, we explored bouncing cosmology within the framework of $f(R,L_m)$ gravity by analyzing two nonlinear models: $f(R,L_m)=\frac{R}{2}+L_m ^\beta +\gamma$, and $f(R,L_m)=\frac{R}{2}+\alpha L_m+\lambda R^2$, under the FLRW background with perfect fluid matter. These models incorporated corrections to the geometric and matter sectors. We then derived the density parameter, pressure, EoS parameter, and energy conditions, providing insights into the initial conditions of the Universe near the bounce. The density parameter increased to a peak before and after the bounce, while the pressure exhibited large negative values near the bounce. Finite values of the density parameter and pressure eliminated the singularity problem. The EoS parameter transitioned through the phantom phase ($\omega<-1$) near the bounce and moved to the quintessence phase ($\omega>-1$) far from it. NEC and SEC violations were observed near the bounce, as shown in Figs. \ref{f5e} and \ref{f6e}, while DEC remained positive. This repeated SEC violation aligned with late-time cosmic acceleration. Stability analysis using the squared speed of sound ($C_s^2$) revealed that Model I was unstable, whereas Model II demonstrated stability, making it a promising candidate for describing a nonsingular cosmic bounce.

In Chapter \ref{Chapter6}, we explored gravitational baryogenesis within the $f(R,L_m)$ gravity framework, focusing on the model $f(R,L_m)=\frac{R}{2}+L_m^\alpha+\zeta$. Assuming that the Universe behaved like a perfect fluid with a constant EoS parameter $\omega$, we assessed the baryon-to-entropy ratio. For a radiation-dominated era ($\beta=\frac{1}{2}$) with a scale factor $a(t)=Bt^\beta$ , our model yielded $\frac{n_B}{s}\simeq7.29\times10^{-11}$, closely aligned with the observed value of $9\times10^{-11}$. Extending to a generalized baryogenesis interaction proportional to $\partial_\mu f(R, L_m)$, we obtained $\frac{n_B}{s}\simeq7.01\times10^{-11}$, which further supported the observational consistency and the viability of our approach.

\section{Future perspectives}\label{sec2g} 
\justifying
Even though the thesis is mostly focused on the $f(R,L_m)$ type of non-minimally coupled gravity with curvature, there is a wide scope for generalizing these to various $f(Q,T)$, $f(\mathcal{T},T)$ or $f(Q,L_m)$, $f(\mathcal{T},L_m)$ types of gravity. In recent years, the modified gravity paradigm has expanded significantly with the inclusion of various non-standard connection-based gravity and its modifications, such as $f(Q)$, $f(\mathcal{T})$, etc., just to name a few. More phenomenologically sound models in this direction are expected to be proposed. In this context, it is worth investigating how non-minimal coupling with matter geometry via $L_m$ or $T$ would affect the overall cosmological scenarios. Additionally, as the current $\Lambda$CDM cosmology is facing various tensions, such as $H_0$ and $S_8$ tensions, non-minimally coupled gravity could be a useful alternative to explore. 

In the future, one can also seek to understand how to quantify these types of gravity or how these models can be reconstructed as an effective theory of some underlying quantum gravity (just as in $f(R)$ gravity). Overall, the future of non-minimally coupled gravity is very promising, and one could extend it in various directions where non-trivial coupling between modified gravity and matter could yield interesting cosmological solutions, potentially resolving the current standing $H_0$ and $S_8$ tensions.